\documentclass[twoside,a4paper,11pt]{oliverthesis}
\usepackage{a4wide,makeidx,t1enc,multido, graphicx, bbm, bm, amsmath, amssymb
}
\newcommand{\rf}[1]{(\ref{#1})}

\newcommand{\hc}{\ensuremath{\mbox{h.c.}} }

\renewcommand{\slash}[1]{\ensuremath{\makebox[0ex][l]{/}#1}}

\newcommand{\myslash}[2]{\ensuremath{\makebox[0ex][l]{\hspace{#2 ex}/}#1}}

\newcommand{\chii}{\raisebox{0.5ex}{$\chi$}}

\newcommand{\newparagraph}{\vspace{3ex}}

\newcommand{\Eq}[1]{Eq.~\rf{#1}}
\newcommand{\Fig}[1]{Fig.~\ref{#1}}
\newcommand{\Sec}[1]{Sec.~\ref{#1}}

\newcommand{\Lag}{\ensuremath{\mathcal{L}}}

\newcommand{\comment}[1]{}

\newcommand{\slA}{\myslash{A}{0.5}}
\newcommand{\sld}{\myslash{\partial}{0.08}}
\newcommand{\slD}{\myslash{D}{0.5}}
\newcommand{\slp}{\myslash{p}{0.08}}

\newcommand{\sla}[1]{\rlap{$#1$}/}

\newcommand{\msb}{\overline{\mathrm{MS}}}
\newcommand{\tr}[1]{\mathrm{Tr}\left\{ #1 \right\} }
\makeindex

\begin{document}
%\title{Ph. D. Thesis\hfill\\ Aspects of universal extra dimensional
 % models and their latticized scenarios}
%\author{Josep F. Oliver}
%\date{\today}
%\maketitle
\pagestyle{empty}
\begin{figure}
\begin{center}
\includegraphics[width=18cm]{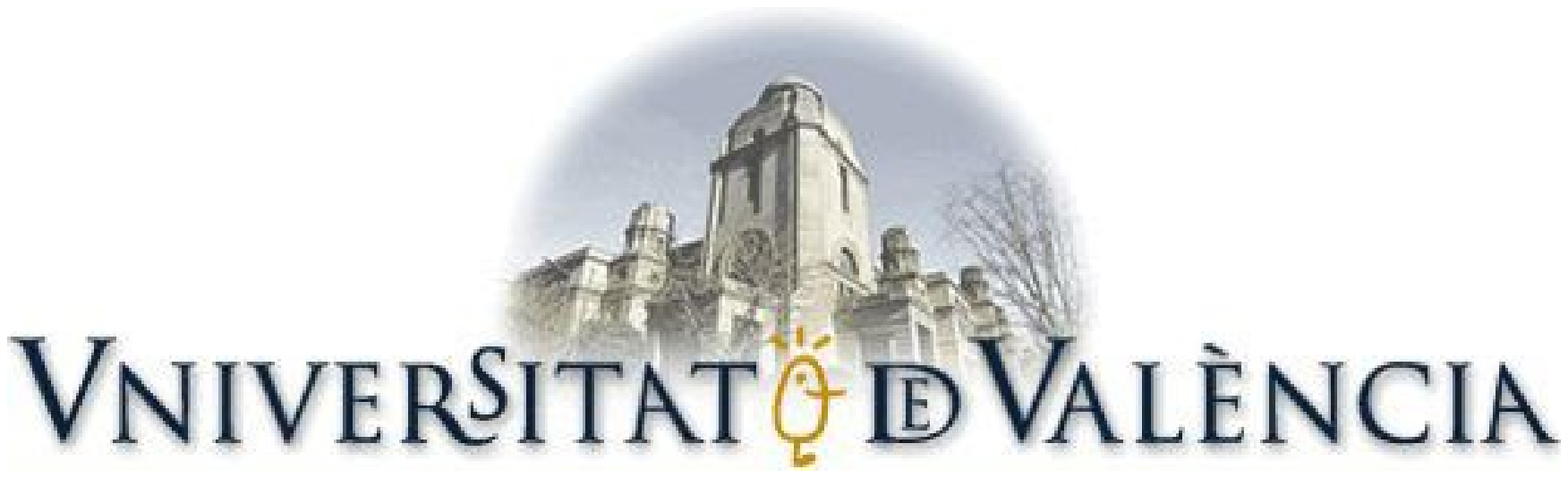} \vspace{3mm} 

\hspace{1.5cm} 

{\Huge \bf \textsf{Departament de Física Teòrica}} 
\includegraphics[width=17cm]{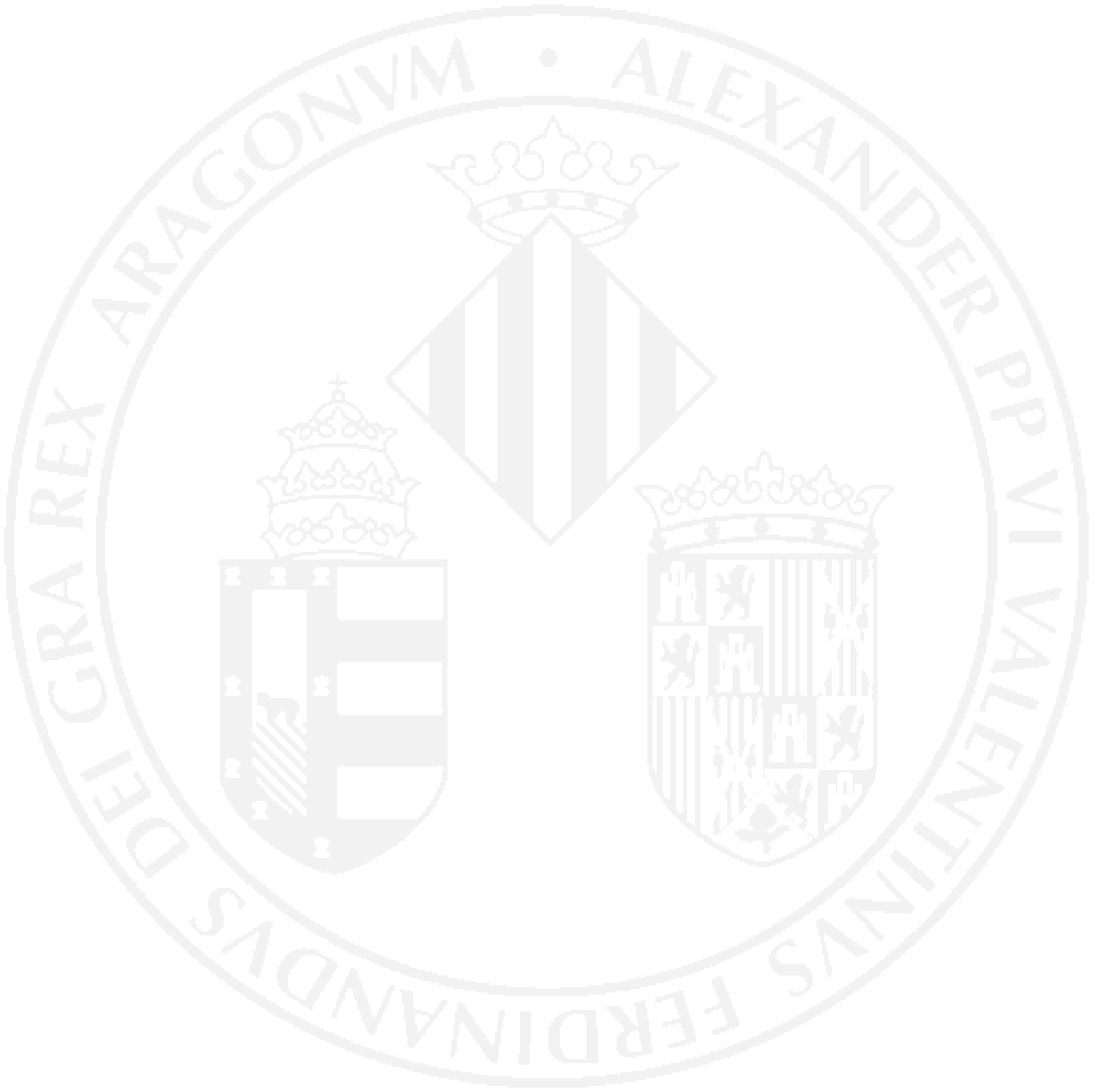}
\setlength{\unitlength}{1cm}
\put(-10.5,9.5){\LARGE Ph. D. Thesis}
\put(-16,6){\Huge \bf Aspects of universal extra dimensional}
\put(-15.5,4){\Huge \bf models and their latticized versions}
\put(-10.,-0.5){\LARGE   Josep F. Oliver}
\end{center}
\end{figure}
\cleardoublepage

\pagestyle{empty}
\mbox{}
\vskip 2.0 cm
{\bf ARCADI SANTAMARIA LUNA}, Profesor titular del Departamento de
F\'\i sica Te\'orica de la Universidad de Valencia y {\bf JOANNIS
PAPAVASSILIOU}, Investigador Contratado del Departamento de F\'\i sica
Te\'orica de la Universidad de Valencia,
\vskip 1.0 cm
{\bf CERTIFICAN} que la presente memoria {\bf ``Aspects of universal
  extra dimensional models and their latticized versions''} ha sido
  realizada bajo su direcci\'on, en el Departamento de F\'\i sica
  Te\'orica de la Universidad de Valencia, por D.~{\bf JOSÉ FRANCISCO
  OLIVER GUILLÉN}, y constituye su Tesis Doctoral para optar al grado
  de Doctor en F\'\i sica.
\vskip 0.5 cm 
Y para que as\'\i\ conste, en cumplimiento de la legislaci\'on vigente,
presenta 
ante la Facultad de F\'\i sica de la Universidad de Valencia la referida
memoria, y firma el presente certificado 
\vskip 0.5 cm
\begin{flushright}
en Burjassot, a 22 de Septiembre de 2003
\vskip 2.5 cm

Arcadi Santamaria Luna \hfill Joannis Papavassiliou
\end{flushright}
\cleardoublepage
\frontmatter
\ 
\vspace{7cm}

\hfill\  \parbox{20ex}{\textsl{\Large A mis abuelos con cariño}}

\cleardoublepage

\section*{\Huge Agraïments}
\thispagestyle{empty}
\newparagraph

\hspace{3.5ex}

Amb aquest treball s'acompleix un dels meus somnis: arribar algun dia
a ser doctor en física teòrica. Des de ben xicotet sempre m'ha
apassionat la física, en particular la física de partícules. El primer
contacte amb ella va ser a través de certs llibres de divulgació on
$W^\pm$ i $Z^0$ eren partícules exòtiques i misterioses fruit d'un
principi, en aquells moments inintel.ligible per a mi, anomenat de
Gauge. Eternament agraït li estaré a Ximo per haver-me'ls
mostrat. Retrospectivament ell és el gran culpable que jo estiga ací
escrivint aquests agraïments.

Vull aprofitar aquestes línies per agrair a tot el departament de
física teòrica tots els recursos que ha ficat al meu abast i el tracte
afectuós que sempre he rebut de tots els seus membres, Armando, Quico,
Toni, Nuria, Vicent, etc \ldots i especialment vull agrair a J. Bernabeu
l'interès que ha mostrat
%, que m'ha permès finalitzar aquesta tesi
%sense la pressió econòmica d'una beca que s'acaba
. També m'agradaria
destacar el treball i ajut de Joannis en aquesta tesi i en especial
l'esforç d'Arcadi al llarg d'aquests quatre anys, ha estat realment
un vertader privilegi poder treballar amb ell.

També vull donar les gracies pel seu acolliment i ajuda a la gent de
Durham: Sacha Davidson, José Santiago, Ignacio, Belen, Clara i
Verónica.

Des que vaig començar en física vells amics han estat al meu costat:
Mario, Javi, Oscar,\ldots, i he fet de nous: Pedro, José, Eva,
Guillem, Raquel, José Enrique, Isa, Juan, Dani, Daniele, Ezequiel,
Miguel, etc, \ldots entre ells voldria destacar a Verónica per la seua
inestimable ajuda i amistat. A Empar i Sergio, company de despatx, amb el que
sempre ha estat un autèntic plaer raonar de física. Especialment vull
mencionar a David i Miguel, amics inseparables de col.legi, institut,
carrera i doctorat, companys d'hores i hores de biblioteca, que ara
comencen noves etapes de les seues vides amb Nuria i Maria. També
voldria agrair a Bea i tota la seua família l'acceptació i suport que
m'han donat al llarg de tot el temps.

Tot aquest treball hagués estat impossible sense el suport
incondicional de la meua família, germans i pares, i des d'aquestes
línies agraïsc als meus pares que m'hagen donat l'oportunitat de saber
fins on podia arribar, en especial a ma mare que a pesar de tot el que
ha hagut de passar en els últims temps sempre ha estat al nostre
costat, servisquen aquestes paraules com a reconeixement. Amb especial
afecte vull agrair a la meua companya de balls, viatges i il.lusions
tot el que ha fet per mi, especialment en aquests últims i negres
anys. Seu ha estat el mèrit d'alçar-me, d'espentar-me en la direcció
correcta per tal que jo pugés arribar a aquest punt on ara em
trobe. Aquestes línies no són sino una pobra compensació.

Finalment, i en agraïment a tot el que ells m'han donat, dedique el
present treball als meus ``iaios''. Desgraciadament, encara que em van
vore començar-lo, ja mai no podran vore com l'acabe. Des d'ací, un
abraç.

%\vfill 

\begin{flushright}
Burjassot, a 23 d'Octubre de 2003

%\scalebox{0.15}{\includegraphics{mifirma.eps}}

\textit{Josep F. Oliver \hspace{50ex}}
\end{flushright}
\cleardoublepage
\pagebreak
\pagestyle{headings}
\tableofcontents
\setcounter{secnumdepth}{4}
%\pagestyle{plain}
%\pagenumbering{roman}
%\setcounter{secnumdepth}{0}
%\include{Titol}
%\include{preambul}

%%%%%%%%%%%%%%%%%%%%%%%%%%%%%%%%%%%%%%%%%%%%%%%%%%%%%%%%%%%%%%%%%%%%%%%
%%%%%%%%%%%%%%%%%%%      CHAPTERS         %%%%%%%%%%%%%%%%%%%%%%%%%%%%%
%%%%%%%%%%%%%%%%%%%%%%%%%%%%%%%%%%%%%%%%%%%%%%%%%%%%%%%%%%%%%%%%%%%%%%%

\mainmatter
\nocite{Oliver:2001eg}%nomes per que aparega el  nostre primer paper.
%Thu Nov 27 09:03:16 CET 2003

\chapter{Introduction}
The possible existence of more dimensions than those directly
accessible to our senses has always fascinated mankind. They have
spurred the imagination of many writers who have speculated about how
to reach them, their nature and possible applications. Interesting as
it would be, we will not take this path in this work; instead we will
divert our attention to a more scientific point of view, namely what
can we say about extra dimensions by using the scientific knowledge we
have accumulated trough time. In the next lines we show, very briefly,
the origin and evolution of the ideas related with extra
dimensions. Finally, we sketch their present state to situate our work
in context\footnote{Part of the contents of this section has been
extracted from \cite{Appelquistllibre}}.

The extra dimensions are a relatively old idea among the scientific
community. Its origins can be traced back to the year 1912 when the
Finnish physicist Gunnar Nordström proposed a relativistic extra
dimensional theory that described simultaneously gravity and
electromagnetism. At that moment the general theory of relativity was
not developed and therefore the geometric origins of gravity were not
unveiled. Before Einstein, Nordström had developed a relativistic
theory of gravitation based on the existence of a scalar potential,
$\phi$ ; he unified this theory with the electromagnetism.

The ideas of Nordström were strongly influenced by Maxwell's theory of
electromagnetism, which unified elegantly electric and magnetic
phenomena by describing the electric and magnetic fields as components
of a single six-component antisymmetric tensor, $F_{\mu\nu}$, while
the corresponding potentials were unified into a four dimensional
vector, $A_\mu$. It became clear through Minkowski's work that the
unification of electricity and magnetism entailed a unification of
space and time in a four dimensional \emph{space-time}. Nordström
followed this reasoning and added an extra dimension to this
space-time; the vectors of this new manifold allowed for one more
scalar field that Nordström proposed to be precisely his gravity
potential, $\phi$. The action was taken to be the five dimensional
version of electromagnetism, now built from the antisymmetric tensor
$F_{\alpha\beta}$, where $\alpha,\beta=0,1,\ldots,4$. Of course, the
fifth dimension needed to be different from the rest, hence it was
assumed to have a topology drastically different compared with the
other four, and it was compactified on a circumference, i.e. the values
of the coordinates in the fifth dimension were restricted. This
topology allowed a Fourier expansion of the fields and when only the
fundamental mode of each field was considered a four dimensional
theory of electromagnetism and gravity emerged. In other words,
Nordström showed that gravitation and electromagnetism could be
understood as two different faces of a five dimensional
electromagnetism.

It was in 1915 when Einstein proposed his awe-inspiring relativistic
theory of gravitation, the "\emph{general theory of relativity}", in
which gravity is understood as geometrical deformations of the
underlying space-time. In 1919, the mathematician Theodor Kaluza
showed that a five dimensional theory of gravity, with the action
taken as the Einstein-Hilbert action, would manifest itself as the
electromagnetism and the gravity in a four dimensional world. In 1926,
the same year Schrödinger proposed his famous equation, Oskar Klein
and the Russian physicist H. Mandel independently rediscovered
Kaluza's theory. They hoped this theory would underlie the quantum
theory that at that moment was still under construction.

These first steps into the fifth dimension were rather hesitant. It
was viewed as a mathematical trick that allowed a more concise
formulation of the laws of Nature but was completely void of any
physical interpretation. In addition, it posed a number of problems:
the correct four dimensional limit was recovered when only the
fundamental Fourier mode was retained, the so called
\emph{cylindricity condition}, and it was unclear the role of the rest
of the modes, the reason why some fields must be constant was also
obscure, just to cite some.

The discovery of new interactions, other than electromagnetism and
gravitation, complicated more the overall picture: using a single
extra dimension as a means of reaching a unified description was
unnatural, because it was not able to accommodate the strong and weak
forces.  The latter were described by a class of theories called
non-Abelian gauge theories, proposed by Yang and Mills in 1954, which
became widely accepted in the seventies.  Yang-Mills theories could be
incorporated within an extra dimensional framework, at the price of
extending the number of additional dimensions; but this could not be
done straightforwardly, and posed a number of difficulties that had to
be overcome. At this point the extra dimensions were still useful for
grouping together equations in a unified mathematical framework, but
had acquired a great degree of complexity, while not being predictive
and presenting serious theoretical problems.

In the next years the interpretation of the extra dimensions changed,
in the sense that they were given a physical meaning. It was due to
the development of new theories; \emph{supergravity} and \emph{string
theory}, where the extra dimensions played a key role. Both kind of
theories provided a promising framework for achieving a quantum
description of gravity. The natural energy scale for these theories is
the Planck mass, $1.2\;10^{19}~\mbox{GeV}\cdot c^2$, that is
completely out of reach for the current particle accelerators.
Nevertheless, in some string scenarios this energy scale can be as low
as a few TeV what suggests that the associated phenomenology can be
more accessible to observation.

In recent years, extra-dimensional quantum field theories have
received a great deal of attention. On one hand, the recent interest
is because the scale at which the extra dimensional effects can be
relevant could be around a few TeV, even hundreds of GeV in some
cases, clearly a challenging possibility for the next generation of
accelerators. On the other hand, this new point of view has permitted
to study many long-standing problems in physics from this new
perspective. These problems cover many different fields of particle
physics: the hierarchy problem, new neutrino physics, the masses of
the fermions, the number of generations in SM, possible modifications
in the running of the coupling constants, new candidates for dark
matter, etc.

Extra dimensional theories offer a wide variety of scenarios and
therefore have a rich phenomenology. For instance, in some scenarios
(\emph{large extra dimensions}) only the gravity field can probe the
extra dimensions. In others gravity is not considered and only boson
fields are allowed to propagate through the extra dimensions. Another
possibility is to allow all the fields present in the theory to feel
the extra dimensions (\emph{universal extra dimensions}). 

One of the reasons why this latter scenario is particularly
interesting is because the corrections to the SM predictions appear
for the first time at the one-loop level. As a consequence, the
modifications to precision observables are small, what implies that
the scale of this theory can be as low as hundreds of GeV. The fact
that these models do not give any tree-level contribution, is due to
the conservation of the so called Kaluza-Klein (KK) number (strictly,
it is not a conserved number in the usual sense). Most of this thesis
is devoted to study the phenomenology of theories with one universal
extra dimension.

In particular, in chapter \ref{toymodels} we show how to treat the
different fields (scalars, fermions and vector bosons) in an
extra-dimensional quantum field theory formalism. The concept of
dimensional reduction is introduced and we show how to obtain a four
dimensional Lagrangian from the expression of the Lagrangian in
$4+\delta$ dimensions. We also show how the extra dimensional fields
are transformed into an infinite number of four-dimensional fields
(the so called KK towers) with the same quantum numbers associated to
all. By studying the low energy limit of these theories we stress the
relevance of selecting a suitable topology for the compactified
dimensions since different topologies have associated different
degrees of freedom in this limit. We show that the orbifold topology
selects the correct low energy degrees of freedom; specifically, it is
possible to obtain chiral fermion fields and to remove the extra
dimensional components of the vector fields in the low energy
spectrum.  Interacting theories are studied to demonstrate explicitly
the KK number conservation, which is related to the local
extra-dimensional Lorentz invariance of the theory and has a deep
impact on the phenomenology of theories with universal extra
dimensions.

In chapter \ref{sec:SMinUED} we use the ideas developed previously to
construct an extra dimensional model that reduces to SM in the
low-energy limit. We study the phenomenology of this model and focus
on the observables that display a strong dependence on the mass of the
top-quark because in this case the deviations from the SM predictions
are more important. We compute the radiative corrections for the $Z\to
b \overline{b}$ decay, $b\to s\gamma$, the $B^0-\overline{B}^0$ mixing
and the $\rho$ parameter, and study their consequences.

In chapter \ref{chap:LUED} we construct in detail the latticized
version of the previous model, i.e. the version in which the extra
dimension is discretized. Latticized as well as deconstructed models
were devised as ultraviolet completions of the extra dimensional
models. The latter are not renormalizable because the coupling
constants have dimensions of mass to some negative power. This is
suggesting that they must be understood as low energy effective
manifestations of a more complete theory. Models with deconstructed
extra dimensions are usual four-dimensional theories, which, due to
the special nature of the interactions present in the Lagrangian,
display an extra dimensional behaviour in certain range of
energies. These kind of models have received a great deal of interest
because in some of their extensions the Higgs boson is a
pseudo-Goldstone boson, what would explain why it is so light and
stable against radiative corrections. This models are very similar to
latticized models. The latter are still non-renormalizable but they
can be understood as usual four-dimensional $\sigma$-models, and all
the known possible ultraviolet completions for the $\sigma$-models can
be applied now. In this thesis we study part of the phenomenology of
the models with one latticized extra dimension.

The next chapter is devoted to investigate the modification of the
running of the coupling constants in models with (continuous) extra
dimensions. It has been pointed out that in extra dimensional theories
the running can be accelerated, i.e. the dependence with the energy
scale is not logarithmic, as usual, but can be power-like.  This
change could have as a major consequence that the unification of the
three interactions could be achieved at a very low scale, of order a
few TeV.  We have studied in some detail, by resorting to simplified
models, how reliably this power corrections can be computed without
knowing the details of the more complete theory, i.e. the theory valid
above the scale at which the extra dimensional theory ceases to be
correct.  We have found that the coefficients that govern this power
corrections are sensitive to the details of the theory in which the
extra dimensional theory is embedded.  This is a completely different
result compared with the situation in usual grand unification
scenarios where the unification of the gauge coupling constants can be
tested without knowing the details of the Grand Unification Theory.

Finally, in chapter \ref{chap:HGLHG} we study a model with a
non-universal extra dimension. In this case only the boson fields are
allowed to propagate through the extra dimension, due to this the
extra-dimensional Lorentz symmetry is broken and the KK number
conservation rule does not apply. The results are compared with those
obtained when the extra dimension is universal to show explicitly the
importance of the KK number conservation. The bound on the
compactification scale is clearly higher for the kind of models that
lack this extra-dimensional Lorentz symmetry because the deviations
from the SM predictions appear already at the tree-level.

%endcomment
%%%%%%%%%%%%%%%%%%%%%%%%%%%%%%%%%%%%%%%%%%%%%%%%%%%%%%%%%%%%%%%%%%%%%%%
%%%%%%%%%%%%%%%%%% Thu Sep 25 11:23:02 CEST 2003%%%%%%%%%%%%%%%%%%%%%%%
%%%%%%%%%%%%%%%%%%%%%%%%%%%%%%%%%%%%%%%%%%%%%%%%%%%%%%%%%%%%%%%%%%%%%%%

\chapter{Quantum field theory with one universal extra dimension}
\label{toymodels}
\comment{
{\sffamily \scshape
\begin{center}
\begin{tabular}{|l|r|}
\hline
Text         & Ok      \\\hline
Gauge Piece  & Ok      \\\hline
Orthography  & Ok      \\\hline
Figures      & Ok      \\\hline
Links        & Ok      \\\hline
Cites        & Ok      \\\hline
Meaning      & --      \\\hline
Makindex     & --      \\\hline
Date         & \today  \\\hline
\end{tabular}
\end{center}
}
}%endcomment
%%%%%%%%%%%%%%%%%%%%%%%%%%%%%%%%%%%%%%%%%%%%%%%%%%%%%%%%%%%%%%%%%%%%%%%
%%%%%%%%%%%%%%%%%%%     Introduction     %%%%%%%%%%%%%%%%%%%%%%%%%%%%%%
%%%%%%%%%%%%%%%%%%%%%%%%%%%%%%%%%%%%%%%%%%%%%%%%%%%%%%%%%%%%%%%%%%%%%%%
In this chapter we study the main features of theories with one
additional space dimension accessible to all fields, called universal
extra dimension. To this end we study a number of toy models which we
will use to show how to treat scalar, spinor and vector fields in five
dimensions. We address the issue of compactification in this theories
and study two different topologies: a sphere, $S^1$, as well as an
\emph{orbifold}, $S^1/Z_2$. It is shown that in the process, called
\emph{dimensional reduction}, one can trade the extra dimension for an
infinite \textit{tower} of fields, called Kaluza-Klein (KK) tower or
KK modes. The different topologies provide different low energy
theories even when one starts from the same five dimensional
Lagrangian. We will show that the advantage of compactifing in an
orbifold is double: on one hand, only four of the five components of
the vector field are present in the low-energy spectrum, on the other,
it can contain chiral fermions. This opens the door to identifying the
SM with the low energy realization of an extra dimensional
theory. Instead of studying possible extra dimensional extensions of
the SM, we first propose some simple interactions to gain some insight
into the properties of these theories while keeping the model as
simple as possible. It is found that a new kind of conserved number
appears, the \emph{KK number}. It cames from the fact that the
theories are locally invariant under the Lorentz group in five
dimensions. Strictly, it is not conserved in the usual sense and
therefore we study it in some detail. Its main contribution is to
screen, to some extent, the impact of the KK towers in the low-energy
effective theory. 

\section{Fields and interactions in five dimensions}
 In particle physics each particle is associated to the quanta of a
field defined in the Minkowski space-time $\mathcal{M}^4$. The
coordinates in this manifold are written as $x^\mu$ where
$\mu=0,\ldots,3$. To extend this formulation to more dimensions one
must define fields that depend on $4+d$ coordinates, say
$\psi(x^\alpha)$, where $\alpha=0,\ldots,3+d$. All the extra
coordinates are supposed to be associated with spatial dimensions,
therefore the metric takes the form
$g_{\alpha\beta}=(+,-,\ldots,-)$. Once this is done, a topology for
the additional dimensions must be selected. This choice has important
consequences in the low-energy spectrum. In the next sections all this
process is performed in detail for different kind of fields.

\subsection{Scalar field with self interaction}
Let us define a complex scalar field $\Phi(x^\alpha)$ that depends on
the $4+d$ coordinates $\alpha = 0,1,\ldots,d$. The action is defined
in the usual way through the standard Klein-Gordon Lagrangian density
\begin{equation}
\label{eq:action} S=\int d^{4+d}x \;\mathcal{L}^{4+d}(x^\alpha).
\end{equation}
For a complex scalar field
\begin{equation}
\label{eq:lagri}
\mathcal{L}^{4+d}=(\partial_\alpha\Phi)^\dagger(\partial^\alpha\Phi) -
m^2\Phi^\dagger\Phi,
\end{equation}
The first consequence is that the canonical dimension of the field
gets modified, now $[\Phi]=E^{1+d/2}$, what will be important when
studying the renormalization of theories with extra dimensions. The
next step to do is to specify the topology of the extra
dimensions. The simplest choice is to associate a circumference,
$S^1$, to each one, i.e. the full manifold is a direct product of the
Minkowski space and $d$ circumferences,
$\mathcal{M}=\mathcal{M}^4\times (S^1)^d$. This means that the extra
coordinates are periodic with a periodicity of $2\pi R$, assuming the
same \emph{radius}, $R$.  From this it follows that $\Phi$ can be
expanded on its Fourier modes
\begin{equation}
\label{eq:Fourierdephi}
\Phi(x^\mu,\vec{x})=\sum_{n_1,\ldots,n_d=-\infty}^{\infty}\phi_{\vec{n}}(x^\mu)e^{i\vec{n}\vec{x}/R}
\end{equation}
where $\vec{x}=(x^4,\ldots,x^{d-1})$ is a vector whose components
are the coordinates in the extra dimensions and
$\vec{n}=(n_1,\ldots,n_d)$ identifies unambiguously each Fourier
mode. By using \Eq{eq:Fourierdephi} the integration over the extra
coordinates in \Eq{eq:action} can be performed
\begin{equation}
\label{eq:dimred}
 S=\int_{-\infty}^{\infty} d^4x\int
d^dx \;\mathcal{L}^{4+d}\equiv\int_{-\infty}^{\infty}
d^4x\;\mathcal{L}.
\end{equation}
This shows that this theory can be described by a four dimensional 
Lagrangian related with the original one by the equation
\begin{equation}
\mathcal{L}=\int d^dx\;\mathcal{L}^{4+d}.
\end{equation}
This process receives the name of \emph{dimensional reduction}. It is
independent of the kind of field(s) (scalar, fermion, vector, etc...)
that $\mathcal{L}^{4+d}$ contains, it only depends on our ability to
perform the integration over the coordinates of the extra dimensions.

In the case we are studying this process leads to
\begin{equation}
  \label{eq:lagdef}
  \mathcal{L} = \sum_{\vec{n} =-\infty}^{\infty} (\partial_\mu
  \phi^{(\vec{n})})^\dagger (\partial^\mu \phi^{(\vec{n})}) -
  (m^2 + m_{\vec{n}}^2)
  \phi^{(\vec{n})\dagger} \phi^{(\vec{n})}\qquad \qquad
  m_{\vec{n}}^2 = \vec{n}^2/R^2.
\end{equation}
To obtain canonical kinetic terms, the original fields,
$\phi^{(\vec{n})}$, must be redefined: $\phi^{(\vec{n})}\to (2\pi
R)^{-1}\phi^{(\vec{n})}$. \Eq{eq:lagdef} shows one of the most
important features of this kind of theories, specifically, the extra
dimensions have been traded for an infinite tower of fields, called
\emph{Kaluza-Klein tower} or \emph{KK tower}, with increasing
masses. The lowest mass, the mass of the fundamental mode,
$\phi^{(\vec{0})}$, is the one appearing initially in $\mathcal{L}^{4+d}$
and in principle is completely independent of $R$ and insensitive to
the compactification procedure. In particular, it could be much lower
than $R^{-1}$ or even zero.  Notice the degeneracy in the spectrum,
except for the fundamental mode.

\comment{
 This has to do with the equality
$m_{\vec{n}}=m_{-\vec{n}}$. There will be additional degenerations
between different modes that satisfy that $\vec{n}^2=\vec{n}^{\prime
2}$.
% For big $\vec{n}$ it can be easily
%estimated that the number of states with mass $m_{\vec{n}}$ is
%approximately given by $N_{\vec{n}}=4\pi \vec{n}^2$, growing as
%$\vec{n}^2$.

It is interesting for seeing the consistency of this model to study
what happens if the initial field, $\Phi(x^\alpha)$, is taken to be
real. The degrees of freedom $\phi^{(\vec{n})\ast}$ and
$\phi^{(-\vec{n})}$ are no longer independent because for a real field
$\phi^{(\vec{n})\ast} = \phi^{(-\vec{n})}$ and therefore
\Eq{eq:lagdef} can be rewritten in terms of the new \emph{real} fields
$\chii^{(\vec{n})}$ and $\eta^{(\vec{n})}$
\begin{equation}
\chii^{(\vec{n})} = \frac{\phi^{(-\vec{n})}+\phi^{(\vec{n})}}{2}\qquad
\eta^{(\vec{n})} = i\frac{\phi^{(-\vec{n})}-\phi^{(\vec{n})}}{2}
\end{equation}
as
\begin{equation}
  \mathcal{L}=\frac{1}{2}[(\partial\phi^{(\vec{0})})^2 -
    m^2\phi^{(\vec{0})2}] + \sum_{\vec{n}=1}^{\infty}
\frac{1}{2}[(\partial\chii^{(\vec{n})})^2-(m^2 + m_{\vec{n}}^2)\chii^{(\vec{n})2}]
    +\sum_{\vec{n}=1}^{\infty}\frac{1}{2}[(\partial\eta^{(\vec{n})})^2
    -(m^2 + m_{\vec{n}}^2) \eta^{(\vec{n})2}].
\end{equation}
This shows explicitly the existence of two degenerate towers of
real scalar fields, of course for the case of $d>1$ there will be more
degeneration within each tower.
}%endcomment

Another important topology for the extra dimensions is the so called
\emph{orbifold}. It is a bit more complicate than $S^1$ and for the
sake of simplicity we will concentrate on the case of a single extra
dimension, which is the important case for this thesis. Its relevance
will be manifest in the next sections when we study the spinor and
gauge fields. From now on, when only a single extra dimension is
present its coordinate, $x^4$, will be denoted by $y\equiv x^4$. The
topology of the space-time is now $\mathcal{M}^4 \times
(S^1/Z^2)$. $S^1$ means that the extra dimension is again periodic and
$Z_2$ reflects the fact that the points $-y$ and $y$ are
identified. The orbifold is schematically represented in the figure.
%\vfill\eject

\begin{minipage}{5cm}
\includegraphics[scale=0.30]{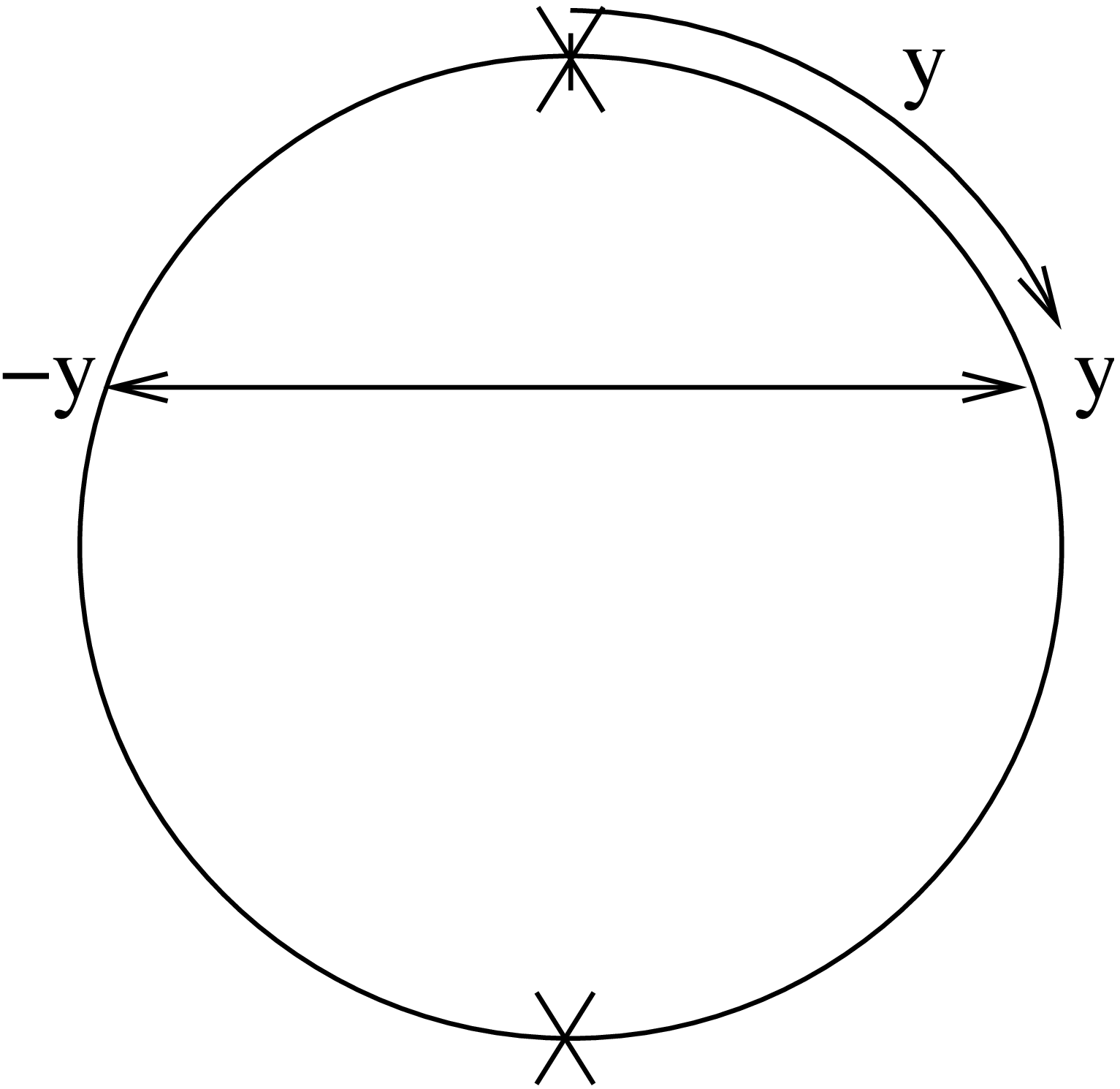}
\end{minipage}
\hfill
\begin{minipage}{9cm}
The crosses in the figure represent two special points, called
\emph{fixed points}, that are mapped onto themselves under the
orbifold symmetry: $y\to -y$. When we say that these points are
identified we mean that the values of the fields in them are related,
i.e. $\Phi(-y)=U\Phi(y)$, where $U$ is a unitary transformation that
is a symmetry in the original Lagrangian $\mathcal{L}^{4+d}$. As a
result, the physics on one side and on the other is exactly the same,
or more formally, the action can be computed restricting the
integration to the interval $y\in [0,\pi R]$
\begin{equation}
S=\int d^4x\int_{0}^{2\pi R}dy\;\mathcal{L}^{5}
=2\int d^4x\int_{0}^{\pi R}dy\;\mathcal{L}^{5}.
\end{equation}
\end{minipage}

If it is further imposed that $U^2=\mathbbm{1}$, then for a scalar
field it is perfectly valid the choice $U=\pm 1$. This extended
symmetry imposes further structure to the fields, for one extra
dimension the $S^1$ topology implies that a field can be expanded
as\footnote{Of course this is the \rf{eq:Psi}, where the exponentials
have been expressed in terms of sines and cosines.}
\begin{equation}
\Phi(x^\mu,y)=
\phi^{(0)}(x^\mu) + \sum_{n=1}^{\infty}
\phi^{(n)+}(x^\mu)\cos\left(\frac{ny}{R}\right) +
\sum_{n=1}^{\infty}
\phi^{(n)-}(x^\mu)\sin\left(\frac{ny}{R}\right).
\end{equation}
The orbifold topology requires that the fields are even or odd under
the orbifold parity transformation ($y\rightarrow -y$), calling
$\Phi^+$ and $\Phi^-$ respectively
\begin{equation}
  \begin{array}{lcr}
    \Phi^+(-y) & = &+\Phi^+(y),\\
    \Phi^-(-y) & = &-\Phi^-(y).
  \end{array}
\end{equation}
Their expansions are now
\begin{eqnarray}
  \label{eq:even}
  \Phi^+(x^\mu,y) & = &\phi^{(0)}(x^\mu) + \sum_{n=1}^{\infty}
  \phi^{(n)+}(x^\mu)\cos\left(\frac{ny}{R}\right), \\ \Phi^-(x^\mu,y) & =
  &\sum_{n=1}^{\infty}\phi^{(n)-}(x^\mu)\sin\left(\frac{ny}{R}\right).
\end{eqnarray}
This is of great importance, since only the even fields have a
fundamental mode. Recall that the mass of the fundamental mode was
only determined by the mass in the original Lagrangian and can be in
principle as low as desired. On the contrary, the lower available mass
for the odd modes is $R^{-1}$. Finally, taking $\mathcal{L}^5 =
(\partial_\alpha \phi)^\dagger(\partial^\alpha \phi)$
\begin{equation}
\mathcal{L}=\int_0^{\pi
R}\mathcal{L}^5(\Phi^+)=\sum_{n=0}^{\infty}\frac{1}{2}[(
\partial_\mu \phi^{(n)})^\dagger  ( \partial^\mu \phi^{(n)})-m_n^2
\phi^{(n)\dagger}
\phi^{(n)}],
\end{equation}
while for $\Phi^-$ the expression is the same without the fundamental
mode, hence the low-energy spectrum is radically different. It is
worth to stress that these theories had a clear separation between
low-energy and high-energy regimes due to the existence of a natural
energy scale, $1/R$, which, excluding the zero mode, is the mass of
the lightest mode. The particles we know could be identified with the
fundamental mode of an even field, the smallness of $R$ would explain
why no KK mode has yet been detected.  If this idea is true then we
can make a rough estimate of the size of the extra dimension,
$R^{-1}>200~\mbox{GeV}$ because this is the highest energy directly
probed by accelerators. Of course, to obtain a serious bound it is
required a more evolved model and the careful study of radiative
contributions to precision observables since these KK modes would also
modify SM predictions via virtual exchanges in loops. This detailed
study is the main aim of this work.

\begin{figure}
\begin{center}
\includegraphics[scale=0.30]{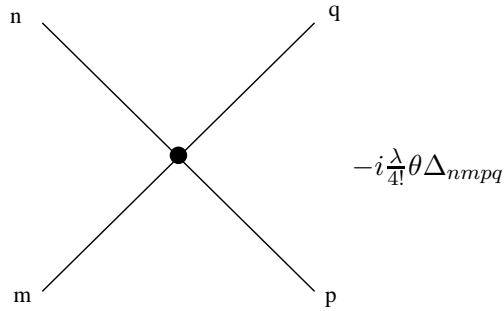}
\begin{picture}(100,100)
%\mygrid{10}{100}{10}{100}
\put(0,50){\mbox{$-i\frac{\lambda}{4!}\theta \Delta_{nmpq}$}}
\end{picture}
\end{center}
\caption{Feynman Rule for $\Phi^4$ theory.}
\label{fig:FR}
\end{figure}

Up to this point only the free part of the theory has been
investigated. As an example of interaction we will use a
five dimensional $\Phi^4$ self-interaction in an orbifold; therefore
we add to the Lagrangian the next interaction term
\begin{equation}
\label{eq:phicuatro}
 \Lag^5_I=-\frac{\tilde{\lambda}}{4!}\Phi^{4},
\end{equation}
where $\Phi$ is assumed to be a real and even field. Notice that
$\tilde{\lambda}$ is not dimensionless, it is easy to derive that
in general $[\tilde{\lambda}]=E^{-d}$. By using the decomposition
given in \Eq{eq:even} one finds after dimensional reduction that
the Lagrangian of this theory can be written as
\begin{equation}
\Lag=\sum_{n=0}^{\infty}\frac{1}{2}[
  \partial_\mu\phi^{(n)} \partial^\mu\phi^{(n)} - m_n^2\phi^{(n)}\phi^{(n)}] -
  \sum_{n,m,p,q=0}^{\infty}
\frac{\lambda}{4!}\;\Theta_{nmpq}\;\phi^{(n)}\phi^{(m)}\phi^{(p)}\phi^{(q)} ,
\end{equation}
where a new dimensionless coupling constant has been defined,
$\lambda=\tilde{\lambda}/\sqrt{\pi R}$. The function $\Theta$ is
decomposed as the product $\Theta_{nmpq}=\theta \Delta_{nmpq}$.  The
first is just a numerical factor due to the fact that the fundamental
mode has different normalization than the rest, it depends on the
number of fundamental modes present in the vertex, $f$, as
$\theta=2^{-\frac{3f+1}{2}}$. The second is more interesting because
it forbids certain combinations of indices
\begin{equation}
\label{defDelta}
\Delta_{nmpq}=\left\{\begin{array}{cl}
1 &
%\mbox{if any of the possible combinations in }
\pm n \pm m \pm p \pm q=0\\
0 & \mbox{otherwise}
\end{array}\right.
\end{equation}
i.e. if any of the possible combinations is zero then the vertex
exists otherwise it is forbidden. The Feynman rule for the vertex of
the theory is otherwise straightforward, see \Fig{fig:FR}.
One of the most interesting properties of the interaction is that
there is no vertex that couples three fundamental modes with one
excited. This means that to create particles with $n\geq 1$ from the
particles of the fundamental mode they must be created in pairs, for
instance through the process $\phi_0 \phi_0\to \phi_n
\phi_n$. Therefore the threshold for creating the new particles is a
least twice the mass of the first mode, $\sqrt{s}\geq 2m_1$. The
situation is reminiscent of other occasions in physics where also new
particles had to be created in pairs, as for instance the creation of
the charm quark. In that case there was a symmetry behind: the charm
quark had to be created via strong interactions which are
flavour-symmetric; thus, in order to create a charm quark it was
necessary to create also an anticharm quark.

In extra dimensions the pair production can also be related to a
symmetry: the local five-dimensional Lorentz symmetry of the tree
level Lagrangian. However, this symmetry is broken by the
compactification. This breaking is a non-local effect.  Since the
vertices describe always local interactions, all of them must conserve
momentum. This is not true for correlators of the theory, as for
instance the propagators or in general any n-point Green function,
because they are extended objects. From \Eq{eq:even} one can see that
each mode $\phi^{(n)}$ is associated with a stationary wave in the
fifth dimension (the cosine in its exponential form contains
$e^{i\frac{n}{R}y}$ and $e^{-i\frac{n}{R}y}$ with the same
amplitude). This is consistent with the fact that the fifth dimension
is compactified because it shows that the momentum in the fifth
direction is quantized, $p^4=\pm n/R$. The term $\Phi^4$ in the
original Lagrangian couples locally this waves. The function
$\Delta_{nmpq}$ in the Feynman rule just checks if there is a choice
among all the possible momenta in the vertex that preserves
momentum. 

Quantization of momentum in the direction of the extra dimension
provides another way of understanding why the masses of the modes are
$m_n$; recall that due to \Eq{eq:lagri} $p^\alpha p_\alpha=m^2$, so
\begin{eqnarray}
p_\alpha p^\alpha&=&m^2\nonumber\\
E^2-|\vec{p}|^2-p_4^2&=&m^2\nonumber\\
p_\mu p^\mu &=&m^2+p_4^2\nonumber\\
p^2&=&m^2 + m_n^2
\end{eqnarray}
As stated, momentum is violated by quantum corrections because these
are inherently non-local, thus the masses, i.e. the spectrum of the
theory, will also get radiative corrections
$m_n=n/R+\delta_n^{1-loop}$, and in general it ceases to be true that
the difference between to consecutive modes is $R^{-1}$, see
Ref.~\cite{Cheng:2002iz}.

\subsection{Spinor field and Yukawa interactions}
%To construct an extra dimensional theory with fermion fields we
%need to look for spinor representations of the Lorentz group,
%$SO(1,d-1)$. These can be found by solving the Dirac equation in
%$d$ dimensions
The Dirac equation in $d$ dimensions reads
\begin{equation}
\label{eq:Dirac}
(i\gamma^{(d)}_\alpha
\partial^\alpha-m)\Psi(x^\alpha)=0
\end{equation}
where $\alpha=0,\ldots,3+d$. The quanta of the field $\Psi$ are
the particles described by it and will obey the dispersion
relation $p^\alpha p_\alpha=m^2$ provided the $\gamma^{(d)}$
matrices obey the Clifford algebra
\begin{equation} \label{eq:gammas}
\{\gamma^{(d)}_\alpha,\gamma^{(d)}_\beta\}=2
g_{\alpha\beta}\mathbbm{1}
\end{equation}
where $g^{\alpha\beta}=diag(+1,-1,\ldots,-1)$. So the problem of
constructing representations of the Lorentz group in $d$ dimensions is
equivalent to looking for a set of matrices that satisfy
\Eq{eq:gammas}. One important outcome is that the number of $\gamma$
matrices is equal to the number of dimensions since they always come
paired with a derivative, \Eq{eq:Dirac}.

In this work we concentrate almost exclusively on the case of five
dimensions, so we will not look for representations in an arbitrary
dimension $d$, this can be found for instance in
Ref.~\cite{Pilaftsis:1999jk}. Instead we will look for a set of five
gamma matrices, $\gamma^{(5)}_\alpha$, denoted for simplicity in the
following by $\Gamma_\alpha$, that fulfil \Eq{eq:gammas}. From now on,
the indices denoted by the first letters in the Greek alphabet will
take the values $\alpha,\beta=0,\ldots,4$, while as usual
$\mu,\nu=0,\ldots,3$. The $\Gamma$'s can be found in terms of the
usual $\gamma$ matrices. It is easy to check that the assignments
$\Gamma_\mu=\gamma_\mu$ and $\Gamma_4=i\gamma^5$ indeed work, where
$\gamma^5=i \gamma^0 \gamma^1 \gamma^2 \gamma^3$.

The Dirac's equation \rf{eq:Dirac} can be obtained from the
Lagrangian density
\begin{equation}
\label{eq:lag5}
\mathcal{L}^5=\overline{\Psi}(i\Gamma^\alpha\partial_\alpha-m)\Psi.
\end{equation}
Of course, it is equivalent to solving the equation of motion of
\Eq{eq:Dirac} and to working with the action obtained from
\Eq{eq:lag5}, but working with $\mathcal{L}^5$ will simplify the
calculations, hence we will use it in the following.

First of all, notice that $\Psi$ is a four component spinor, even
if it is associated with a five dimensional representation. If we
assume that the extra dimension is compactified on a sphere,
$S^1$, then the field must be periodic in $y$ and can be expanded
in its Fourier modes 
\begin{equation}
\label{eq:Psi}
\Psi(x^\mu,y)=\psi^{(0)}(x^\mu)+\sum_{n=1}^{\infty}\eta^{(n)}(x^\mu)
\cos\left(\frac{ny}{R}\right)+
\varepsilon^{(n)}(x^\mu) \sin\left(\frac{ny}{R}\right)
\end{equation}

Integrating on the extra dimension and rescaling the fields
\begin{equation}
\psi^{(0)}\to \frac{1}{\sqrt{\pi R}}\psi^{(0)}\qquad \eta^{(n)}\to
\sqrt{\frac{2}{\pi R}}\;\eta^{(n)} \qquad \varepsilon^{(n)}\to
\sqrt{\frac{2}{\pi R}}\;\varepsilon^{(n)},
\end{equation}
the four-dimensional Lagrangian density reads
\begin{equation}
\mathcal{L}= \overline{\psi}^{(0)} ( i \sld-m ) \psi^{(0)} +
\sum_{n=1}^{\infty} \overline{\eta}^{(n)} ( i \sld-m )
\eta^{(n)} + \overline{\varepsilon}^{(n)}( i \sld-m )
\varepsilon^{(n)}  +
 m_n (\overline{\eta}^{(n)} \gamma^5 \varepsilon^{(n)} -
\overline{\varepsilon}^{(n)} \gamma^5 \eta^{(n)})
\end{equation}
the fields $\eta$ and $\varepsilon$ are four-component spinors
because so it was $\Psi$. Using this, one can write them in terms
of their chirality components, $\eta=\eta_R+\eta_L$ and similarly
for $\varepsilon$.
\begin{equation}
\label{eq:step2} \mathcal{L}=\overline{\psi}^{(0)} ( i
\sld-m ) \psi^{(0)} + \sum_{n=1}^\infty \overline{\eta}^{(n)}
i \sld \eta^{(n)} + \overline{\varepsilon}^{(n)}  i
\sld \varepsilon^{(n)} -
[\begin{array}{cc}\overline{\eta}^{(n)}_{L} &
\overline{\varepsilon}^{(n)}_{L}\end{array}]
\left[\begin{array}{cc} m & -m_n\\
m_n  &  m
\end{array} \right]
\left[\begin{array}{c}\eta^{(n)}_{R}\\\varepsilon^{(n)}_{R}
\end{array}\right]+\hc
\end{equation}

The mass matrix has to be diagonalized with a bi-unitary
transformation. If we call $M$ the mass matrix in \Eq{eq:step2}
then $U^\dagger M V=m_D$. It turns out that $U=\mathbbm{1}$ because $M^\dagger
M$ is 
diagonal, therefore only the right-handed fields are changed by
$V$:
\begin{equation}
\left[\begin{array}{c}\eta^{(n)}_{R} \\
\varepsilon^{(n)}_{R}\end{array}\right] = \frac{1}{\sqrt{m^2+m_n^2}}
 \left[\begin{array}{cc} \;m &
m_n \\ -m_n  & m
\\
\end{array}\right]
\left[\begin{array}{c}\eta_{R}^{\prime(n)} \\
\varepsilon_{R}^{\prime(n)}\end{array}\right].
\end{equation}
Finally, if we define the Dirac (or vector-like) fields $\psi^{(n)}
\equiv \eta^{\prime (n)}_{R} + \eta^{(n)}_{L}$ and $\xi^{(n)}\equiv
\varepsilon^{\prime (n)}_{R}+\varepsilon^{(n)}_{L}$ the Lagrangian is
written
\begin{equation}
\label{eq:ptu}
\mathcal{L}=\overline{\psi}^{(0)} ( i \sld-m ) \psi^{(0)} +
\sum_{n=1}^{\infty} \overline{\psi}^{(n)} ( i \sld-m_n^\prime )
\psi^{(n)} + \overline{\xi}^{(n)} ( i \sld-m_n^\prime ) \xi^{(n)},
\end{equation}
where as in the case of the boson field $m_n^\prime = +\sqrt{m^2+m_n^2}$.
From the above result one can see that there are two infinite KK
towers formed by vector-like spinors, $\psi^{(n)}$ and $\xi^{(n)}$, with
masses $m_n$. \Eq{eq:ptu} also shows that the fundamental
low-energy spectrum is formed by a vector-like field $\psi^{(0)}$
which mass is the one appearing in the original Lagrangian.

This poses a serious problem if one wants to identify the fundamental
mode with any of the known particles. Specifically, we would want to
identify the fundamental modes of a set of fields as the fields
appearing in the SM Lagrangian. But to achieve this, it is essential
that the fundamental modes are chiral. In four dimensions, a chiral
field can be defined as the field that fulfils simultaneously
$\gamma^5\Psi=\pm\Psi$ as well as the Dirac equation $i\sld\Psi =
0$. One could try to impose the same definition in five dimensions,
but now the situation is completely different because in four
dimensions $\gamma^5$ anticommutes with all the $\gamma$ matrices, a
fact that ceases to be true in five dimensions. This is because, in
the former case, the representation of the Lorentz group that comes
from the Dirac's equation is not irreducible, but it is a direct sum
of two irreducible representations that can be distinguished by their
different eigenvalues under the action of $\gamma^5$. In five
dimensions $\gamma^5$ is one of the $\gamma$ matrices, hence it no
longer anticommutes with \emph{all} the $\gamma$ matrices, and as a
consequence the two equations can not be fulfilled simultaneously. To
see this in detail, let $\Psi$ be a four-component spinor field that
fulfils the five dimensional Dirac's equation. Then the transformed
field $\gamma^5\Psi$ does not obey Dirac's equation
\begin{eqnarray}
\label{eq:demos}
i\Gamma^\alpha\partial_\alpha(\gamma^5\Psi) & = & 0\\
\gamma^5(-i\sld+i\Gamma^4\partial_y)\Psi &  = & 0\\
(i\sld-i\Gamma^4\partial_y)\Psi &  = & 0.
\end{eqnarray}
Notice that it is the relative sign between $\Gamma^4$ and the rest
of $\gamma$ matrices what prevents $\gamma^5\Psi$ to be a valid
solution. But this sign can be reabsorbed by the derivative
$\partial_y$ if the action of $\gamma^5$ is accompanied by a
parity transformation in the fifth direction, i.e. $\Psi^\prime(y)
= \gamma^5\Psi(-y)$ will be a solution of the Dirac's equation if
previously $\Psi(y)$ is a solution. The presence of a mass term in
\Eq{eq:demos} would invalidate this last conclusion.

This result can be exploited to obtain chiral fundamental
modes. Recall that we had assigned the topology of a circumference to
the fifth dimension $S^1$; now suppose that impose in addition that
the action computed with the values of the fields in one side $y\in
[0,\pi R]$ is the same as in the other $y\in [-\pi R,0]$, then to
extract the physics one only needs to look for extremals of the action
$S$ in only one side. This means that the value of a field in one side
must be related with the values it takes in the other side by a
transformation that is a symmetry of the original Lagrangian,
$\Psi(-y) = U\Psi(y)$. For a spinor field we choose
$U=\pm\gamma^5$, or what is the same we impose that the combined
action of $\gamma^5$ and $y\to -y$ should leave $\Psi$ invariant
$\Psi(y) = \pm \gamma^5\Psi(-y)$, with this, the Fourier modes are all
chiral. Take for concreteness the minus sign
\begin{equation}
\Psi(x^\mu,y)=\psi_{L}^{(0)}(x^\mu)+
\sum_{n=1}^{\infty}\eta^{(n)}_{L}(x^\mu)
\cos\left(\frac{ny}{R}\right)+ 
\varepsilon^{(n)}_{R}(x^\mu) \sin\left(\frac{ny}{R}\right).
\end{equation}
So the modes that are even under $y\to -y$ have the same chirality as
the fundamental one while the ones that are odd have opposite
chirality. Since one only need to know the values that the field takes
on one side, say $y\in[0,\pi R]$, this means that the new topology is
no other than an orbifold, $S^1/Z_2$, see previous section. The
Lagrangian in five dimensions is taken to be
\begin{equation}
  \label{eq:5dferlg}
  \mathcal{L}^5_F = \overline{\Psi}i\Gamma^\alpha\partial_\alpha\Psi,
\end{equation}
and it has no mass term because it breaks the orbifold symmetry. After
the dimensional reduction the four-dimensional Lagrangian reads
\begin{equation}
  \label{eq:5dimreducted}
  \mathcal{L}_F=\overline{\psi}^{(0)}_{L} i \sld \psi^{(0)}_{L} +
  \sum_{n=1}^{\infty} \overline{\psi}^{(n)} [i \sld-m_n ]
  \psi^{(n)},
\end{equation}
where $\psi^{(n)} = \eta^{(n)}_{L} + \varepsilon^{(n)}_{R}$. So we
have finally 
achieved a chiral fundamental mode, $ \psi^{(0)}_{L}$. On top of it, it has
appeared a KK-tower of vector-like fields, $\psi^{(n)}$, with masses
$m_n$.

If instead of the minus sign we had taken the plus, then we would have
ended with a similar equation but with sign of the mass reversed,
i.e. for each mode we would have obtained $\overline{\psi}^{(n)} [i
\sld + m_n ] \psi^{(n)}$. Of course, the sign of the mass is
unobservable.
%, it depends on what one calls particle and antiparticle
%in the theory. What it is called particle when one sign is selected
%becomes the antiparticle when the other one is taken. The change of
%base that goes from one to the other is $\psi^{(n)}\to \gamma^5
%\psi^{(n)}$. 
So in addition of the required field redefinitions to
obtain a canonical kinetic term, the modes of the fields with a
right-handed fundamental mode must be further redefined to get the
right sign of the mass.

Now that we have succeeded in constructing a theory with chiral
fundamental modes, the next step is to study possible interactions
that may involve this kind of fields. We will study in the following
the Yukawa couplings for two reasons: it is the easiest interaction
between fermions and bosons and because it is present in the SM. The
interacting fields will be a spinor field $\Psi$ with a right-handed
fundamental mode, another spinor field $\chii$ with a left-handed
fundamental mode and a boson field $\phi$ even under $Z_2$. The
orbifold topology is assumed. The Lagrangian is taken to be
\begin{equation}
\mathcal{L}^5_Y =\widetilde{Y}\:\overline{\chii}\phi\Psi+\hc
\end{equation}
which, as we will see, will provide a four-dimensional 
 Yukawa interaction for the zero modes.
%A term like the one in \Eq{eq:phicuatro} can be included, in fact, in
%four dimensions it would have been necessary to ensure the
%renormalizability of the theory. Although here this argument does not
%apply (extra-dimensional theories are not renormalizable in general)
%its inclusion would give a theory with a renormalizable low-energy
%effective Lagrangian. Nevertheless, it has been studied in the
%previous section, and here we will concentrate on the Yukawa coupling.

The canonical dimensions of the Yukawa coupling constant are
$[\widetilde{Y}] = E^{\frac{1}{2}}$; this is the reason why this
theory is not renormalizable. Now, after dimensional reduction and
field redefinitions to get canonical kinetic terms, the
Lagrangian reads
\begin{equation}
\mathcal{L} = Y \phi^{(0)}
\overline{\chii}^{(0)}_{R}\Psi^{(0)}_{L}+Y \sum_{n=1}^{\infty}[\phi^{(0)}
\overline{\chii}^{(n)}\psi^{(n)} + \phi^{(n)}\overline{\chii}^{(0)}\psi^{(n)}
+\phi^{(n)}\overline{\chii}^{(n)}\psi^{(0)} ] + \frac{Y}{\sqrt{2}} \sum_{n,p,q=1}^\infty
\Delta_{npq} \phi^{(n)}\overline{\chii}^{(p)}\psi^{(q)} + \hc
\end{equation}
where $y= \widetilde{y}/\sqrt{\pi R}$ and $\Delta_{npq}$ is
defined in \Eq{defDelta}. Notice that there is no vertex that
couples the fundamental modes $n=0$ with only one single mode of
the KK tower $n>0$. The last ones appear at least paired, and as a
consequence the energy threshold to produce them using only the
fundamental fields is twice the mass of the lightest KK mode,
$\sqrt{s}\geq 2 m_1$. This feature also appeared when we studied
the case of the boson field, and the same reasoning we gave there
applies now here.

\subsection{Vector bosons and gauge theories}
\label{sec:vectorbosonsinUED}
In the previous section we have constructed a scenario in which the
fundamental modes of the extra dimensional scalar and spinor fields
can be identified with the fields in the SM. To achieve the full SM as
a low-energy realization of a five-dimensional theory we still need to
do a similar construction for the vector fields. In five dimensions
these fields have five components, $A^\alpha$, corresponding to the
five possible polarizations. Therefore, to associate $A^\alpha$ to one
of the vector fields in the SM we need to remove one of its components
from the low-energy spectrum. As we will see, this can be done when we
compactify the fifth dimension on the orbifold. The free Lagrangian in
five dimensions is taken to be
\begin{equation}
  \label{eq:5dgaugeLagrangian}
  \mathcal{L}^5_G = -\frac{1}{4}F^{\alpha\beta}F_{\alpha\beta} =
  -\frac{1}{4}(\partial^\alpha A^\beta-\partial^\beta A^\alpha)
  (\partial_\alpha A_\beta-\partial_\beta A_\alpha) .
\end{equation}
To perform the dimensional reduction we impose that the fifth
component of the field $A_4$ is odd under the action of $Z_2$ while
the rest $A_\mu$ are even. On the following the fifth components will
be denoted by $A_5$ because it is done so in the literature.
\begin{eqnarray}
  \label{eq:Ainorbifold1}
  A_\mu(x^\mu,-y) & = &  \;\;  A_\mu(x^\mu,y),\\
  \label{eq:Ainorbifold2}
  A_5(x^\mu,-y) & = &    -A_5(x^\mu,y).
\end{eqnarray}
The above prescription is not gauge invariant, it breaks gauge
symmetry in five dimensions but preserves it in four. Different
components of the strength tensor transform differently:
\begin{eqnarray}
F^{\mu\nu}   & \to &  \;\;F^{\mu\nu},\\
F^{\alpha\nu}& \to &  -F^{\alpha\nu}.
\end{eqnarray}
Despite of this, $\mathcal{L}^5_G$ is invariant under
Eq.(\ref{eq:Ainorbifold1},\ref{eq:Ainorbifold2}), as it should, since
both sides of $S^1$ must be physically equivalent. After dimensional
reduction the Lagrangian reads
\begin{equation}
  \label{redprimera}
  \mathcal{L}_G =
  -\frac{1}{4}F^{(0)}\cdot F^{(0)} +
  \sum_{n=1}^{\infty}
  -\frac{1}{4}F^{(n)}\cdot F^{(n)}   +  \frac{1}{2}m_n^2A^{(n) \mu} A^{(n)}_{\mu}
  + \frac{1}{2}\partial_\mu A^{(n)}_{5}\partial^\mu A^{(n)}_{5}
  + m_n^2 \partial_\mu A^{(n)}_{5} A^{(n)\mu},
\end{equation}
where $F^{(n)}_{\mu\nu} \equiv \partial_\mu A^{(n)}_{\nu}-
\partial_\mu A^{(n)}_{\nu}$ and $m_n$ is defined in
\Eq{eq:lagdef}. $\mathcal{L}_G$ shows that the $A_5$ components are
not present in the low-energy spectrum. Moreover, the modes
$A^{(n)}_{\mu}$ have acquired a mass , $m_n$, $A^{(n)}_{5}$ being the
Goldstone boson eaten by $A^{(n)}_{\mu}$. The appearance of a
tree-level mixing between $A^{(n)}_{\mu}$ and $A^{(n)}_{5}$ suggest a
gauge fixing term of the form:
\begin{equation}
  \label{eq:gaugefixing}
  \mathcal{L}_{gf} = -\frac{1}{2\xi}(\partial_\mu A^{(n)\mu} - \xi m_n
  A^{(n)}_{5})^2.
\end{equation}
This is very similar to the usual $R_\xi$ type of gauge fixing, known
from the SM. When $\mathcal{L}_{gf}$ is taken in this way the propagators are
\begin{equation}
  \begin{array}{lcl}
    A^{(0)}_{\mu}\; :&
    \includegraphics[scale=0.6]{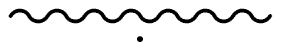}\hspace{3ex}
    & =\frac{-i}{k^2}\left[g^{\mu\nu} -(1-\xi) \frac{k^\mu
    k^\nu}{k^2}\right],\\ A^{(n)}_{\mu}\; :&
    \includegraphics[scale=0.6]{propphoton.eps}\hspace{3ex}
    & =\frac{-i}{k^2-m_n^2}\left[g^{\mu\nu} -(1-\xi) \frac{k^\mu
    k^\nu}{k^2-\xi m_n^2}\right],\\ A^{(n)}_{5} :&
    \includegraphics[scale=0.6]{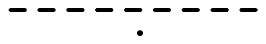}\hspace{3ex}
    & =\frac{i}{k^2-\xi m_n^2}.
  \end{array}
\end{equation}
\Eq{eq:gaugefixing} fixes the gauge \emph{after} compactification.
One may wonder if it is equivalent to fixing it \emph{before}
compactification. We will not study this issue in detail here, we
refer the interested reader to Ref.\cite{Papavassiliou:2001be}, where
it is treated. It is shown there that the choices of the Feynman gauge
and unitary gauge in \Eq{eq:gaugefixing} can be obtained through
suitable choices in the gauge fixing terms before compactification,
while there is no such possibility for the Landau gauge, thus to avoid
any problem we will work always in the Feynman gauge.

Notice that the aim of the orbifold topology is to remove from the
low-energy spectrum the zero mode of the fifth component of the gauge
field, $A_5^{(0)}$, but as we will see, its presence would offer an
interesting possibility. It could be present as a \emph{massless}
scalar if we compactify in a circumference $S^1$. The extra
dimensional gauge symmetry forbids a mass term for it. Nevertheless,
we have seen that the compactification breaks this symmetry, therefore
radiative corrections will provide it mass. When these corrections are
studied they are found to be finite. The UV divergences are not
present because at very high energies, very small distances (compared
to the compactification radius), all the dimensions are equivalent and
$A_5^{(0)}$ is just one component of a gauge field and therefore it
can not receive any contribution to its mass because of gauge
symmetry. All the contributions to its mass came from the low
energy. This situation is very similar to the way in which SUSY
protects the mass of Higgs. In this case the Higgs is associated via
the SUSY symmetry with a fermionic field and the chiral symmetry
protects the mass terms for both superpartners. Basically, this is the
idea behind the work done in
\cite{Hatanaka:1998yp,Dvali:2001qr,Arkani-Hamed:2001nc,Masiero:2001im}
. More specifically, the Higgs field, a scalar field, can be
associated to a gauge field $A_\alpha$ and with this, its mass can be
protected by gauge symmetry. Despite the interest of these kind of
models we will not follow them. Instead we will continue using the
orbifold topology without the $A_5^{(0)}$ field in our spectrum.

The vector fields appear in general in gauge theories. As an example
of a five dimensional gauge theory we develop briefly in the next
lines a theory that reduces to a version of QED after
compactification. Its low energy consists in one massless chiral
fermion coupled with a massless photon via a gauge interaction, a more
detailed derivation is given in Ref.\cite{Papavassiliou:2001be}. The
Lagrangian is written as
\begin{equation}
  \mathcal{L}^5 = \mathcal{L}_G^5 +
  \overline{\Psi}i\slD\Psi ,
\end{equation}
where $\mathcal{L}_G^5$ is defined in \Eq{eq:5dgaugeLagrangian} and
the five dimensional covariant derivative is defined as $D_\alpha
=
\partial_\alpha -i \tilde{e} A_\alpha$. After compactification the Lagrangian
can be decomposed in the sum of three terms $\mathcal{L} =
\mathcal{L}_G + \mathcal{L}_F + \mathcal{L}_I$, where
$\mathcal{L}_G$ and $\mathcal{L}_F$ are defined in \Eq{redprimera}
and \Eq{eq:5dimreducted} respectively, and the interaction term
can be divided in three pieces $\mathcal{L}_I = \mathcal{L}_I^0 +
\mathcal{L}_I^{0K} + \mathcal{L}_I^{K}$
\begin{eqnarray}
  \mathcal{L}_I^0 & = & e\; \overline{\psi}_L^{(0)} \slA^0\psi_L^{(0)}\\
  \mathcal{L}_I^{0K} & = & e \sum_{n=1}^\infty \overline{\psi}^{(n)}
  \slA^{(0)}\psi^{(n)} + e \sum_{n=1}^\infty \left[ \overline{\psi}^{(0)}_L
   \slA^{(n)}\psi^{(n)}_L + i \overline{\psi}^{(0)}_L A_5^{(n)} \psi^{(n)}_R + \hc
  \right]\\
  \mathcal{L}^K &= & \frac{e}{\sqrt{2}}\sum_{n,m=1}^\infty
  \left[\overline{\psi}^{(n+m)} \slA^{(m)} \psi^{(n)} - i
  \overline{\psi}^{(n+m)}
  A_5^{(m)} \psi^{(n)} + \hc \right]\\
  & & \frac{e}{\sqrt{2}}\sum_{n,m=1}^\infty \left[\overline{\psi}^{(m)}
  \slA^{(n+ m)} \gamma^5 \psi^{(n)} + i \overline{\psi}^{(m)}
  A_5^{(n+m)} \gamma^5 \psi^{(n)}  \right],
\end{eqnarray}
where we have rewritten the five-dimensional coupling $\tilde{e}$ in terms
of the four dimensional $e$ as $e = \tilde{e}/\sqrt{\pi R}$. Notice that, as
mentioned, the low-energy (below $R^{-1}$) effective theory
corresponds exactly to QED with one chiral fermion. In addition, all
the vertices conserve the KK number as it happened in the previous
examples, hence KK modes must be created in pairs from the fundamental
modes what reduces its impact on the low-energy phenomenology because
the effective Lagrangian only receives contributions at the one-loop
level. The Feynman rules can be easily read from the above formulae.

% LocalWords:  orbifold compactify compactification superpartners

%%%%%%%%%%%%%%%%%%%%%%%%%%%%%%%%%%%%%%%%%%%%%%%%%%%%%%%%%%%%%%%%%%%%%%%
%%%%%%%%%%%%%%%%%%Thu Sep 25 11:23:02 CEST 2003 %%%%%%%%%%%%%%%%%%%%%%%
%%%%%%%%%%%%%%%%%%%%%%%%%%%%%%%%%%%%%%%%%%%%%%%%%%%%%%%%%%%%%%%%%%%%%%%
\comment{
{\sffamily \scshape
\begin{center}
\begin{tabular}{|l|r|}
\hline
Text         & Ok      \\\hline
Gauge Piece  & Ok      \\\hline
Orthography  & Ok      \\\hline
Figures      & Ok      \\\hline
Links        & Ok      \\\hline
Cites        & Ok      \\\hline
Meaning      & --      \\\hline
Makindex     & --      \\\hline
Date         & \today  \\\hline
\end{tabular}
\end{center}
}
}%endcomment
\chapter{SM with one universal extra dimension }
\label{sec:SMinUED}
Until now we have constructed a number of toy models that have helped
us to study some features of the theories with one extra dimension. In
particular we have shown that the fundamental modes of the different
fields remain in the low-energy spectrum after compactification. If
the topology of the extra dimension is taken to be an orbifold then
the fundamental modes of the spinor fields are chiral and the ones of
the vector fields have only four components. We exploit these results
here to build a five dimensional model, which was initially proposed
in Ref.~\cite{Appelquist:2000nn}, that after compactification reduces
to the SM.  We study the phenomenology and use the results to set
bounds on the compactification scale, $R$, or what is the same, to the
mass of the first KK mode, to be denoted by $M = R^{-1}$.  We
concentrate on the observables with a strong dependence on the mass of
the top-quark, $m_t$, for which the corrections to SM will be
enhanced.  Although the process $b\to s \gamma$ has no $m_t$
enhancement, the relative impact of the new physics is also important
because it is one-loop suppressed in the SM due to gauge invariance,
hence we will also study it.

\comment{
To get
an estimate we do not need to compute the whole modifications to the
SM results but instead the dominant ones will be enough. In the next
lines we will explain how to build a model that reduces to the usual
SM in the low-energy limit but only the those sectors relevant to us
will be described in detail. We expect that the dominant contributions
will be proportional to the mass of the top quark because of two
reasons:
\begin{itemize}
\item The first corrections to the SM results will be suppressed by the
  scale of the new physics. By dimensional arguments this scale must
  be compensated by another mass in the numerator, and the heaviest
  masses in the SM are the top and Higgs masses.
\item As we will see, in this kind of models the diagrams that
  contribute to a given process are basically the same as the ones in
  SM, therefore we expect the structure of the radiative corrections
  to be the same. In SM the dominant corrections are the ones
  proportional to the top mass, the corrections that depend on the Higgs mass
  happen to be suppressed.
\end{itemize}
}%endcomment

\section{The model}
We  use the  above considerations  to simplify  the Lagrangian  of the
model,  and  in what  follows  all mass  scales  below  $m_t$ will  be
neglected.  On the other hand,  considering that we are not interested
in  strong   processes,  we  concentrate  only  on   the  gauge  group
$SU(2)_L\times U(1)_Y$. The Lagrangian is separated in four pieces
\begin{equation}
\label{eq:primera}
\mathcal{L}^{\mathrm{UED}}=\int_0^{L=\pi R} dy (\mathcal{L}_G + \mathcal{L}_{H} +
\mathcal{L}_{F} + \mathcal{L}_Y).
\end{equation}

The gauge piece is defined to be
\begin{equation}
  \label{eq:segunda}
  \mathcal{L}_G=-\frac{1}{4}F^{a} \cdot F^a  -\frac{1}{4}F\cdot F
\end{equation}
where $F^{a}_{\alpha\beta}$ is the five dimensional gauge field
strength associated with the $SU(2)_L$ gauge group and
$F_{\alpha\beta}$ is the one of the $U(1)_Y$ group
\begin{eqnarray}
  F_{\alpha\beta}^a&=&\partial_\alpha W_\beta^a - \partial_\beta
  W_\alpha^a+ \tilde{g} \epsilon^{abc}W_\alpha^b W_\beta^c \\
  F_{\alpha\beta}&=&\partial_\alpha B_\beta - \partial_\beta B_\alpha
\end{eqnarray}

The Higgs piece is
\begin{eqnarray}
\mathcal{L}_H= (D_\alpha H)^\dagger D^\alpha H -V(H),
\end{eqnarray}
and the covariant derivative is defined as $D_\alpha \equiv
\partial_\alpha - i\widetilde{g}W_\alpha^a T^a-i \widetilde{g}^\prime
B_\alpha Y$, where $\widetilde{g}$ and $\widetilde{g}^\prime$ are
the (dimension-full) gauge coupling constants of $SU(2)_L$ and
$U(1)_Y$ respectively in the five dimensional theory, $T^a$ and
$Y$ are the generators of these groups.

The fermionic piece is
\begin{equation}
  \mathcal{L}_F = \overline{Q}(i\Gamma^\alpha D_\alpha)Q +
  \overline{U}(i\Gamma^\alpha D_\alpha)U+\overline{D}(i\Gamma^\alpha
  D_\alpha)D,
\end{equation}
where $\Gamma_\alpha$ are the five dimensional gamma matrices. The
fields $Q$, $D$ and $U$ carry a generational index that is not
explicitly written here.

Finally, the Yukawa piece reads
\begin{equation}
  \label{eq:yukdefUED}
  \mathcal{L}_Y = -\overline{Q} \widetilde{Y}_u H^c  U - \overline{Q}
  \widetilde{Y}_d H D + \hc
\end{equation}
where $H^c=i\sigma^2 H^\ast$ is the usual charge conjugated field and
the $\widetilde{Y}_u$ are the Yukawa matrices in the five dimensional
theory, which, as usual, mix different generations.

We use the topology of the extra dimension, assumed to be an orbifold
$S^1/Z_2$, and expand the fields in a Fourier series
\begin{eqnarray}
  \label{eq:uno}
  G_\mu & = &  \frac{1}{\sqrt{\pi R}}G_\mu^{(0)} +
  \sqrt{\frac{2}{\pi R}}\sum_{n=1}^{\infty} G_\mu^{(n)} \cos\left(
  \frac{n y}{R}\right)\\
  \label{eq:two}
  G_5 & = &\sqrt{\frac{2}{\pi R}}\sum_{n=1}^{\infty} G_5^{(n)}
  \sin\left( \frac{n y}{R}\right)\\
  \label{eq:three}
  Q &=& \frac{1}{\sqrt{\pi R}} Q_L^{(0)} + \sqrt{\frac{2}{\pi R}}
    \sum_{n=1}^{\infty} \left[ Q_L^{(n)} \cos\left( \frac{n
    y}{R}\right) + Q_R^{(n)} \sin\left( \frac{n
    y}{R}\right)\right]\\
  \label{eq:five}
  U & = & \frac{1}{\sqrt{\pi R}} U_R^{(0)} + \sqrt{\frac{2}{\pi R}}
    \sum_{n=1}^{\infty} \left[ U_R^{(n)} \cos\left( \frac{n
    y}{R}\right) + U_L^{(n)} \sin\left( \frac{n
    y}{R}\right)\right]
\end{eqnarray}
where the expansion for $G_\mu$ is valid for each component of the
gauge fields as well as for the Higgs doublet, the one for $G_5$ is
valid for the fifth component of the gauge fields. Analogously, the
expansion for $U$ is valid also for $D$. We have included
different normalization for the modes to obtain canonical kinetic
terms after compactification.

\subsection{The spectrum of the model}

To make any calculation we require the spectrum of the model. In
order to extract it, the Higgs sector has to be studied. It will
undergo SSB, hence it will contribute to the masses of the
different particles. The Higgs doublet is parametrized as
\begin{equation}
H= \left[
\begin{array}{c}
\Phi^+ \\
\Phi^0
\end{array}
\right] = \frac{1}{\sqrt{2}}\left[
\begin{array}{c}
-i(\phi_1 - i \phi_2) \\
\phi_0 + i \phi_3
\end{array}
\right],
\end{equation}
where $\phi_i$ are real fields. As a Higgs potential it is chosen
\begin{equation}
  \label{eq:Higgspot}
  V(H)=-\mu^2H^\dagger H+\frac{\widetilde{\lambda}}{4!}(H^\dagger H)^2,
\end{equation}
where $\mu^2$ is a real positive parameter with mass dimensions and
$\widetilde{\lambda}$ is a real parameter with dimension
$E^{-1}$. After dimensional reduction the potential contains a number
of couplings between the different KK modes. Here we show only those that are
relevant
\begin{eqnarray}
V & = & -\mu^2 H^{(0)\dagger}H^{(0)} +
\frac{\lambda}{4!}(H^{(0)\dagger}H^{(0)})^2 + \sum_{n=1}^\infty
(-\mu^2+ m_n^2) H^{(n)\dagger}H^{(n)} + \\
\label{eq:secon}
 & + & \sum_{n=1}^\infty \frac{\lambda}{4!} \left[
  (H^{(0)\dagger}H^{(0)})(H^{(n)\dagger}H^{(n)}) +
  (H^{(0)\dagger}H^{(n)})^2 +
  (H^{(0)\dagger}H^{(n)})(H^{(n)\dagger}H^{(0)}) + \hc\right],\qquad\qquad
\end{eqnarray}
where $\lambda \equiv \widetilde{\lambda}/(\pi R)$. The first thing to
notice is that the potential for the fundamental mode induces SSB only
for the zeroth mode, $H^{(0)}$, since for $n>0$ we expect reasonably
$m_n^2 > \mu^2$. This is consistent with the
association of the fundamental mode of $H$ to the SM Higgs
doublet. Following with this idea, the neutral component of the
doublet will get a VEV, specifically, $\langle\phi_0^{(0)}\rangle_0 =
v$, what implies
\begin{equation}
  \phi_0^{(0)}(x^\mu) = v + h(x^\mu), \qquad \langle H^{(0)}\rangle_0
  = \frac{v}{\sqrt{2}} \left[\begin{array}{c} 0 \\1\end{array}
  \right].
\end{equation}
On the contrary, the modes of the KK tower ($n>0$) do not undergo SSB
because their masses before SSB are positive.  Nevertheless, the terms
in \Eq{eq:secon} will modify their masses after SSB. At the end, the
masses of the Higgs field and its associated KK tower are
\begin{equation}
  m^2(h) = 2 \mu^2 \equiv m_h^2, \qquad m^2(\phi_0^{(n)})= m_h^2 +
  m_n^2, \qquad n \geq 1.
\end{equation}
Recall that the fields $\phi^{(n)}_0$, for $n \geq 1$ do not get a
VEV. For the rest of the fields the masses are
\begin{equation}
  \label{eq:SSBlast}
  m^2(\Phi^{\pm (n)}) = m^2(\phi_3^{(n)}) = m_n^2, \qquad n\geq 0.
\end{equation}
If our interpretation is correct, the fields $\Phi^{\pm (0)}$ and
$\phi_3^{(0)}$ will be precisely the SM Goldstone bosons absorbed by
$W^\pm$ and $Z$, the fact that they are massless is also consistent
with the identification of $H^{(0)}$ with the SM Higgs doublet.

\comment{and we see
that they are massless, consistently with the assumption that the
fundamental mode of $H$, $H^{(0)}$ is the SM Higgs doublet.}

In the gauge sector, this SSB is also relevant. In addition, it
is easy to convince oneself that this model coincides exactly with the
SM when only zero modes are taken into account, what accounts for
taking the limit $m_n \to \infty$. Thus,
%at tree level
retaining only the zero modes, all goes much in the same way as in SM:
the neutral component of the Higgs doublet gets a VEV and induces
mixing between $W_{\mu 3}^{(0)}$ and $B_{\mu}^{(0)}$ to give a
massless gauge boson, the photon $A_\mu^{(0)}$, and a massive one, the
$Z$ boson $Z_\mu^{(0)}$, the mixing being parametrized by the weak
mixing angle \label{eq:weakangle}
\begin{eqnarray}
Z_\mu^{(0)}&=&\cos\theta_w\;W_{\mu 3}^{(0)}-\sin\theta_w\;
B_{\mu}^{(0)}\\
A_\mu^{(0)}&=&\sin\theta_w\;W_{\mu
3}^{(0)}+\cos\theta_w\;B_{\mu}^{(0)}
\end{eqnarray}
and the same mixing is generated for $A_\mu^{(n)}$ and
$Z_\mu^{(n)}$. But we will concentrate on the charged gauge bosons
because they will appear in our calculations. After the dimensional
reduction, the bilinear terms relevant for the gauge sector are
\begin{eqnarray}
-\frac{1}{4}F^{a}_{\alpha \beta}F^{a \alpha \beta} + \left(D^\alpha H
 \right)^\dagger \left(D_\alpha H \right)
  \stackrel{D.R.}{\longrightarrow} &  - & \frac{1}{4}F^{(n)}_{\mu\nu}F^{\mu \nu (n)} + (m_n^2 + M_W^2)
 W_\mu^{+(n)}  W^{\mu -(n)} \\
& + & \partial_\mu W_5^{+(n)} \partial^\mu W_5^{-(n)} - M_W^2
 W_5^{+(n)}W_5^{-(n)} \\
& + & \partial_\mu \Phi^{+(n)} \partial^\mu \Phi^{-(n)} - m_n^2
 \Phi^{+(n)}\Phi^{-(n)}\qquad  \\
\label{eq:derivativemixing}
 & + & W^{\mu (n) -} \partial_\mu (iM_W \Phi^{+(n)} + m_n W_5^{+(n)}) + \hc
 \\
\label{eq:mixing}
& + & i M_W m_n W_5^{-(n)} \Phi^{+(n)} + \hc,
\end{eqnarray}
where the sum on $n$ is implicit and we have used the tree level
relation $gv/2=M_W$. The previous equations can be understood as
follows: the first two show that the vector bosons $W_\mu^{(n)}$ are
now massive with mass $\sqrt{M_W^2 + m_n^2}$, while the fifth
components KK modes, $W_5^{(n)}$, have became massive charged scalars
with masses $M_W$. Nevertheless, \Eq{eq:derivativemixing} and
\Eq{eq:mixing} show that $W_5^{(n)}$ have not diagonal mass terms,
because they mix with the modes of the charged component of the Higgs
doublet, $\Phi^{+(n)}$. In particular, \Eq{eq:derivativemixing} points
out the combination that defines the Goldstone field, $\Phi_G^+$, that
is been absorbed by the $W_\mu^{+(n)}$ fields to get masses. The
orthogonal combination, $\Phi_P^{+(n)}$, is a physical scalar. The
expressions that relate those fields are
\begin{eqnarray}
\Phi_G^{+(n)} & = & \frac{m_n W_5^{+(n)}+i M_W
\Phi^{+(n)}}{\sqrt{m_n^2 +
M_W^2}} \stackrel{M_W\to 0}{\longrightarrow} W_5^{+(n)}, \\
\label{eq:physicalHiggs}
\Phi_P^{+(n)} & = & \frac{i M_W W_5^{+(n)} + m_n
\Phi^{+(n)}}{\sqrt{m_n^2 + M_W^2}} \stackrel{M_W\to
0}{\longrightarrow} \Phi^{+(n)}.
\end{eqnarray}
These formulae are valid only for $n\ge 1$. In the limit of neglecting
all mass scales below $m_t$, the mixing is not important and
$W_5^{+(n)}$ can be identified with the Goldstone bosons,
$\Phi_G^{+(n)}$, and $\Phi^{+(n)}$ with the physical scalars,
$\Phi_P^{+(n)}$.

We pass now to study the quark sector, in particular, the third
generation because it contains the top. To distinguish between the up
and down components of the five-dimensional doublet $Q$ in
\Eq{eq:three} we will use subindices that will carry also information
about the generation, $e.g.$ when we write
\begin{equation}
Q = \left[
  \begin{array}{c}
    Q_t \\
    Q_b
  \end{array}
\right]
\end{equation}
we are denoting by $Q_t$ ($Q_b$) the up (down) component of the weak
doublet in the \emph{third} generation. Analogously by $U_t$ we denote
the weak singlet of the third generation.  After dimensional reduction
$Q_t^{(n)}$ acquires a mass $m_n$ and $U_t^{(n)}$ a mass $-m_n$,
$n>0$. These are Dirac fermions defined as $Q_t^{(n)} = Q_{tR}^{(n)} +
Q_{tL}^{(n)}$. The masses receive also contributions coming from the
couplings with $H^{(0)}$, which are contained in the Yukawa sector,
\Eq{eq:yukdefUED}. Here we extract the relevant terms
\begin{eqnarray}
  \mathcal{L}_Y & = & -Y_u \overline{Q}^{(0)}_t H^{(0)c} U_t^{(0)} - Y_u
  \sum_{n=1}^\infty\overline{Q}^{(n)}_t H^{(0)c} U_t^{(n)} + \hc + \ldots\\
  & = & -\frac{Y_u v}{\sqrt{2}} \overline{Q}^{(0)}_t U_t^{(0)} -
  \frac{Y_u v}{\sqrt{2}}\sum_{n=1}^\infty\overline{Q}^{(n)}_t
  U_t^{(n)} + \hc + \ldots ,
\end{eqnarray}
where $Y_u \equiv \widetilde{Y}_u/\sqrt{\pi R}$, and we have used
\Eq{eq:uno}. Since the fundamental modes are, by construction,
identified with the SM fields, then $Y_u$ must be the SM Yukawa
matrix, what implies that the KK modes $Q^{(n)}_t$ and $U_t^{(n)}$
have a mixing proportional to the mass of the top-quark, $m_t = Y_u v
/ \sqrt{2}$. So the bilinear terms for these fields may be written as
\begin{equation}
\mathcal{L}_t \equiv \overline{t}(i \sld - m_t) t + \sum_{n=1}^\infty
\overline{Q}^{(n)}_t i\sld Q^{(n)}_t + \overline{U}^{(n)}_t i\sld
U^{(n)}_t -
\left[\begin{array}{cc}
    \overline{Q}^{(n)}_t & \overline{U}^{(n)}_t
\end{array}\right]
\left[\begin{array}{cc}
    m_n & \;m_t \\
    m_t & -m_n
\end{array}\right]
\left[\begin{array}{c}
Q^{(n)}_t \\
U^{(n)}_t
\end{array}\right], \qquad \qquad
\end{equation}
We denote by $t$ the top-quark, $t = Q^{(0)}_{tL} +U^{(0)}_{tR}$. The
mass eigenfields, denoted by a prime, are
\begin{equation}
  \left[
    \begin{array}{c}
      Q^{(n)}_t \\
      U^{(n)}_t
    \end{array}
    \right] =
  \left[\begin{array}{cc}
      \cos(\alpha_n ) & -\sin(\alpha_n) \\
      \sin(\alpha_n) & \;\;\cos(\alpha_n )
    \end{array}
    \right]
  \left[\begin{array}{cc}
      1 & 0\\
      0 & \gamma^5
    \end{array}
    \right]
  \left[\begin{array}{c}
      Q^{\prime(n)}_t \\
      U^{\prime(n)}_t
    \end{array}
    \right],
\end{equation}
where $\tan(2\alpha_n) \equiv m_t/m_n$ and the $\gamma^5$ is included to
obtain a positive mass of the $ U^{\prime(n)}_t$. Finally, the masses
are
\begin{equation}
m(Q_t^{\prime (n)}) = m(U_t^{\prime (n)}) = \sqrt{m_n^2 +
  m_t^2} \equiv m_Q  \qquad  \qquad n > 0.
\end{equation}

In the calculations of this work, the degrees of freedom $Q_t^{\prime
(n)}$ and $U^{\prime(n)}_t$ will appear inside loops, hence it will be
advantageous to work with the fields $Q_t^{(n)}$ and $U^{(n)}_t$, we
call the latter the \emph{interaction base}. Since they have not
definite mass, the inverse of their quadratic forms, i.e. their
propagators, are not diagonal. In this base, the expression of the
couplings is simpler, but the propagators are non-canonical, and the next
expressions must be used for them
\begin{displaymath}
\left[
\begin{array}{cc}
\includegraphics[scale=0.6]{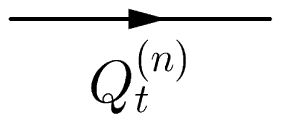}\hspace{3ex}
&
\includegraphics[scale=0.6]{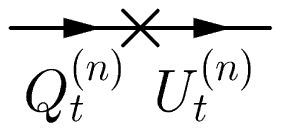}\hspace{3ex}\\
\includegraphics[scale=0.6]{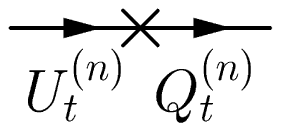}
&
\includegraphics[scale=0.6]{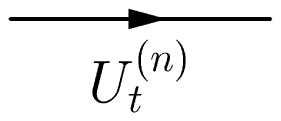}\hspace{3ex}
\end{array}
\right]
=
\left[
\begin{array}{cc}
  \;i \frac{\slp+m_n}{p^2-m_Q^2} &i \frac{m_t}{p^2-m_Q^2}\\
i \frac{m_t}{p^2-m_Q^2}&\;i \frac{\slp-m_n}{p^2-m_Q^2}
\end{array}
\right]
\end{displaymath}

\subsection{Couplings}
We will compute the dominant corrections to the modifications of the
$\rho$ parameter, which can be found by calculating the \comment{ We
will use the measurements of the $\rho$ parameter to set bounds to the
compactification scale. The dominant contributions to it come through}
radiative corrections to the self energies of the gauge bosons $W_{\mu
1}^{(0)}$ and $W_{\mu 3}^{(0)}$. To this end, we need to extract the
couplings of $W_{\mu 1(3)}^{(0)}$ with the KK modes of the $Q$ fields
because the $m_t$ proportional contributions come exclusively from
them.
%Thus, we will extract the couplings
%of these gauge bosons and among them the ones that couple $W_{\mu
%1(3)}^{(0)}$ to the KK modes of the $Q$ fields because the $m_t$
%proportional contributions come exclusively from them.
This is so because the mass of the top, $m_t$, only appears in the
propagators of the $Q$ and $U$ fields and in the vertices proportional
to the Yukawa matrices, but neither this vertices contribute at one
loop in UED\footnote{Because in the limit of $M_W\to~0$ the Higgs
doublets are not eaten by the gauge bosons, in contrast with what
happens in SM.}, nor do the $U$ fields, which do not even couple
directly to $W_{\mu 1(3)}^{(0)}$. Therefore, the relevant couplings
are
\begin{equation}
\label{Lagrho}
\mathcal{L}_\rho = \frac{g}{2}W_{\mu 1}^{(0)}
\left[\overline{Q}^{(n)}_{t} \gamma^\mu Q^{(n)}_b+
\overline{Q}^{(n)}_b \gamma^\mu Q^{(n)}_t \right] +
\frac{g}{2}W_{\mu 3}^{(0)} \left[\overline{Q}^{(n)}_t \gamma^\mu
Q^{(n)}_t \right],
\end{equation}
where $g=\widetilde{g}/\sqrt{\pi R}$. Notice that for simplicity we
have not considered the CKM mixing matrix\footnote{We have explicitly
checked that the calculations give exactly the same result when
$V_{CKM}$ is taken into account and that this is numerically so
because in our approximation all quark masses are zero except $m_t$.}.

The couplings of the physical scalar $\Phi^{+(n)}$ with the modes of
the top quark are also important because they are proportional to
$m_t$.
\begin{equation}
\label{eq:higgscouplings} \mathcal{L}_Y = \frac{\sqrt{2}}{v} m_t
V_{tj} \overline{U}^{(n)}_{R} Q^{(0)}_{j\;L} \Phi^{(n)+}+ \hc
\end{equation}
In this case, it has been maintained the CKM mixing matrix because we
will need it to cast the modifications in the $B^0 -\overline{B}^0$
mixing in a standard form, which is defined to bound any new physics
affecting this mixing. Notice that although $\Phi^{(n)+}$ are physical
degrees of freedom their couplings are exactly the same as the
Goldstone bosons of the SM.

We are also interested in the radiative corrections to the $Z\to b
\overline{b}$ decay, therefore we need to know the couplings of 
$Z_\mu = Z_\mu^{(0)}$.
\begin{equation}
\label{eq:couplingZ}
\mathcal{L}_Z =
\frac{g}{2c_w}Z_\mu[J^{\mu}_{SM} + J^{\mu(n)} + J^{\mu(n)}_{\Phi}],
\end{equation}
where the $J^{\mu}_{SM}$ is the usual SM neutral current
\begin{equation}
\label{eq:consm}
 J^{\mu}_{SM} = \overline{Q}_L \gamma^\mu
2[T^3-s_w^2 Q]Q_L + \overline{U}_R \gamma^\mu 2[T^3-s_w^2 Q]U_R .
\end{equation}
Analogously we find for $J^{\mu(n)}$, $n\ge 1$:
\begin{equation}
J^{\mu(n)} = \overline{Q}^{(n)} \gamma^\mu 2[T^3-s_w^2 Q]Q^{(n)} +
\overline{U}^{(n)} \gamma^\mu 2[T^3-s_w^2 Q]U^{(n)}.
\end{equation}
If we indicate explicitly the charges
\begin{equation}
\begin{array}{lcl}
\begin{array}{lcl}
T^3 Q^{(n)}_t&=& + \frac{1}{2} Q^{(n)}_t\\
T^3 Q^{(n)}_b&=& - \frac{1}{2} Q^{(n)}_b \\
T^3 U^{(n)}&=& 0
\end{array}
&
%\hspace{0ex}
&
\begin{array}{lcl}
Y Q^{(n)}_t&=& + \frac{1}{6}Q^{(n)}_t\\
Y Q^{(n)}_b&=& + \frac{1}{6}Q^{(n)}_b  \\
Y U^{(n)}&=&+ \frac{2}{3} U^{(n)}
\end{array}
\end{array}
\end{equation}
the current reads
\begin{equation}
\label{eq:conKK}
 J^{\mu(n)} = \left(+1-\frac{4}{3}
s_w^2\right)\overline{Q}^{(n)}_t \gamma^\mu Q^{(n)}_t +
\left(-1+\frac{2}{3} s_w^2\right)\overline{Q}^{(n)}_b \gamma^\mu
Q^{(n)}_b  + \left(-\frac{4}{3} s_w^2\right)\overline{U}^{(n)}
\gamma^\mu U^{(n)}.
\end{equation}
Finally, the couplings with the KK modes of the Higgs doublet charged
components are
\begin{equation}
\label{eq:couplingZHiggs}
J^{\mu(n)}_{\Phi}=  (-1 +2s_w^2)\Phi^{+(n)}i\partial^\mu\Phi^{-(n)}
+\hc
\end{equation}
From \Eq{eq:consm}, \Eq{eq:conKK} and \Eq{eq:couplingZHiggs} it is now
straightforward to extract the correspondent Feynman rules for the
couplings with the $Z$. The couplings of the photon can be derived
similarly.

\section{Phenomenology}
The detection of the first members of the KK towers would be a
compelling signature in favor of extra dimensions.
%One of the most compelling signals of the extra dimensions would be
%the detection of the KK towers 
But until now, there is no direct detection of any member of these
towers in the experiments.  This means that we have to look for their
contributions to observables through virtual production. Since these
are expected to be small, the best places to look for them are
processes that have been experimentally measured with high degree of
precision or that can only proceed through radiative corrections in
the SM. Among the formers we will study in the next sections the $Z\to
b\bar{b}$ decay, the radiative corrections to the $\rho$ parameter and
the $B^0-\overline{B}^0$ mixing, and among the latter the process
$b\to s \gamma$.

\subsection{Radiative corrections to the $Z\to b \overline{b}$
decay} 
\label{sec:Zbb}
%In this section we will compute the corrections to the
%effective \( Zb\bar{b} \) coupling. 
Shifts in the \( Zb\bar{b} \) coupling due to radiative corrections,
either from within the SM or from new physics, affect observables such
as the branching ratio $R_{b}=\Gamma _{b}/\Gamma _{h}$, where $\Gamma
_{b}=\Gamma (Z\rightarrow b\bar{b})$ and $
\Gamma _{h}=\Gamma (Z\rightarrow \mathrm{hadrons})$, or the left
right asymmetry $A_b$. These type of corrections can be treated
uniformly by expressing them as a modification to the tree level
couplings $g_{L(R)}$ defined as
\begin{equation}
\label{eq:definitions} \frac{g}{c_W}\overline{b} \gamma^\mu(g_L
P_L +g_R P_R)b Z_\mu~.
\end{equation}
$Z$ and $b$'s are SM fields, $P_{L(R)}$ are the chirality
projectors and
\begin{eqnarray}
g_L& = & -\frac{1}{2}+\frac{1}{3}s_W^2+\delta g_L^{\mathrm{SM}}
+\delta g_L^{\mathrm{NP}}~,\\
g_R & = &\frac{1}{3}s_W^2 + \delta g_R^{\mathrm{SM}} +\delta
g_R^{\mathrm{NP}}~,
\end{eqnarray}
where we have separated radiative corrections coming from SM
contributions and from new physics, (NP). It turns out that, both
within the SM as well as in most of its extensions,  only $g_L$
receives corrections proportional to $m_t^2$ at the one loop
level, due to the difference in the couplings between the two
chiralities. In particular, a shift $\delta g_L^{NP}$ in the value
of $g_L$
 due to new physics
translates into a shift in $R_{b}$ given by
\begin{equation}
\label{eq:rbab} \delta R_b = 2 R_b(1-R_b)\frac{g_L}{g_L^2+g_R^2}
\delta g_L^{\mathrm{NP}}~,
\end{equation}
and to a shift in the left-right asymmetry $A_b$ given by
\begin{equation}
\label{eq:asym} \delta A_b = \frac{4 g_R^2 g_L}{(g_L^2+g_R^2)^2}
\delta g_L^{\mathrm{NP}}~.
\end{equation}
These equations, when compared with experimental data, will be
used to set bounds on the compactification scale.

By far the easiest way to compute the leading top-quark-mass
dependent one-loop corrections to $\delta g_L$ from the SM itself,
$\delta g_L^{\mathrm{SM}}$, is to resort to the gaugeless limit of
the SM~\cite{Lytel:1980zh}, $e.g.$ the limit where the gauge
couplings $g$ and $g^\prime$, corresponding to the gauge groups
$SU(2)_L$ and $U(1)_Y$ respectively, are switched off. In that
limit the gauge bosons play the role of external sources and the
only propagating fields are the quarks, the Higgs field, and the
charged and neutral Goldstone bosons $G^{\pm}$ and $G^0$. As
explained in \cite{Barbieri:1993dq,Barbieri:1992nz} one may relate
the one-loop vertex $Z b \bar{b}$ to the corresponding $G^0 b
\bar{b}$ vertex by means of a Ward identity; the latter is a
direct consequence of current conservation, which holds for the
neutral current before and after the Higgs doublet acquires a
vacuum expectation value $v$.
\begin{figure}
\begin{center}
\includegraphics[scale=0.6]{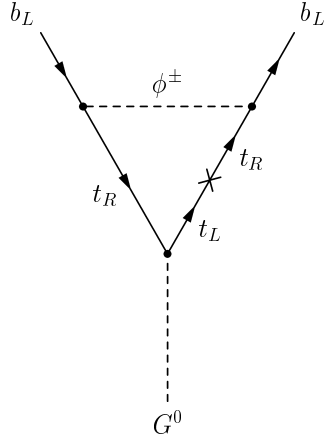}
\end{center}
\caption{The diagram contributing
to the SM $G^0 b \bar{b}$ vertex in the gaugeless limit for massless
$b$-quarks.}
\label{fig:gaugelessZbbinSM}
\end{figure}

In practice, carrying out the calculation in the aforementioned
limit amounts to the elementary computation of the one-loop
off-shell vertex $G^0 b \bar{b}$. In the gaugeless limit and for
massless $b$-quarks the only contribution to this vertex is
depicted in Fig.~\ref{fig:gaugelessZbbinSM}, where the cross in
the top-quark line represents a top-quark mass insertion needed to
flip chirality (the diagram with an insertion in the other top-quark line is
assumed). This diagram gives a derivative coupling of the
Goldstone field to the $b$-quarks which can be gauged (or related
to the $Z$ vertex through the Ward identity) to recover the $Z b
\bar{b}$ vertex. Then, one immediately finds
\begin{equation}
\label{eq:deltasm} \delta g_{L}^{\mathrm{SM}}\approx
\frac{\sqrt{2}G_{F}m_{t}^{4}}{(2\pi )^{4}}\, \int \frac{i
d^{4}k}{(k^{2}-m_{t}^{2})^{2}k^{2}}
=\frac{\sqrt{2}G_{F}m_{t}^{2}}{(4\pi )^{2}}~,
\end{equation}
where $G_F$ is the Fermi constant. The $m_t^4$ dependence
(coming from three Yukawa couplings and one mass insertion) is
partially compensated by the $1/m_t^2$ dependence coming from the
loop integral.

\begin{figure}
\begin{center}
\includegraphics[scale=0.6]{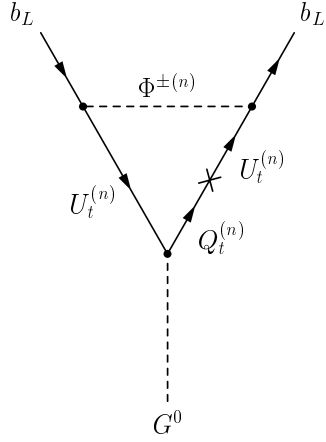}
\caption{\label{fig:gaugelessZbbinUED1}The dominant diagram
contributing to the UED $G^0 b \bar{b}$ vertex in the gaugeless
limit for massless $b$-quarks.}
\end{center}
\end{figure}

In the case of a single UED this argument persists: one must
simply consider the analog of diagram in
Fig.~\ref{fig:gaugelessZbbinSM}, where now the particles inside
the loop have been replaced by their KK modes, as shown in
Fig.~\ref{fig:gaugelessZbbinUED1}. If we denote by $\delta
g_L^{UED}$ the new physics contributions in the UED model (the SM
contributions are not included) the result is
\begin{eqnarray}
\delta g_{L}^{\mathrm{UED}} &\approx&
\frac{\sqrt{2}G_{F}m_{t}^{4}}{(2\pi )^{4}}\,  \sum_{n=1}^\infty
\int
\frac{i d^{4}k}{(k^{2}-m_{Q}^{2})^{2}(k^{2}-m_{n}^{2})}\nonumber \\
&=& \frac{\sqrt{2}G_Fm_t^4}{(4\pi)^2} \sum_{n=1}^{\infty}\int_0^1
\frac{dx x}{x m_t^2 + m_n^2} \approx
\frac{\sqrt{2}G_{F}m_{t}^{4}}{(4\pi )^{2}} \frac{\pi^2 R^2}{12}~,
\label{eq:deltaUED}
\end{eqnarray}
and depends {\it quartically} on the mass of the top quark. Notice
that there are several differences with respect to the SM: (i) The
cross now represents the mixing mass term between $Q_t^{(n)}$ and
$U^{(n)}_t$, which is proportional to $m_t$; (ii) The $\Phi^{\pm(n)}$,
for $n\not=0$, are essentially the physical KK modes of the charged
Higgses as shown in Eq.\rf{eq:physicalHiggs}; (iii) From the virtual
momentum integration one obtains now a factor $1/m_n^2$, instead of
the factor $1/m_t^2$ of the SM case.

\begin{figure}
\begin{displaymath}
\begin{array}{ccc}
\includegraphics[scale=0.5]{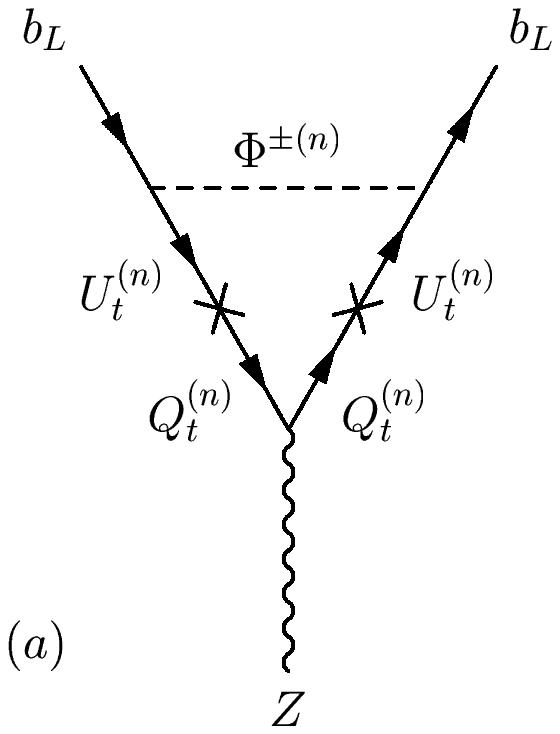} &
\includegraphics[scale=0.5]{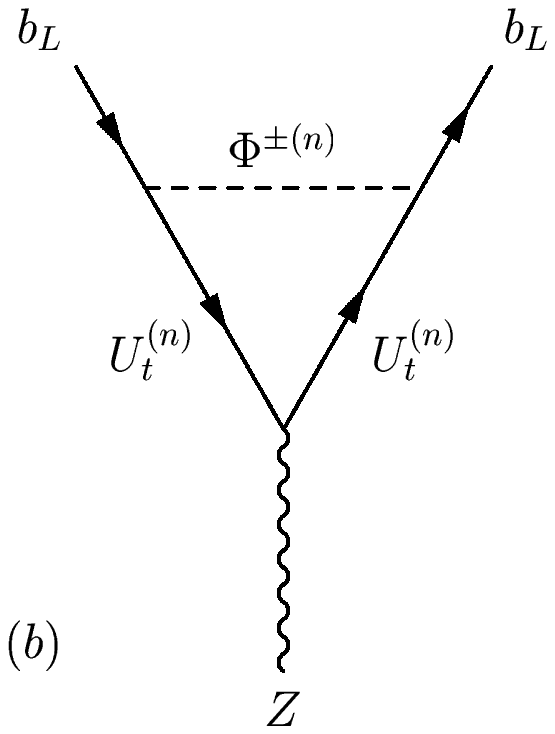} &
\includegraphics[scale=0.5]{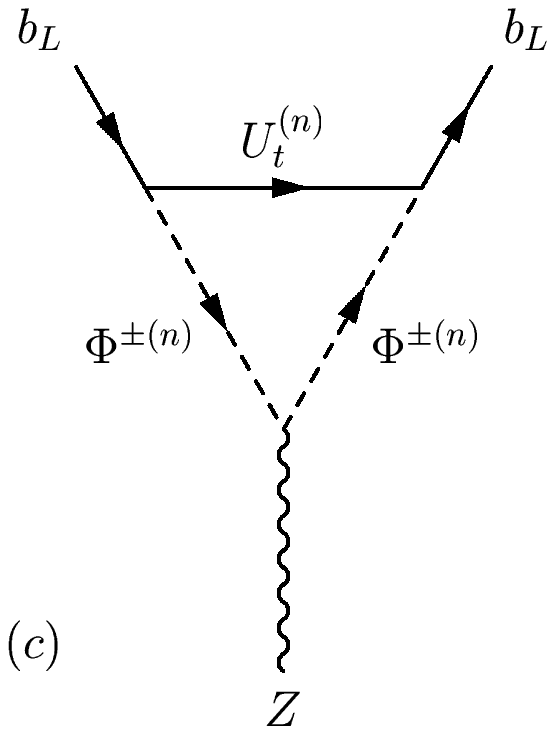}
\end{array}
\end{displaymath}
\begin{displaymath}
\begin{array}{cc}
\includegraphics[scale=0.5]{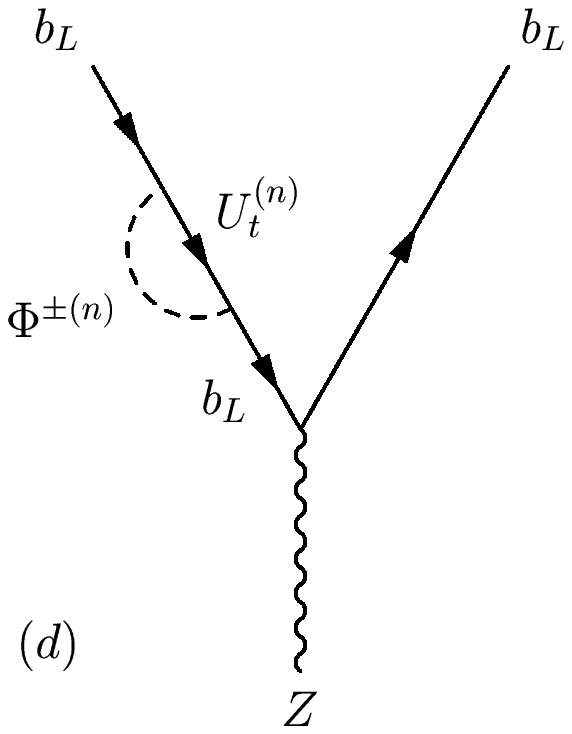} &
\includegraphics[scale=0.5]{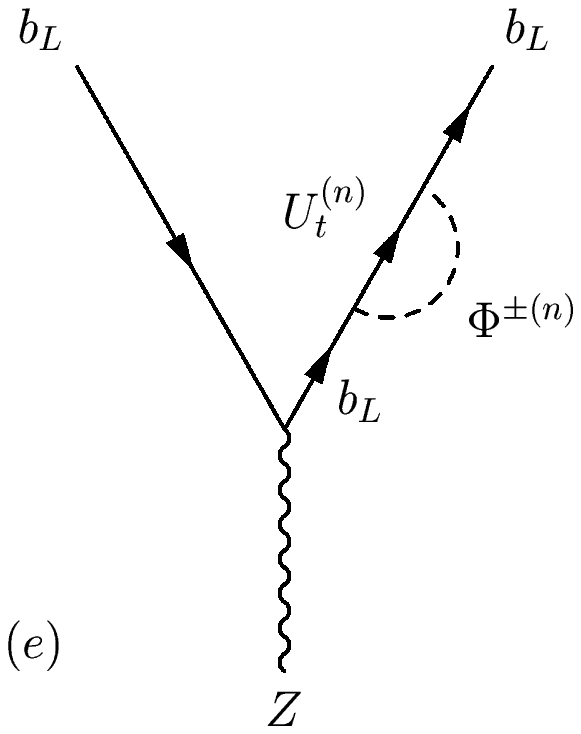}
\end{array}
\end{displaymath}
\caption{Dominant UED contributions to the $Zb\bar{b}$ vertex.}
\label{fig:ZbbinUED}
\end{figure}

This simple calculation allows us to understand easily the leading
corrections arising from extra dimensions. A more standard
calculation of the $Zb\bar{b}$ vertex in UED yields exactly the
same result. In this case the radiative corrections to the
$Zb\bar{b}$ vertex stem from the diagrams of
Fig.~\ref{fig:ZbbinUED}.

If we neglect the $b$-quark mass and take $M_Z\ll R^{-1}$, the
result, for each mode, can be expressed in terms of a single
function, $f(r_n)$, defined as
\begin{equation}
i {\mathcal M}^{(n)}= i \frac{g}{c_w} \frac{\sqrt{2}G_F
m_t^2}{(4\pi)^2} f(r_n) \overline{u}^\prime \gamma^\mu P_L u
\epsilon_\mu~,
\end{equation}
where $u$ and $u^\prime$ are the spinors of the $b$ quarks and
$\epsilon_\mu$ stands for the polarization vector of the $Z$
boson. The argument of the function $f$ is $r_n = m_t^2/m_n^2$.

Although the complete result is finite, partial results are
divergent and are regularized by using dimensional regularization.
The contributions of the different diagrams in
Fig.~\ref{fig:ZbbinUED} are
\begin{eqnarray}
f^{(a)}(r_n)& =& \left(1-\frac{4}{3}s_w^2 \right)
\left[\frac{r_n-\log(1+r_n)}{r_n}\right]~,\nonumber\\
f^{(b)}(r_n)& = &\left(-\frac{2}{3}s_w^2 \right)\left[\delta_n -1
+ \frac{2r_n+r_n^2-2(1+r_n^2)\log(1+r_n)}{2r_n^2}\right]
 ~,\nonumber\\
f^{(c)}(r_n) & = & \left(-\frac{1}{2}+s_w^2\right) \left[ \delta_n
+ \frac{2r_n+3 r_n^2-2(1+r_n)^2 \log(1+r_n)}{2r_n^2}\right]
 ~,\nonumber\\
f^{(d)}(r_n)+f^{(e)}(r_n)&=&\left(\frac{1}{2}-\frac{1}{3}s_w^2\right)\left[
\delta_n + \frac{2r_n+3 r_n^2-2(1+r_n)^2
\log(1+r_n)}{2r_n^2}\right]~, \label{eq:Zbbresults}
\end{eqnarray}
%%%%%%%%%%%%%%%%%%%%%%%%%%%%%%%%%%%%%%%%%%%%%%%%%%%%%%%%%%%%%%%%%%%%%%
%%%%%%%%%%%%%%%%%%%%%%%%%%%%%%%%%%%%%%%%%%%%%%%%%%%%%%%%%%%%%%%%%%%%%%%%%
where $ \delta_n \equiv
 2/\epsilon - \gamma + \log(4\pi) + \log(\mu^2/m_n^2),
$ and $\mu$ is the 't Hooft mass scale. From
Eq.~(\ref{eq:Zbbresults}) it is straightforward to verify that all
terms proportional to $ \delta_n$ cancel, and so do all terms
proportional to $s_w^2$, as expected from the gaugeless limit
result. Thus, finally, the only term which survives is the term in
$f^{(a)}(r_n)$ not proportional to $s_w^2$, yielding the following
(per mode) contribution to $g_L$:
\begin{equation}
\delta g_L^{(n)} = \frac{\sqrt{2}G_F m_t^2}{(4\pi)^2}
\left[\frac{r_n-\log(1+r_n)}{r_n}\right]~,
\end{equation}
which is precisely the one obtained from the gaugeless limit
calculation, $e.g.$ \Eq{eq:deltaUED} with the elementary
integration over the Feynman parameter $x$ already carried out.
 Notice also that
the above result is consistent with the decoupling theorem since
the contribution for each mode vanishes when its mass is taken to
infinity, i.e. $r_n\to 0$.

In order to compute the effect of the entire KK tower, it is more
convenient to first carry out the sum and then evaluate the
Feynman parameter integral; this  interchange is mathematically
legitimate since the final answer is convergent. Thus,
\begin{equation}
\label{eq:forLUED} \delta g_L^{\mathrm{UED}} =
\sum_{n=1}^{\infty}\delta g_L^{(n)} = \frac{\sqrt{2}G_F
m_t^2}{(4\pi)^2} \int_0^1 dx \sum_{n=1}^{\infty} \frac{r_n
x}{1+r_nx} = \frac{\sqrt{2}G_F m_t^2}{(4\pi)^2}
F_{\mathrm{UED}}(a)~,
\end{equation}
where $a=\pi R m_t$, and the function $F(a)$ is defined in general as
\begin{equation}
\label{eq:deffa}
F(a)\equiv \frac{\delta g_L^{\mathrm{NP}}}{\delta g_L^{\mathrm{SM}}}.
\end{equation}
In the case of UED 
\begin{equation}
F_{\mathrm{UED}}(a) = - \frac{1}{2} + \frac{a}{2} \int_0^1\;dx\;
\sqrt{x}\coth(a\sqrt{x})\approx \frac{a^2}{12}-\frac{a^4}{270}+\mathcal{O}(a^6)~.
\label{eq:faued}
\end{equation}
It is instructive to compare the above result with the one obtained in
the context of models where the extra dimension is not
universal\footnote{In Chapter \ref{chap:HGLHG} we study some of these
models.}. In particular, in the model considered in
Ref.~\cite{Papavassiliou:2000pq} the fermions live in four dimensions,
and only the gauge bosons and the Higgs doublet live in five
\cite{Pomarol:1998sd}. In this case there is no KK tower for the
fermions, and therefore, in the loop-diagrams appear only the SM
quarks interacting with the KK tower of the Higgs fields. The result
displays a logarithmic dependence on the parameter $a$, which gives
rise to a relatively tight lower bound on $R^{-1}$, of the order of 1
TeV. Specifically, the corresponding $F(a)$ is given by\footnote{Note
that, unlike in Ref.~\cite{Papavassiliou:2000pq}, the $F(a)$ does not
include the SM contribution.}
\begin{equation}
F(a)=-1+2a\int_0^\infty\;dx \frac{x^2}{(1+x^2)^2} \coth(ax)\approx
\left(\frac{2}{3}\log(\pi/a)-\frac{1}{3}-\frac{4}{\pi^2}\zeta^\prime(2)\right)
a^2~, \label{eq:fahg}
\end{equation}
where the expansion on the second line holds for small values of $a$,
and \( \zeta' \) is the derivative of the Riemann Zeta function. The
appearance of the $\log(a)$ in $F(a)$ and its absence from
$F_{\mathrm{UED}}(a)$ may be easily understood from the effective
theory point of view: due to the KK-number conservation in UED models,
the tree-level low energy effective Lagrangian when all KK modes are
integrated out is exactly the Standard Model; there are no additional
tree-level operators suppressed by the compactification scale. Since
the one-loop logarithmic contributions in the full theory can be
obtained in the effective theory by computing the running of operators
generated at tree level, it is clear that in the UED no $\log(a)$ can
appear at one loop in low energy observables. The situation is
completely different if higher dimension operators are already
generated at tree level, as is the case of the model considered in
Ref.~\cite{Papavassiliou:2000pq}, where the leading logarithmic
corrections can be computed by using the tree-level effective
operators in loops.

We next turn to the bounds on $R^{-1}$. We will use the SM
prediction \mbox{$R_b^{\mathrm{SM}}=0.21569 \pm 0.00016$} and the
experimentally measured value \mbox{$R_b^{\mathrm{exp}}=0.21664 \pm
0.00068$}. Combining Eq.~\rf{eq:rbab} and Eq.~\rf{eq:deffa}, we
obtain $F(a)= -0.24 \pm 0.31$, and making a weak
signal treatment \cite{Feldman:1998qc} we arrive at the 95\% CL
bound $F(a)<0.39$. The results for a single UED can
be easily derived from \Eq{eq:faued}, yielding
\begin{equation}
R^{-1}> 230\;\mbox{GeV}~\qquad 95 \% \mbox{CL}.
\end{equation}
The SM prediction for the left-right asymmetry $A_b^{SM}=0.9347\pm
0.0001$, and the measured value $A_b^{exp}=0.921\pm 0.020$ gives a
looser bound.

\comment{Above we have computed only the leading contribution,
which goes as $G_F m_t^4 R^2$. There are also formally subleading
contributions, suppressed by (at least) an additional factor
$M_W^2/m_t^2$; given that this factor is not so small such
corrections could be numerically important, and should be
estimated. The dominant contributions of this type come from
diagrams with $W_\mu^{\pm(n)}$ and $W_5^{\pm(n)}$ running in the
loops. Since these corrections are still proportional to $m_t^2$
they can be estimated using again the Ward identity that relates
the $G^{0}$ couplings to the $Z$ couplings. The relevant diagrams
are shown in Fig.~\ref{fig:subleading}.
\begin{figure}
\begin{displaymath}
\begin{array}{cc}
\includegraphics[scale=0.6]{gaugelessZbbinUED2.eps}&
\includegraphics[scale=0.6]{gaugelessZbbinUED3.eps}
\end{array}
\end{displaymath}
\caption{Diagrams giving subleading contributions to the
$G^{0}b\bar{b}$ vertex.} \label{fig:subleading}
\end{figure}
Their contribution modify the value of $\delta g_L^{\mathrm{UED}}$
as follows:
\begin{equation}
\delta g_L^{\mathrm{UED}} = \frac{\sqrt{2}G_F m_t^2}{(4\pi)^2}
F_{\mathrm{UED}}(a) \left(1+3\frac{M_W^2}{m_t^2}\right)~.
\end{equation}
Taking these corrections into account leads to a slight
modification of the bound on the compactification scale, $R^{-1}>
300\;\mbox{GeV}$. Evidently, this bound is absolutely comparable
to the one obtained from the $\rho$ parameter.}

\subsection{Radiative corrections to $b\to s \gamma$} The
experimental observable is the semi-inclusive decay
$\mbox{Br}(B\to X_s\gamma)$. Using heavy quark expansion it is
found that, up to small bound state corrections, this decay agrees
with the parton model rates for the underlaying decays of the $b$
quark \cite{Falk:1994dh,Neubert:1994ch}, $b\to s \gamma$. This
flavor violating transition is a very good place to look for new
physics because in the SM it is forbidden at tree level due to
gauge symmetry, it can though proceed through radiative
corrections. From an effective field theory point of view the
transition can be understood as due to the generation via
radiative corrections of the next effective Hamiltonian
\begin{equation}
\label{eq:effhamiltonian}
\mathcal{H}_{eff}=\frac{4G_F}{\sqrt{2}}V_{ts}^\ast
V_{tb}\sum_{i=1}^8 C_i\mathcal{O}_i
%\frac{e}{(4\pi)^2} m_b \overline{s}\sigma^{\mu\nu} P_R b F_{\mu\nu}
\end{equation}
where $\mathcal{O}_7$ operator is the one that drives the
transition $b\to s \gamma$
\begin{equation}
\mathcal{O}_7= \frac{e}{(4\pi)^2}m_b \overline{s}\sigma^{\mu\nu}
P_R b F_{\mu\nu}.
\end{equation}
Notice that the presence of the strength tensor, $F_{\mu\nu}$,
guarantees the gauge invariance of $\mathcal{O}_7$. As usual, the
operators encode the low energy physics while the high energy physics
is contained in the coefficients, in this case $C_7$.

In the SM and at the scale of the $W$ mass,
$C_7^{SM}(M_W)=-1/2~A(m_t^2/M_W^2)$, where
\begin{equation}
\label{eq:defa} A(x)=x\left[ \frac{\frac{2}{3}x^2 + \frac{5}{12}x
-\frac{7}{12}}{(x-1)^3} - \frac{\left(\frac{3}{2}x^2-x\right)\ln
x}{(x-1)^4} \right]
\end{equation}

The leading logarithmic contributions of the two loop calculations are
important, these come from standard QCD running from $M_W$ to
$m_b$. The RGE reads
\cite{Agashe:2001xt,Buchalla:1996vs}
\begin{equation}
\label{eq:running} 
C_7(m_b)\approx
0.698\;C_7(M_W)-0.156\;C_2(M_W)+0.086\;C_8(M_W),
\end{equation}
where $C_2$ and $C_8$ are the coefficients of the operators
$\mathcal{O}_2$ and $\mathcal{O}_8$ respectively, which are
defined as
\begin{eqnarray}
\mathcal{O}_2 & = &[\overline{c}_{L\alpha} \gamma_\mu
b_{L\alpha}][\overline{s}_{L\beta} \gamma^\mu c_{L\beta}] ,\\
\mathcal{O}_8 & = & \frac{g_s}{(4\pi)^2} m_b
\overline{s}_{L\alpha}\sigma^{\mu\nu} T^a_{\alpha\beta} b_{R\beta}
G^a_{\mu\nu}.
\end{eqnarray}
$\alpha$ and $\beta$ are color indexes. In the case of SM, the
contribution of $\mathcal{O}_8$ is negligible, $C_8(M_W)=-0.097$
\cite{Buchalla:1996vs}, but the one of $\mathcal{O}_2$,
$C_2(M_W)=1$, turns out to be important.

\begin{figure}
\begin{displaymath}
\begin{array}{c}
  \begin{array}{ccc}
    \includegraphics[scale=0.6]{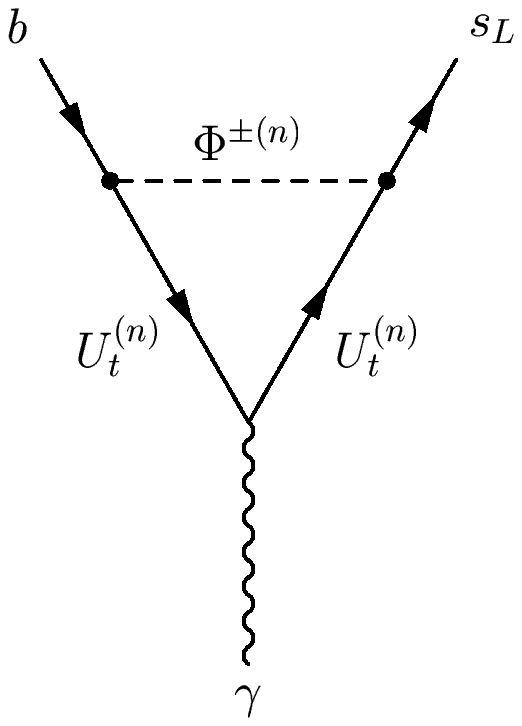}   &
    \includegraphics[scale=0.6]{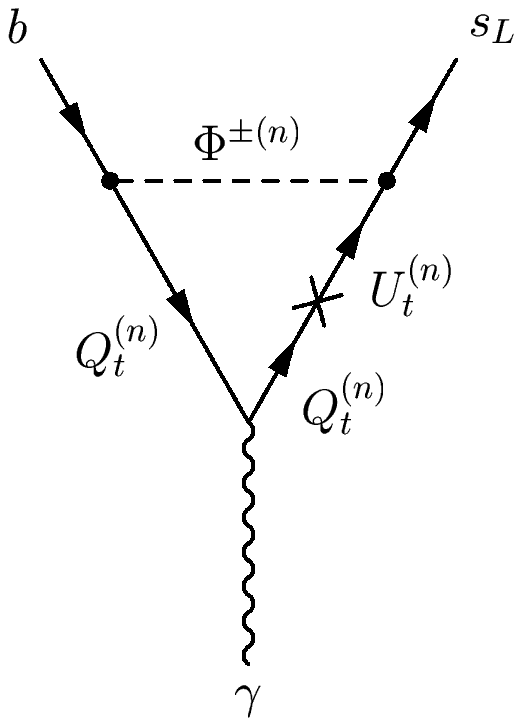}  &
    \includegraphics[scale=0.6]{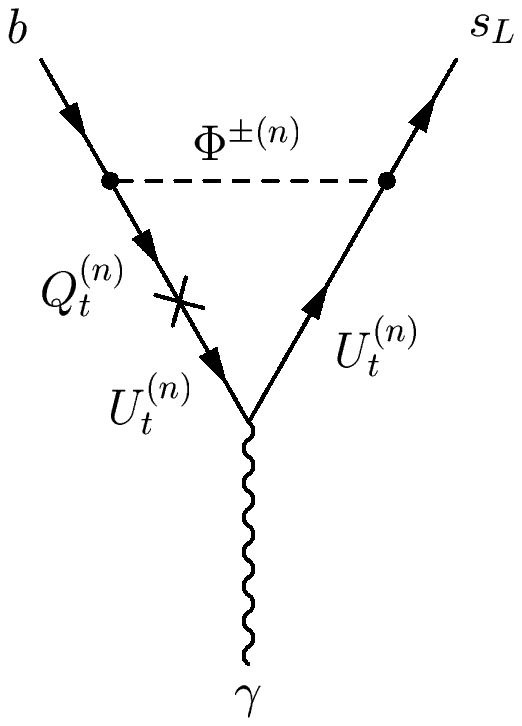}
  \end{array}
  \\
%\end{displaymath}
%%%%%%%%%%%%%%%%%%%%%%%%%%%%%%%%%%%%%%%%%%%%%%%%%%%%%%%%%%%%%%%%%%%%%%%%%
%%%%%%%%%%%%%%%%%%%%%%%%%%%%%%%%%%%%%%%%%%%%%%%%%%%%%%%%%%%%%%%%%%%%%%%
%\begin{displaymath}
  \begin{array}{cc}
    \includegraphics[scale=0.6]{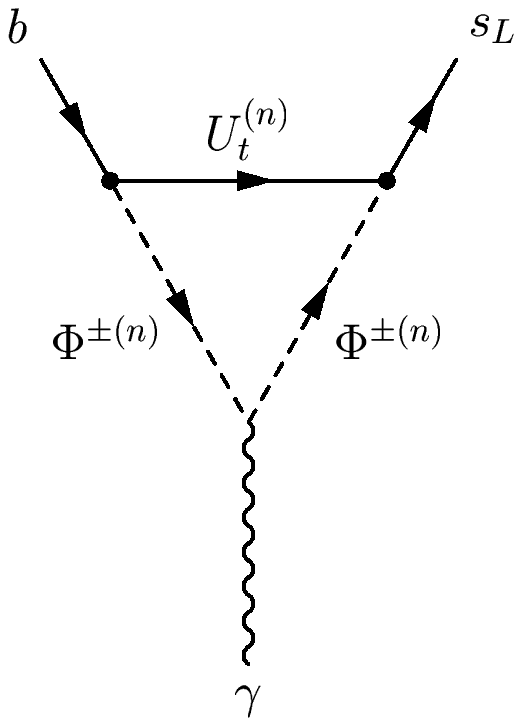}&
    \includegraphics[scale=0.6]{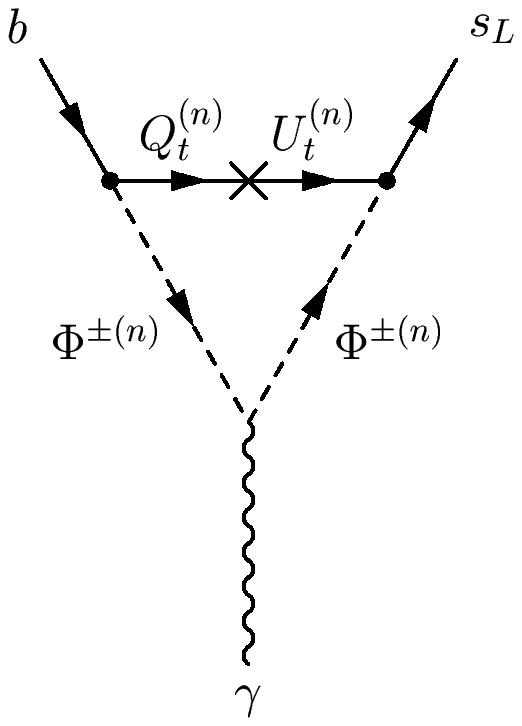}
  \end{array}
\end{array}
\end{displaymath}
\caption{Diagrams that contribute to $\mathcal{O}_7$ in UED.}
\label{fig:bsginUED}
\end{figure}

In UED the transition also proceeds through the same effective
Hamiltonian but the coefficient $C_7$ receives new contributions aside
the ones coming from SM, and the running could be different for each
case because $C_2$ and $C_8$ are also modified. The value of $C_7$ at
the scale $M_W$ stems from the diagrams in
\Fig{fig:bsginUED}

There are also diagrams in which the $\Phi^{\pm (n)}$ are replaced by
$W_{\mu}^{(n)}$ and by the non physical 
degrees of freedom $W_{5}^{(n)}$ but since the couplings of these
are reduced by a factor $(M_W/m_t)^2\approx~0.22$ we will ignore
them and work at this level of precision.

Observe that we have not considered the self energies of the external
legs, in opposition to what we did with $Z\to b\overline{b}$ because
these diagrams do not contribute to $\mathcal{O}_7$. In addition, they
are proportional to $m_t$ and its structure is of the form
$\overline{u}\gamma^\mu u$; when all the contributions with this
structure are added, they cancel exactly. Hence, the $Z_\mu
\overline{b} \gamma^\mu b$ vertex does not
appear as it should be since it is forbidden by gauge invariance.

The contribution of the n-th mode to the $C_7$ coefficient can be
written in the form \cite{Agashe:2001xt}
\begin{equation}
C_{7}^{(n)}= \frac{m_t^2}{m_t^2+m_n^2} \left[
B\left( \frac{m_t^2+m_n^2}{m_n^2}\right) - \frac{1}{6} A \left(
\frac{m_t^2+m_n^2}{m_n^2} \right) \right]
\end{equation}
where $B(x)$ is given by
\begin{equation}
\label{eq:defb} B(x)= \frac{x}{2} \left[ \frac{\frac{5}{6} x -
\frac{1}{2}}{(x-1)^2} - \frac{\left( x - \frac{2}{3}\right)\log
x}{(x-1)^3} \right]
\end{equation}

This result can be obtained by direct calculation but the more elegant
reasoning given in Ref.~\cite{Agashe:2001xt} is also possible.  An
expansion of $C_{7}^{(n)}$ reveals that it is free of logarithms that
relate the two different mass scales: $m_n$ and $m_t$.
\begin{equation}
\label{eq:freelogs}
  C_{7}^{(n)}=\frac{23}{144}\frac{m_t^2}{m_n^2} -
  \frac{13}{120}\frac{m_t^4}{m_n^4} + \mathcal{O}\left(\frac{m_t^6}{m_n^6}\right).
\end{equation}
This result can be understood by using effective field theory ideas:
when the heavy degrees of freedom are integrated out, the tree level
effective Lagrangian is exactly the SM, there are no additional tree
level operators suppressed by powers of $m_n^{-1}$. It is well known
that the dominant logarithms that may appear relating the two
different scales can be recovered from the running of the operators in
the low energy effective Lagrangian induced by the presence of the
additional operators, since in our case they are not present no logs
can appear in \Eq{eq:freelogs}.  Finally, all contributions must be
put together
\begin{equation}
C_{7}^{\mathrm{UED}}(M_W)=C_7^{\mathrm{SM}}(M_W)+\sum_{n=1}^{\infty}
C_{7}^{(n)}(M_W),
\end{equation}
where we have neglected the running between $m_t$ and $M_W$, i.e.
$C_{7}^{\mathrm{UED}}(m_t)\approx~C_{7}^{\mathrm{UED}}(M_W)$.

\comment{The fact that one of the external legs has a fixed chirality
implies that there is no extra factor two between the zero mode,
SM, and the modes of the KK tower.}

Once we have determined $C_7(M_W)$ the next step is to apply the
RGE given in \Eq{eq:running} and compute $C_7(m_b)$. To this end
we need $C_2(M_W)$ and $C_8(M_W)$, $C_2(M_W)=1$, i.e. it
takes the SM value because it is a contribution at tree level
while the UED contributions are at the one-loop level. In
addition, the small coefficient of $C_8$ in the equation
\rf{eq:running} and the fact that $C_8$ is expected to be small
allows us to neglect again this term in the running. The
modifications to $b\to cl\nu$, the necessity of which will be
explained later, are also negligible because UED corrects it again
at the one loop level and in the SM it is already corrected at
tree level.

To extract the bounds this process sets, it is used the observable
\begin{equation}
\widetilde{\Gamma}=\frac{\Gamma(b\to s\gamma)}{\Gamma(b\to cl\nu)}
\end{equation}
that lacks the $m_b$ dependence and therefore presents smaller
uncertainty \cite{Kagan:1998ym}: 10\% for the theoretical value
and 15\% for the experimental determination (both at $1\sigma$).
When compared, it is found that if a 95\% CL is required the
current bounds allow a modification as big as a 36\% with respect
to the SM value \cite{Agashe:2001xt}, i.e.
%\begin{equation}
$|\widetilde{\Gamma}^{total}/\widetilde{\Gamma}^{SM}-1|\leq 0.36$.
%\end{equation}
 Since the process $b\to c l \nu$ is not modified by the new
physics the previous equation can be easily translated into the
more useful one
\begin{equation}
\label{eq:boundonc7}
\left| \frac{|C_7^{total}(m_b)|^2}{|C_7^{SM}(m_b)|^2} - 1
\right|<0.36 \qquad 95\%\;\mbox{CL},
\end{equation}
and from this the bound can be found to be
\begin{equation}
R^{-1} \leq 300\;\mbox{GeV}    \qquad 95\%\;\mbox{CL}.
\end{equation}

\subsection{Radiative corrections to the $\rho$ parameter} 
The $\rho$ parameter can be defined as the ratio of the relative
strength of neutral to charged current interactions at low momentum
transfer. In the SM, and at tree level, it is predicted to be unity as
a consequence of the custodial symmetry of the Higgs potential:
\begin{equation}
\rho\equiv\frac{G_{NC}(0)}{G_{CC}(0)}\approx \frac{M_W^2}{c_W^2
M_Z^2}=1~.
\end{equation}
Because the SM contains couplings that violate the symmetry (the
Yukawa couplings and the U(1) coupling $g'$) radiative corrections
modify the tree-level value of $\rho$. At one loop, $\rho$ receives
corrections from vertex, box and gauge-boson self-energy diagrams;
however the dominant contributions, proportional to $m_t^2$, come from
the top-quark loops inside the gauge boson self-energies. Keeping only
these contributions, one has
\begin{equation}
\label{eq:defrho}
\rho=1+\frac{\Sigma_W(0)}{M^2_W}-\frac{\Sigma_Z(0)}{M^2_Z} \approx
1+\frac{1}{M_W^2}\bigg(\Sigma_1(0)-\Sigma_3(0)\bigg) \approx 1+N_c
\frac{\sqrt{2}G_F m_t^2}{(4 \pi)^2}.
\end{equation}
$\Sigma_W(0)$ and $\Sigma_Z(0)$ are co-factors of the $g^{\mu\nu}$
in the one-loop self-energies of the $W$ and $Z$ bosons, evaluated
at $q^2=0$, and $\Sigma_1(0)$ and $\Sigma_3(0)$ are the equivalent
functions for the $W_1$ and $W_3$ components of the SU(2) gauge
bosons. In arriving at the above formula one uses the fact that
the photon-$Z$ self-energy $\Sigma_{AZ}^{\mu\nu}$ is transverse,
i.e.  $\Sigma_{AZ} (0)=0$; this last property holds  {\it only}
for the subset of graphs containing fermion-loops, but is no
longer true when gauge-bosons are considered inside the loops of
$\Sigma_{AZ}$~\cite{Degrassi:1993kn,Papavassiliou:1996hj}.
Finally, $N_c$ is the number of colors.

\begin{figure}
\begin{displaymath}
\begin{array}{c}
  \begin{array}{cc}
    \includegraphics[scale=0.6]{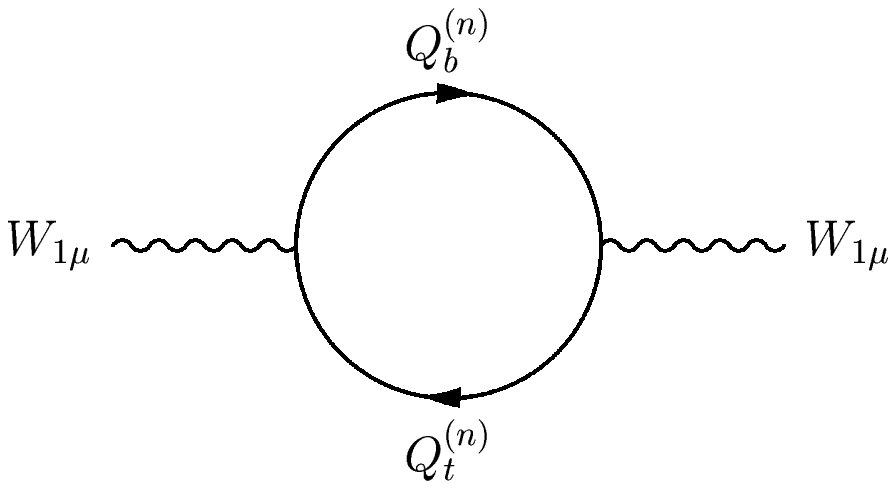}   &
    \includegraphics[scale=0.6]{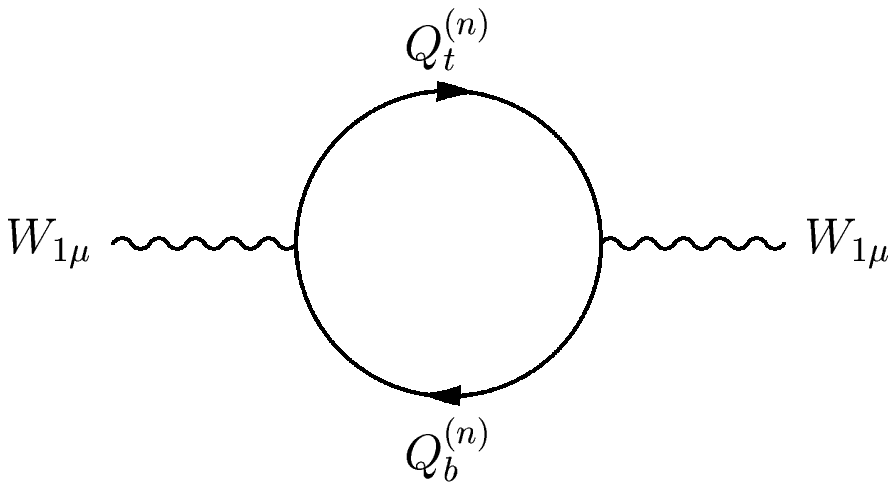}
  \end{array}
  \\
  \begin{array}{c}
    \includegraphics[scale=0.6]{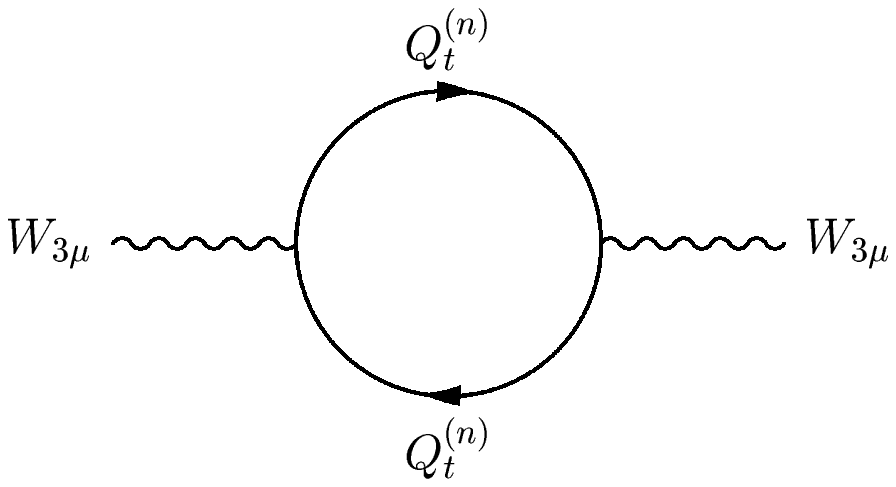}
  \end{array}
\end{array}
\end{displaymath}
\caption{Diagrams that contribute to the $\rho$ parameter in UED.}
\label{fig:rhoparaUED}
\end{figure}

In UED the tree-level value is the same as in SM because, as we have
seen, the first corrections appear at the one-loop level. The
relevant diagrams are shown in \Fig{fig:rhoparaUED}.

As we are using the interaction fields, only the $Q$ fields
contribute. Since these are not the mass eigenfields there are more
diagrams with the propagators that mix $Q$ and $U$ fields, but the
latter are singlets under $SU(2)$, and, therefore, do not couple to
$W_\mu^{(n)}$. No diagram with $U$ fields gives any contribution,
which is the reason why in this base the calculus is much easier. Had
we chosen to work with the mass eigenstates, then we would had to
consider all fields inside the loop because all of them have
projection on the $Q$ fields and the vertices would have been more
complicate containing combinations of $\sin(\alpha_n)$ and
$\cos(\alpha_n)$.
% that at the end of the day what they are doing is
%just projecting the mass fields onto the prime ones.

The self-energy corrections due to a single mode are
\begin{eqnarray}
\label{eq:selfenergies}
i \Sigma_{1}^{(n)}(q^2)&=& i \Sigma_{1}^{(a)} + i \Sigma_{1}^{(b)}
= 2 g^2 N_c m_t^2 \frac{i}{(4 \pi)^2}
\left[\frac{1}{2\hat{\epsilon}} - b_1(m_n^2,m_Q^2,q^2)
\right]\\ i \Sigma_{3}^{(n)}(q^2) &=& i \Sigma_{3}^{(c)}
= 2 g^2 N_c m_t^2 \frac{i}{(4 \pi)^2}
\left[\frac{1}{2\hat{\epsilon}} - b_1(m_Q^2,m_Q^2,q^2) \right],
\nonumber
\end{eqnarray}
where
\begin{equation}
\label{eq:defepsilon} m_Q^2 = m_n^2+m_t^2 \qquad
\frac{1}{\hat{\epsilon}}=
 \frac{2}{\epsilon} - \gamma + \log(4\pi) + \log(\mu^2)
\end{equation}
$\mu$ is an arbitrary mass scale introduced in dimensional
regularization. The function $b_1$ is defined as
\begin{equation}
\label{eq:defbs} b_1(m_1^2, m_2^2, q^2) \equiv \int_0^1\;dx\;
x\log\left(\frac{\Delta(m_1^2, m_2^2, q^2)}{\mu^2}\right)
\end{equation}
and
\begin{equation}
\Delta(m_1^2, m_2^2, q^2) = xm_2^2+(1-x)m_1^2-x(1-x)q^2.
\end{equation}

The total contribution is found by summing up the whole KK tower.
This can be done with a bit of care: first consider the
contribution of the first $N$ modes which can be obtained from
\Eq{eq:selfenergies} and \Eq{eq:defbs}. After a bit of numerics, it
can be expressed in the following way
\begin{eqnarray}
\label{eq:forsumming1} i \Sigma_{1} &=& \sum_{n=1}^N i
\Sigma_{1}^{(n)}=  g^2 N_c m_t^2 \frac{i}{(4 \pi)^2}\nonumber
%\\
%& &
\left[\frac{N}{\hat{\epsilon}} - \sum_{n=1}^N \log(m_n^2 )
-\log(1+r_n )+ \int_0^1 dx\; x^2
 \sum_{n=1}^N\frac{r_n}{1+r_n x}\right] \qquad\qquad\\\label{eq:forsumming2}
i \Sigma_{3} &=& \sum_{n=1}^N i \Sigma_{3}^{(n)}= g^2 N_c m_t^2
\frac{i}{(4 \pi)^2}\nonumber
%\\
%& &
\left[\frac{N}{\hat{\epsilon}} - \sum_{n=1}^N \log(m_n^2 )
-\log(1+r_n )\right].
\end{eqnarray}
From the point of view of the five dimensional theory these equations
%\Eq{eq:forsumming1} and \Eq{eq:forsumming2} 
can be interpreted as the regularized
integrations over the five components of momentum using a
\emph{mixed} regularization scheme: dimensional regularization to
render finite the integral over the four momentum and cutoff
regularization for the integral over 
the fifth component. 
%It is easy to convince oneself that the sum
%of the first logarithm is divergent, for large $N$
%\begin{equation}
%\sum_{n=1}^{N}\log(m_n^2 ) =2 N \log N + \mathcal{O}(N)
%\end{equation}
%On the contrary, 
The second logarithm is convergent, since term by term is smaller than
the general term of the harmonic series
\begin{equation}
 \sum_{n=1}^\infty\log\left(1+r_n \right) <\frac{(m_t R
 \pi)^2}{6}\qquad\;\; (r_n>0)
\end{equation}
The last term in \rf{eq:forsumming1} can be also easily summed
when $N \to \infty$ and then integrated by using the identity
\begin{equation}
\label{eq:forcotan} \sum_{n=1}^\infty  \frac{1}{(n\pi)^2+\alpha^2}
= \frac{\alpha \coth \alpha-1}{2\alpha^2}.
\end{equation}
%therefore these terms constitute the finite contribution of this
%mixed regularization scheme.

Observe that the divergences related with the limits $d\to 4$
($\hat{\epsilon}\to 0$) and $N\to \infty$ cancel when the subtraction
is performed, so we can safely take these limits. It should be this
way because divergences present the same symmetries than the tree
level terms, and at this level $W_{\mu 1}$ and $W_{\mu 3}$ have the
same mass. Thus the contribution of each mode is perfectly finite and
reads
\begin{equation}
\label{eq:finresl} 
\Delta\rho^{(n)} =\frac{4}{g^2
v^2}\left[\Sigma_{1}^{(n)}(0)-\Sigma_{3}^{(n)}(0)\right] = 2 N_c
\frac{\sqrt{2} G_F m_t^2}{(4\pi)^2}
\left[1-\frac{2}{r_n}+\frac{2}{r_n^2}\log(1+r_n)\right],
\end{equation}
where $r_n=m_t^2/m_n^2$. The factor two of difference with respect the
SM can be understood in the following way: take the limit in
\Eq{eq:finresl} $m_n\rightarrow 0$ which corresponds to gain the
fundamental mode contribution, but this limit has to predict twice
the actual contribution of the fundamental mode because in
\Eq{eq:finresl} each mode has left and right contributions coming
from $Q_{t\;L(R)}^{(n)}$ and $Q_{b\;L(R)}^{(n)}$ running inside the
loops, so the proposed limit would coincide with the SM prediction if
the fundamental mode has had left and right contribution, which is not
the case. In addition to this, the contribution of a diagram with a
given set of fields and the diagram with all the chiralities reversed
is the same.

\Eq{eq:finresl} is consistent with the decoupling theorem
\cite{Appelquist:1975tg}; if the mass of an individual mode,
$m_n$, is taken to infinity, i.e. $r_n\to 0$, its contribution
vanishes
\begin{equation}
\Delta\rho^{(n)}=2 N_c \frac{\sqrt{2} G_F
m_t^2}{(4\pi)^2}\left[\frac{2}{3}r_n+ \mathcal{O}(r_n^2)\right].
\end{equation}
Finally
\begin{equation}
\label{eq:rhotoT}
\Delta\rho^{\mathrm{UED}} = \sum_{n=1}^\infty\Delta\rho^{(n)}= \frac{4}{g^2
v^2}\left[\Sigma_{1}^{(n)}-\Sigma_{3}^{(n)}\right] = 2 N_c
\frac{\sqrt{2} G_F m_t^2}{(4\pi)^2} \int_0^1 dx\; x \left[ a
\sqrt{x} \coth(a \sqrt{x})-1\right],
\end{equation}
where we have used \Eq{eq:forcotan}. This can be expressed in a
more compact form
\begin{equation}
\frac{\Delta\rho^{SM}+\Delta\rho^{UED}}{\Delta\rho^{SM}}= 2\int_0^1 dx\; x \left[
a \sqrt{x} \coth(a \sqrt{x})\right] \comment{ \approx
1+\frac{2}{9}a^2-\frac{1}{90}a^4+\frac{4}{4725}a^6+\ldots
}%endcomment
 ,
\end{equation}
where $a=m_tR\pi$.

With the previous results we can extract the bounds coming from the
experimental measures. In order to discriminate between the
corrections coming from SM and the ones coming exclusively from new
physics we work with the $T$ parameter as defined in PDG
\cite{Hagiwara:2002pw}. It contains only the corrections to $\rho$
coming from new physics. We will adopt the PDG definitions.
%\begin{equation}
%\label{eq:rhopara} \rho_0= \frac{M_W^2}{M_Z^2 \widehat{c}_w^2
%\widehat{\rho}} \qquad \alpha (M_Z) T = \rho_0-1
%\end{equation}
%note that in the absence of new physics $\rho_0=1$.

The contribution to $T$ can be extracted from \Eq{eq:rhotoT}, for
small $m_t R$ it can be expanded as
\begin{equation}
\label{eq:resR} T \approx 2.85 (m_t R)^2 \left[1-0.49 (m_t R)^2 +
0.37(m_t R)^4 \right],
%+\mathcal{O}\left(m_t^6 R^6\right)\right]
\end{equation}
since the contributions to the T parameter coming from new physics
are bounded to be $T<0.4$ at 95 \% CL, the lower
bound is
\begin{equation}
R^{-1}_{UED}> 450~\mbox{GeV},
\end{equation}
which at the end, will be the best of all bounds. This calculation was
firstly done in Ref.~\cite{Appelquist:2000nn} and later on corrected
in Ref.~\cite{Appelquist:2002wb}. Our result is in agreement with the
latter.

%%%%%%%%%%%%%%%%%%%%%%%%%%%%%%%%%%%%%%%%%%%%%%%%%%%%%%%%%%%%%%%%%%%%
%%%%%%%%%%%%%    B0 B0
\subsection{Radiative corrections to the $B^0 - \overline{B}^0$
system}
The models we are studying in this paper fall within the so called
\emph{Minimal flavor violating} (MFV) models because they fulfill
the next two conditions:
\begin{itemize}
\item Only the SM operators in the effective weak Hamiltonian are
relevant \item Flavor violation is entirely governed by the CKM
matrix
\end{itemize}

The two Higgs doublet model and the MSSM at low $\tan\beta$ (and of
course the SM) belong also to the MFV class of models. The interesting
virtue of the MFV models is that with respect to $B^0-\overline{B^0}$
mixings and the CP-violating parameter $\varepsilon_K$, they all can
be parametrized by a single function $S(x_t)$ \cite{Buras:2002yj}. In
the literature the $S(x_t)$ function appears also under the name
$F_{tt}$, and in general ceases to be only function of
$x_t=m_t^2/M_W^2$. In our case, at the level of precision we are
working $S$ is only function of $x_t$. $S(x_t)$ can be defined as
\begin{equation}
\label{eq:sxtdef}
\mathcal{H}_{eff}= \frac{M_W^2 G_F^2 (V_{tb}
V_{td}^\ast)^2}{(4\pi)^2} S(x_t)
[\overline{d}\gamma^\mu(1-\gamma_5)b]
[\overline{d}\gamma_\mu(1-\gamma_5)b].
\end{equation}

The dominant contributions to $S(x_t)$ will be proportional to $x_t$,
hence proportional to $m_t$ proportional corrections. In SM, this
function is dominated by the box diagrams with longitudinal $W$
exchanges and the quark top running inside the loop
\begin{equation}
\label{eq:S} S_{SM}(x_t)=\frac{x_t}{4} \left[ 1 + \frac{9}{1-x_t}
- \frac{6}{(1-x_t)^2} - \frac{6 x_t^2 \log(x_t)}{(1-x_t)^3}\right].
\end{equation}

The measured top quark mass $m_t=175\;\mbox{GeV}$ implies
$S_{SM}(x_t)\approx 2.5$. In UED, the radiative corrections from new
physics can be encoded into a function, which we call $G(a)$ that is
defined as
\begin{equation}
\label{eq:defga}
S(x_t)=S_{SM}(x_t)+\delta S(x_t), \qquad 
\delta S(x_t)=\frac{x_t}{4}(G(a)-1).
\end{equation}

\begin{figure}
\begin{displaymath}
  \begin{array}{cc}
    \includegraphics[scale=0.6]{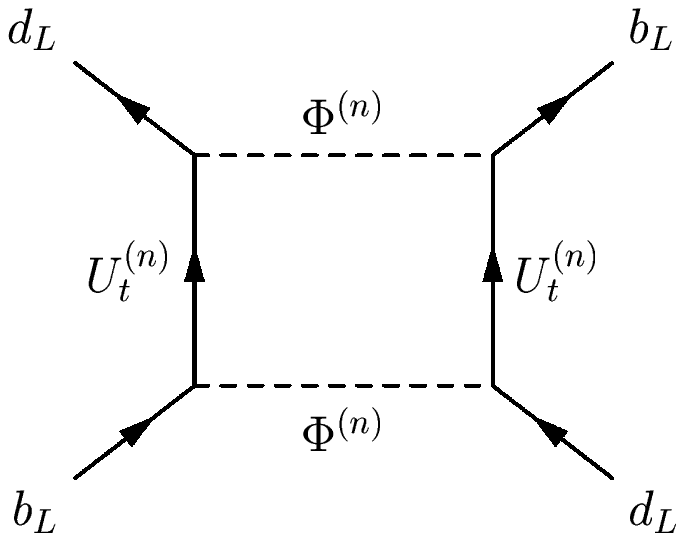}   &
    \includegraphics[scale=0.6]{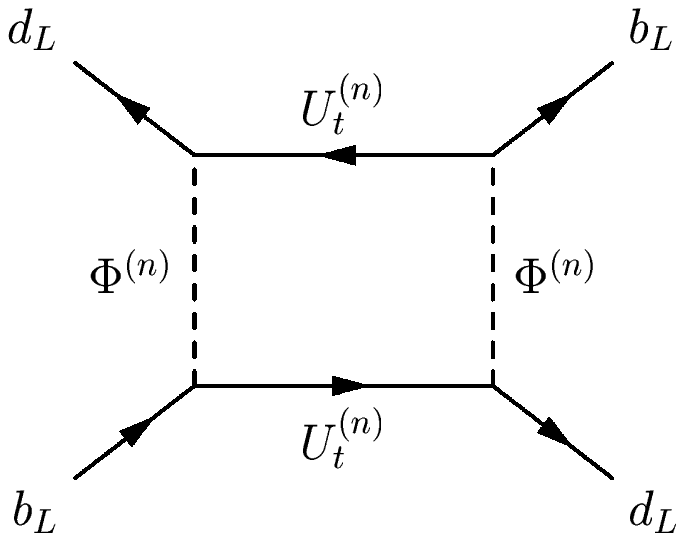}
  \end{array}
\end{displaymath}
\caption{Diagrams that contribute to the $B^0-\overline{B}^0$ mixing in UED.}
\label{fig:BBmixinUED}
\end{figure}

The contributions to this function come from the amplitude of the
diagrams shown in \Fig{fig:BBmixinUED}.
In contrast with what happens in the computation of the modifications
to $\rho$, only the $U$ fields are important in this case because the
couplings of $Q$ fields with the $\Phi$ are proportional to the mass
of the $b$ quark. After the usual Fierz reordering and a bit of
combinatorics, done also in the SM, the result is:  
%Again the fact that the external legs have a fixed
%chirality implies that there is no extra factor two between the zero
%mode, SM, and the modes of the KK tower.
\begin{equation}
\label{eq:sxtued}
 G_{\mathrm{UED}}(a) = \int_0^1dx\;(1-x)\left[ a \sqrt{x}
\coth(a \sqrt{x} )+1\right] \approx 1+\frac{a^2}{18}- \frac{a^4}{540}+\ldots,
\end{equation}
where $a=m_t\pi R$. The last experimental determinations \cite{Buras:2002yj}
agree 
with the SM expectations
\begin{equation}
\label{eq:Sxtexp} 1.3\le S(x_t)\le 3.8  \qquad  95\%\;\mbox{CL}.
\end{equation}
The possible positive contributions have been lowered with respect
to previous determinations \cite{Bergmann:2001pm} allowing better
bounds
\begin{equation}
R^{-1}_{UED} >
40~\mbox{GeV} \qquad 95\%\;\mbox{CL}.
\end{equation}
This bound is an order of magnitude worse than the obtained using
$\rho$. Conversely, if the bound coming from the $\rho$ parameter is
taken, contributions to $S(x_t)$ compatible with experiment are found
to be relatively big.
%contributions to the $S(x_t)$ function compatible with
%experiments. That would be hard to explain within the SM, in other
%words, a signature of the extra dimensions would be an appreciable
%disagreement between the experimental value and the SM prediction to
%$S(x_t)$ and the absence of such disagreements for the other
%observables.

Given the future experimental improvements on the determinations of
$\sin 2\beta$ by BaBar and BELLE and in particular of the mass
splitting $\Delta M_s$ for the B sector in LHC and FNAL one may use
this observable to predict possible deviations from the SM
predictions. It turns out that to an excellent accuracy
\cite{Buras:2002ej} the mentioned deviations in the case of $\Delta
M_s$ are governed by $G(a)$
\begin{equation}
G_{NP}(a)=\frac{(\Delta M_s)_{NP}}{(\Delta M_s)_{SM}}>1
\end{equation}
The value of $G(a)$ is too small to be discriminated
experimentally. This result was initially derived in
Ref.~\cite{Buras:2002ej}, where it is explained that the possible
existence of extra dimensions will not pollute the extraction of the
CKM matrix parameters from the future improvements in the
determination of the unitarity triangle.

\section{Outlook and conclusions}
In this chapter we have studied an extra dimensional extension of the
SM, in which one single extra dimension is accessible to all the
fields. This scenario is called universal extra dimension, or UED. As
explained in the previous chapter, one consequence of the universal
scenarios is the ``conservation'' of KK number, what implies that the
corrections to the SM results are at least one-loop suppressed. Given
the smallness of these corrections we have studied precision
observables that display a strong dependence on the mass of the
top-quark, $m_t$, because this dependence enhances the relative
importance of the new physics with respect to the SM predictions. In
particular, we have studied the radiative corrections for the $Z\to b
\overline{b}$ decay, the $\rho$ parameter and the $B^0-\overline{B}^0$
mixing. The $b\to s \gamma$ process has been also studied. Although it
is not enhanced by a large mass, the relative impact of the new
physics is important for it because in the SM it is one-loop
suppressed, hence the new contributions can be competitive with the SM
radiative corrections.

By comparing these different observables with data, bounds on the
compactification scale, $R$, can be set. Equivalently, these results
can be translated into bounds on the mass of the first member in the
KK tower, $M=R^{-1}$. The table~\ref{thetable} summarizes the results.
\begin{table}
 \begin{center}\begin{tabular}{|c|c|c|c|c|}
            \hline
            &$Z\to b \overline{b}$ & $b \to s + \gamma$ & $\overline{B}^0-B^0$ & $\rho$\\
            \hline
            M(GeV) & 230 & 300 & 40 & 450 \\
            \hline
 \end{tabular}\end{center}
\caption{Bounds coming from the different observables.}
\label{thetable}
\end{table}
The scale of the new physics can be as low as $450~\mbox{GeV}$ without
contradicting any of the experimental determinations. This is a
relatively low value for $M$ because precision observables, in
general, tend to establish the scale of any new physics around or
above the TeV. The reason why the scale for the universal extra
dimension can be so low without affecting too much precision
observables is the above mentioned one-loop suppression due to the KK
number conservation.

%endcomment

\chapter{SM with one latticized universal extra dimension}
\label{chap:LUED}
\comment{
{\sffamily \scshape
\begin{center}
\begin{tabular}{|l|r|}
\hline
Text        & Ok \\\hline
Orthography & Ok  \\\hline
Figures     & Ok  \\\hline
Links       & Ok  \\\hline
Cites       & Ok  \\\hline
Meaning     & --  \\\hline
Makindex    & --  \\\hline
Date        &  \today  \\\hline
\end{tabular}
\end{center}
} }%endcomment 

In the previous chapter we have studied models with one extra
dimension and with all SM fields propagating in it. The models
displayed two different energy regimes: the low energy, below the
compactification scale, reduced to the SM while the high energy regime
described the couplings among the modes of the KK towers. Because of
the presence of the extra dimension the coupling constants are
dimension full, in particular Yukawa couplings and gauge couplings,
that are dimensionless quantities in the SM, have dimensions of energy
raised to some negative power, which is the reason why these theories
are non-renormalizable. Hence, they must be understood as effective
field theories that in their low energy limit reduce to the SM. A
possible ultraviolet completion was proposed in
Ref.~\cite{Arkani-Hamed:2001ca}, which eventually treated the extra
dimensions as if they were discontinuous. The idea of a discretized
dimension was also simultaneously suggested in
Ref.~\cite{Hill:2000mu,Cheng:2001vd}. The latter is not really an
ultraviolet completion because the Lagrangian is described by a number
of $\sigma$-models, but for this class of models there are some known
possible renormalizable extensions. The aim of this chapter is to
investigate how the phenomenology is modified in models with
discretized, sometimes also called \emph{latticized}, extra
dimensions; specifically, we will study the latticized version of UED,
called in the following LUED.

%In this chapter we describe in detail a possible extension of the
%SM in this scenario, which we will refer to as LUED on the
%following. The spectrum is derived as well as the couplings and
%after that the bounds to it coming from experiments will be set.

\section{The model}
The Lagrangian is divided, as usual, in four pieces
\begin{equation}
\mathcal{L}^{\mathrm{LUED}}=\mathcal{L}_G+\mathcal{L}_F+\mathcal{L}_H+\mathcal{L}_Y.
\end{equation}
The gauge piece, $\mathcal{L}_G$, is the one associated to the
gauge group $G=\Pi_{i=0}^{N-1}
SU(2)_{i}\times U(1)_{i}$ and it also contains some scalars fields
which will be necessary on the following, their role will be
clarified later
\begin{eqnarray}
\mathcal{L}_G&=&\sum_{i=0}^{N-1}-\frac{1}{4}F_{i\mu\nu}^{a}F^{\mu\nu
a}_i- \frac{1}{4}F_{i\mu\nu}F^{\mu\nu}_i\\
\label{eq:kinterm}
  & +&\sum_{i=1}^{N-1}\mbox{Tr}\{(D_\mu\Phi_i)^\dagger (D^\mu \Phi_i) \}+
(D_\mu\phi_i)^\dagger (D^{\mu}\phi_i) - V(\Phi,\phi),
\end{eqnarray}
where $F_{i\mu\nu}^a$ is the strength tensor associated with the
gauge field of the i-th $SU(2)_{i}$ and $F_{i\mu\nu}$ is the one
for $U(1)_{i}$. $\Phi_i$ and $\phi_i$ are the elementary scalars
that will acquire a VEV independent of $i$ due to the potential
term $V(\Phi,\phi)$. Each of them become effectively nonlinear
$\sigma$ model fields that can be parametrized as usual in terms
of the scalar fields $\pi_i$ and $\pi^a_i$
\begin{equation}
\phi_i=\frac{v_1}{\sqrt{2}}e^{i\pi_i/v_1}\qquad \Phi_i=v_2
e^{i\pi^a_i \sigma^a/2 v_2}
\end{equation}
$v_1$ and $v_2$ are the VEVs of $\phi_i$ and $\Phi_i$ respectively and
$\sigma^a$ are the Pauli matrices. In this work we will concentrate in
the so called ``\emph{aliphatic model}'' \cite{Hill:2000mu} in which
the $\Phi_i$ fields are assumed to transform as
$(\mathbf{2},\mathbf{\overline{2}})$ under the groups $SU(2)_{i}$ and
$SU(2)_{i-1}$ and as singlets for the rest, they carry no $U(1)_i$
charge. On the other hand, the $\phi_i$ fields are singlets under all
the $SU(2)$ groups and they are charged only under $U(1)_i$ and
$U(1)_{(i-1)}$ with hypercharges $(Y_i,-Y_{i-1})$, later on, every
$Y_i$ will be set to $Y_i=1/3$ \cite{Cheng:2001vd}. With this, the
covariant derivative reads
\begin{equation}
D_\mu \Phi_i = \partial_\mu \Phi_i - i \mathcal{W}_{\mu,i} \Phi_i
+ i \Phi_i \mathcal{W}_{\mu,i-1}
\end{equation}
where $\mathcal{W}_{\mu,i}=\tilde{g} W_{\mu\;i}^a T^a_i$, $T^a_i$
are the generators of the $SU_{i}(2)$ and $\tilde{g}$ is the
dimensionless gauge coupling constant that is assumed to be the
same for all the $SU(2)$ groups. The covariant derivative for
$\phi_i$ can be constructed similarly. The gauge coupling constant
for all the $U(1)$ groups will be called $\tilde{g}^\prime$.

 The next piece is the fermionic one, $\mathcal{L}_F$, it
contains the following fields (generational indices assumed)
\begin{equation}
\label{eq:fielddefinition}
Q_i=\left[\begin{array}{c}Q_{ui}\\
Q_{di}\end{array}\right] \qquad U_i \qquad D_i \qquad
i=0,\ldots,N-1,
\end{equation}
where we have used a similar notation than in the continuous case.
$Q_i$ transforms as a doublet under $SU(2)_i$ and as a singlet for
the rest of $SU(2)$ groups and among the $U(1)$ fields it is only
charged under $U(1)_i$ with hypercharge $Y_Q=1/3$. On the contrary,
$U_i$ and $D_i$ are only charged under $U(1)_i$ with hypercharges $Y_U=4/3$ and
$Y_D=-2/3$.
%Although not explicitly written
%in \rf{eq:fielddefinition} also generational indexes are assumed
%ranging trough three generations.
They are all vector fields with right and left handed chiral
components except for $i=0$. In this case they are chiral fields,
$Q$ is left-handed and $U$ and $D$ are right-handed, which is
equivalent to impose
\begin{equation}
\label{eq:chirality} Q_{0R}=0\qquad U_{0L}=0 \qquad D_{0L}=0
\end{equation}
With this we can split the fermionic piece in:
$\mathcal{L}_F=\mathcal{L}_Q+\mathcal{L}_U+\mathcal{L}_D$, where
\begin{equation}
\label{eq:fer5derifa}
\mathcal{L}_{Q} = \sum_{i=0}^{N-1}
\left[\overline{Q}_{iL}
i\slD Q_{iL} +\overline{Q}_{iR} i\slD Q_{iR} \right]
 -\sum_{i=0}^{N-1} M_f
\overline{Q}_{iL}
\left(\frac{\Phi_{i+1}^{\dagger}\phi_{i+1}^{\dagger}}{v_2
(v_1/\sqrt{2})} Q_{i+1R}-Q_{iR}\right)+\hc\qquad
%\\
%\label{eq:fer5derifc}
%&-&\sum_{i=0}^{N-1}
%%M_f\overline{Q}_{iR}\left(Q_{iL}-\frac{\Phi_i\phi_i}{v_2
%(v_1/\sqrt{2})}Q_{i-1L} \right)
\end{equation}
and 
\begin{equation}
\label{eq:fer5deri2a} \mathcal{L}_{U} = \sum_{i=0}^{N-1}
\left[\overline{U}_{iR} i\slD U_{iR} +\overline{U}_{iL}
i\slD
U_{iL} \right] + \sum_{i=0}^{N-1} M_f
\overline{U}_{iR}
\left(\frac{\phi_{i+1}^{4\dagger}}{(v_1/\sqrt{2})^4}
U_{i+1L}-U_{iL}\right)+\hc\qquad
\end{equation}
$\mathcal{L}_D$ can be extracted from $\mathcal{L}_U$ making the next
substitutions, $U\to D$, $\phi_i\to \phi_i^{\dagger}$ and the exponent
should be replaced $4\to 2$. In the previous formulae $\slD$ is the
usual covariant derivative associated with the gauge group $G$ and
$M_f$ is a generic mass that in principle could depend on $i$ but for
simplicity it is set independent of $i$. The exponent of the $\phi$
fields must be adjusted in each case because the terms must be
invariant under $G$ by construction. In addition, when the index of a
field runs out of bounds it is understood as a zero, for instance in
\Eq{eq:fer5derifa} it must be set $Q_{NR}=0$, and so on.

The next piece in the Lagrangian, $\mathcal{L}_H$, is the one
associated with the Higgs doublet \cite{Cheng:2001vd}

\begin{equation}
\mathcal{L}_H = \sum_{i=0}^{N-1}(D_\mu H_i)^\dagger(D^\mu H_i) -
M_0^2\left|H_{i+1}-\left(\frac{\Phi_{i+1}\phi_{i+1}^3}{(v_1/\sqrt{2})^3
v_2}\right)H_i\right|^2-V(H_i),
\end{equation}
where $H_i$ is a doublet under $SU(2)_i$ and singlet for $SU(2)_{j\neq
i}$ with hypercharges $Y_i=1$ and $Y_{j\neq i}=0$. We parametrize its
components as
\begin{equation}
H_i= \left[
 \begin{array}{c}
   \Phi^+_i\\
   \Phi^0_i
 \end{array}
\right]\qquad i=0,1,\ldots,N-1.
\end{equation}
As a potential it is chosen
\begin{equation}
\label{eq:HiggsPotential} V(H_i)=-m^2H_i^\dagger
H_i+\frac{\tilde{\lambda}}{2} (H_i^\dagger H_i)^2
\end{equation}
Finally the Yukawa sector, $\mathcal{L}_Y$, will be taken with the
Yukawa matrices independent of $i$
\begin{equation}
\label{eq:yukawas} \mathcal{L}_Y=\sum_{i=0}^{N-1}\overline{Q_i}
\widetilde{Y}_u H_i^c U_i +\sum_{i=0}^{N-1}\overline{Q_i}
\widetilde{Y}_d H_i D_i+\hc
\end{equation}
where $H_i^c\equiv i\tau^2H_i^\ast$ is the usual Higgs doublet
conjugate.

\subsection{Relation with continuous extra dimensions}
\begin{figure}
\begin{center}
\includegraphics[scale=0.9]{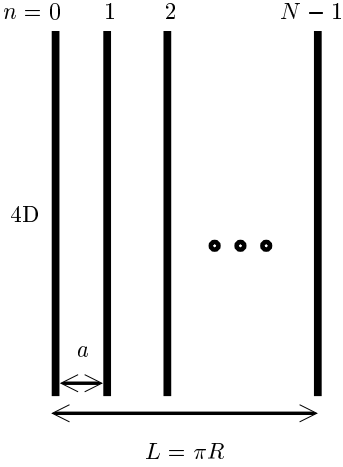}
\caption{Schematic representation of the discretization in the extra
  dimensions.}
\label{fig:lattice}
\end{center}
\end{figure}

The fields and couplings proposed in the above lines are set to
describe a situation in which the full SM is contained in a five
dimensional space-time with four spacial dimensions but with the extra
fifth dimension latticized. In LUED the length of the new dimension,
$L$, is taken to be finite and a new variable $R$ is defined through
the relation $L=\pi R$, this will simplify the comparison of the
results with the continuous situation. The extra volume is filled with
4D surfaces equally spaced by a distance $a$, the first one situated
in $x^5=0$, see \Fig{fig:lattice}.
Therefore, these magnitudes fulfill the trivial relation
\begin{equation}
a=\frac{\pi R}{N-1}.
\end{equation}
In this picture every field that propagates in the fifth dimension
will be represented by the $N$ values that it takes in the different
surfaces, i.e. $\psi_i(x^\mu)=\psi(x^\mu,x^5=ia)$. This is the reason
why we have $N$ copies for each of the SM fields.

But the $\Phi$ and $\phi$ fields were not previously present in SM.
They are related with the component of the five dimensional gauge
fields polarized in the direction of the extra dimension, generically
denoted by $W_5$. In the continuous theory these components are
necessary in order to define a consistent covariant derivative. In our
case the relation between these fields is \cite{Hill:2000mu}
\begin{equation}
\label{eq:relationphiw5}
\Phi_i(x^\mu)=\exp\left[i\tilde{g}\int_{(i-1)a}^{ia}\!\!\!\!\!\!\!\!\! dx^5\;
\mathcal{W}_5(x^\mu,x^5)\right], \qquad  i=1,\ldots,N-1.
\end{equation}
In other words the $\Phi$ fields are Wilson lines connecting two
adjacent surfaces, that is the reason why there are only $N-1$
copies of them, instead of the $N$ that have the rest of the
fields. \Eq{eq:relationphiw5} enables us to reinterpret
the interaction terms between the different fields present in the
Lagrangian with the $\Phi$'s as a lattice approximation of their
covariant derivatives provided that the general masses obey
$M_f=M_0=1/a$.

We have exploited this picture when writing the interaction terms of
the fermions. Note the relative sign between the $Q$ doublets and the
$U$ and $D$ singlets, compare \Eq{eq:fer5derifa} with
\Eq{eq:fer5deri2a}. Written in this form it will be easier to extract
the spectrum of the theory and both possibilities are equally
legitimate since they are lattice approximations, to the same order,
of the fifth covariant derivative.

When studying continuous extra dimensions sometimes some of the fields
are not allowed to propagate through the extra volume, there are many
variations in the literature, but in particular we will study in
chapter \ref{chap:HGLHG} the case when only fermion fields do not
propagate in the extra dimension, LHG. The Lagrangian for this case
can be obtained by setting in the above described one, $\psi_{iR}=
\psi_{iL} = 0$ for $i\neq 0$, where $\psi$ stands for any fermion
field.

As a last remark, the assumption that the coupling constants are
independent of the position translates in the latticized scenario as
an independence on the index $i$ of each group,
i.e. $\tilde{g}_i=\tilde{g}(x^5=ia)=\tilde{g}$.

\subsection{The spectrum of the model}
\comment{In the first part of this section we will concentrate on the
LUED scenario and afterwards on LHG. }Before extracting the bilinear
terms from $\mathcal{L}$ it is interesting to study which fields can
be removed from the spectrum by exploiting the gauge freedom. Under a
generic transformation of G the $\Phi_i$ transforms as
\begin{equation}
\Phi_i=U_i\Phi_i U^\dagger_{i-1} \qquad U_i\in SU(2)_{i}.
\end{equation}
Since the vacuum configuration is by construction
$\Phi_i=v_2\mathbbm{1}$, it is clear that only the gauge
transformation defined by $U_0=U_1=\ldots=U_{N-1}$ leaves invariant
the vacuum, i.e. the diagonal group $SU(2)_D$ remains unbroken after
the SSB, analogously for $\phi_i$. Therefore \mbox{$G\to SU(2)_D\times
U(1)_D$}. This procedure removes the $\pi$ fields from the spectrum
and leaves massless only one combination of the gauge bosons.  This
amounts to work in the unitary gauge and it is indeed the approach
followed in Ref.~\cite{Hill:2000mu,Cheng:2001vd}, where $SU(2)_D\times
U(1)_D$ is identified with the SM electroweak gauge group which is
further broken by the usual SM Higgs VEV. We will not follow those
steps and instead we will work in an arbitrary $R_\xi$-covariant
gauge, in the same line as it was done in Ref.~\cite{Herrero:1994tj},
therefore maintaining explicitly the $\pi$ fields.

Extracting the bilinear terms is lengthy and cumbersome, but
straightforward otherwise. From $\mathcal{L}_G$ it can be obtained

\begin{equation}
\label{tralari}
\mathcal{L}_G^{(2)}  =
\sum_{i=0}^{N-1}-\frac{1}{4}\widetilde{F_i}\cdot\widetilde{F_i}+M_i^2\widetilde{W}_{\mu i}^a\widetilde{W}_{i}^{\mu
a} +  \sum_{i=1}^{N-1}\frac{1}{2}\partial_\mu\widetilde{W}^a_{5i}\partial^\mu\widetilde{W}^a_{5i}
+M_i\partial^\mu\widetilde{W}^a_{5i}\widetilde{W}_{\mu i}^a.
\end{equation}
Here we only show the $SU(2)$ piece but the $U(1)$ one is similar. The
new tilde fields are related with the initial ones by the next change
of base
\begin{equation}
W_{\mu i}^a\equiv\sum_{i=0}^{N-1} a_{ij} \widetilde{W}_{\mu j}^a
\qquad \pi_i^a\equiv\sum_{i=1}^{N-1} b_{ij} \widetilde{W}_{5 j}^a
\end{equation}
where the $a$ and $b$ matrices are
\begin{equation}
\label{eq:defch1}
a_{ij}=\left\{ \begin{array}{rcl}j=0&\ & \sqrt{1/N}
%\frac{1}{\sqrt{N}}
%\cos\left[\frac{2j+1}{2}\frac{\pi}{N}\right]
\\ j \neq 0 & &
\sqrt{{2}/{N}}\cos\left(\frac{2i+1}{2}\frac{j\pi}{N}\right)\end{array}\right.
\end{equation}
and
\begin{equation}
b_{ij}=\sqrt{\frac{2}{N}}\sin\left(ij\frac{\pi}{N}\right)
\end{equation}
$\widetilde{F}$ is understood to be the usual kinetic term for gauge
bosons expressed now in terms of $\widetilde{W}_{\mu i}^a$. Finally,
the masses $M_i$ are
\begin{equation}
\label{eq:massdefinition} M_i=2 \tilde{g}v_2 \sin\left(\frac{i
\pi}{2N}\right)=2 \frac{N-1}{\pi R}\sin\left(\frac{i
\pi}{2N}\right),
\end{equation}
where we have chosen as usual $\tilde{g}v_2=1/a$ to guarantee that the
large $N$ limit reproduces the continuous scenario\footnote{
\Eq{eq:massdefinition} could also have been written in the form $M_i=2
N/\pi R\sin(i \pi/2N)$, as it is done in Ref.~\cite{Hill:2000mu}; the
large $N$ limit is the same. This will not be a problem because the
bounds that we will obtain are completely independent of which of the
two definitions is taken.}\cite{Hill:2000mu}. The massless vector
bosons $\widetilde{W}_{\mu 0}^a$ and $\widetilde{B}_{\mu 0}$ are
associated to the SM model gauge bosons. They are the gauge vector
bosons of the unbroken diagonal group and consequently this last is
identified with the SM gauge group.

On the other hand, the bilinear terms for the fermions are
\begin{equation}
\mathcal{L}^{(2)}_Q +\mathcal{L}^{(2)}_U =
\overline{\widetilde{Q}}_{0L}i\sld\widetilde{Q}_{0L} +
\overline{\widetilde{U}}_{0R}i\sld\widetilde{U}_{0R}\label{eq:bilifermi}
+\sum_{i=1}^{N-1}\overline{\widetilde{Q}}_{i}(i\sld-M_i)\widetilde{Q}_{i} +
\overline{\widetilde{U}}_{i}(i\sld+M_i)\widetilde{U}_{i},
\end{equation}
where the vector like fields are defined as
$\widetilde{Q}_{i}=\widetilde{Q}_{iR}+\widetilde{Q}_{iL}$ and similarly
for $\widetilde{U}_i$. The tilde fields are given by
\begin{equation}
\label{eq:QU}
\begin{array}{cc}
Q_{iL}=a_{ij}\widetilde{Q}_{jL} & U_{iR}=a_{ij}\widetilde{U}_{jR}\\
Q_{iR}=b_{ij}\widetilde{Q}_{jR} & U_{iL}=b_{ij}\widetilde{U}_{jL}\\
\end{array}
\end{equation}
the mass $M_i$ appearing in \Eq{eq:bilifermi} is the one defined in
\Eq{eq:massdefinition} provided $M_f=1/a$. 
%The reason why
%$\widetilde{Q}$ and $\widetilde{U}$ are defined with $a$ and $b$
%interchanged is due to the different definitions we have taken in
%\Eq{eq:fer5derifa} and \Eq{eq:fer5deri2a}. 
This will be advantageous when studying the Yukawa terms. The relative
sign in the mass terms is a common feature after dimensional reduction
in theories with universal (continuous) extra dimensions, where it is
due to the definition of the fifth gamma matrix, $\Gamma^4=i\gamma^5$
\cite{Appelquist:2000nn}. Here it has been explicitly introduced in
\Eq{eq:fer5derifa} and \Eq{eq:fer5deri2a}, precisely to reproduce this
feature.

Finally, in the Higgs sector one must perform the change of basis
$H_i=a_{ij} \widetilde{H}_j$ and the masses are now
\begin{equation}
\label{eq:beforevev}
M^2(\widetilde{H}_i)=4\frac{(N-1)^2}{\pi^2 R^2}\sin^2\left(\frac{i
  \pi}{2N}\right)-m^2
\end{equation}
This equation shows that $\widetilde{H}_0$ will break spontaneously
symmetry, hence it is identified with the SM Higgs doublet, i.e.
$\langle\widetilde{H}_0\rangle_0=v/\sqrt{2}$ with
\mbox{$v=246~\mbox{GeV}$}. This means that new contributions to the
masses came from the Yukawa piece $\mathcal{L}_Y$ and the covariant
derivative of $H$.  The Yukawa piece in terms of the tilde fields can
be written as
\begin{equation}
\label{eq:yukpi}
\mathcal{L}_Y=\sum_{i=0}^{N-1}\overline{\widetilde{Q}}_i\frac{\widetilde{Y}_u}{\sqrt{N}}\widetilde{H}^c_0\widetilde{U}_i
+
\overline{\widetilde{Q}}_i\frac{\widetilde{Y}_d}{\sqrt{N}}\widetilde{H}_0\widetilde{D}_i+\hc
\end{equation}
where we have concentrated on the terms containing the Higgs
doublet. It has been used \Eq{eq:QU}. From the first term in the sum
of \Eq{eq:yukpi} is easy to convince oneself that
$Y_u\equiv\widetilde{Y}_u/\sqrt{N}$ is the SM Yukawa matrix. When the
Higgs doublet acquires a VEV one must diagonalize $Y_u$ using the same
field redefinitions as in SM, $\widetilde{Q}_{ui}\to U_u^\dagger
\widetilde{Q}_{ui}$, $\widetilde{U}_i\to V_u^\dagger
\widetilde{U}_i$. At the end the mass matrix for fermions will
be\footnote{We do not study explicitly the term containing $Y_d$
because it does not contain the mass of the top-quark, $m_t$, but its
treatment would be completely similar.}
\begin{equation}
\left(
  \begin{array}{cc}
    \overline{\widetilde{U}}_{if} & \overline{\widetilde{Q}}_{if}
  \end{array}
\right)
\left(
  \begin{array}{cc}
-M_i & m_f \\
m_f  & M_i
  \end{array}
\right)
\left(
  \begin{array}{c}
    \widetilde{U}_{if} \\\widetilde{Q}_{if}
  \end{array}
\right)
\end{equation}
where $f$ is the index of the generation, in this case $f=u,c,t$. This
mass matrix is exactly the same obtained in the continuous scenario
with one extra dimension by making the substitution $m_n\to M_n$
\cite{Appelquist:2000nn}, where $m_n=n/R$ is the mass of the n-th
Kaluza-Klein mode of the field in the absence of Yukawa couplings. As
a consequence the mixing between the $\widetilde{Q}$ and
$\widetilde{U}$ is the same as in that scenario as well as the masses
$M(Q^\prime_{if})=\sqrt{M_i^2+m_f^2}$, prime denotes mass eigenfields,
for later reference $m_Q=M(Q^\prime_{if})$
\begin{equation}
\left(
  \begin{array}{c}
    \widetilde{U}_{if} \\
    \widetilde{Q}_{if}
  \end{array}
\right) =
\left(
  \begin{array}{cc}
    -\gamma^5\cos\alpha_{if} & \sin\alpha_{if} \\
    \;\gamma^5\sin\alpha_{if} & \cos\alpha_{if}
  \end{array}
\right)
\left(
  \begin{array}{c}
    U_{if}^\prime \\ Q_{if}^\prime
  \end{array}
\right)
\end{equation}
where $\tan(2\alpha_{if})=m_f/M_i$. As in the continuous situation we
are specially interested in the case $f=t$.

Notice that the zero-th modes have exactly the same masses as in
SM, all of them coming purely from the Yukawa piece in \Eq{eq:yukpi}
which, as said, for the zero-th modes coincides exactly with the
SM Yukawa sector. In fact, the same happens for the rest of the
pieces in the Lagrangian and one can safely identify the zero-th
tilde fields, $\widetilde{Q}_{0L}$, $\widetilde{U}_{0R}$ and
$\widetilde{D}_{0R}$, with the SM fields.

We will not show it explicitly but the SSB of the Higgs doublet will
cause the usual mixing between $W_{\mu i}^3$ and $B_{\mu i}$ with
$\theta_w$ as the weak mixing angle \cite{Cheng:2001vd}, instead we
will concentrate on the mixing of the charged bosons. Coming from
$\mathcal{L}_H$ and due to the VEV of $\widetilde{H}_0$ the next terms
arise
\begin{equation}
\label{eq:biliaftervev}
\mathcal{L}_{\widetilde{H}_0}^{(2)} = i M_W \widetilde{W}^-_{\mu
  i}\partial^\mu\widetilde{\Phi}^+_i+M_W^2\widetilde{W}^-_{\mu
  i}\widetilde{W}^{\mu +}_{i}\\  - M_W^2\widetilde{W}^-_{5
  i}\widetilde{W}^{+}_{5i} + i M_W M_i
  \widetilde{W}_{5i}^-\widetilde{\Phi}^+_i+\hc
\end{equation}
In addition, due to the quartic couplings in \Eq{eq:HiggsPotential}
the masses for $\widetilde{H}_i$ are shifted from \Eq{eq:beforevev} to
\Eq{eq:massdefinition}. This implies, jointly with \Eq{tralari}, that
there is a combination of fields, $\Phi^{\pm}_{Gi}$, that act as a
Goldstone field absorbed by $\widetilde{W}_{\mu i}^\pm$ which acquires
in the process a mass $M(\widetilde{W}_{\mu
i}^\pm)=\sqrt{M_W^2+M_i^2}$. The orthogonal combination is a physical
scalar, $\Phi^{\pm}_{Pi}$, with the same mass.
\begin{eqnarray}
\label{eq:bosonsmixing}
\Phi^+_{Gi} & = & \frac{M_i \widetilde{W}_{5i}^+ + i M_W
  \widetilde{\Phi}^+_i}{\sqrt{M_i^2+M_W^2}} \stackrel{M_W\to 0}{\longrightarrow} \widetilde{W}_{5i}^+\\
\Phi^+_{Pi} & = & \frac{i M_W \widetilde{W}_{5i}^+ +  M_i
  \widetilde{\Phi}^+_i}{\sqrt{M_i^2+M_W^2}}  \stackrel{M_W\to
  0}{\longrightarrow} \widetilde{\Phi}_{i}^+
\end{eqnarray}
In the limit in which all the mass scales below $m_t$ are neglected,
the Goldstone bosons and the physical scalars can be directly
identified with $\widetilde{W}_{5i}^\pm$ and
$\widetilde{\Phi}_{i}^\pm$ respectively. There will be terms that
cross the massive vector bosons $\widetilde{W}_{\mu i}^\pm$ with the
derivatives of their Goldstone bosons, $\Phi^{\pm}_{Gi}$; these can be
removed using a convenient $R_\xi$ gauge as done in
\cite{Buras:2002yj,Muck:2001yv}.

With this, we have concluded the demonstration that, at least for the
degrees of freedom we will be interested in, the spectrum of this
model is equivalent to one continuous universal extra dimension with
$m_n$ replaced by $M_n$.

\comment{But recall that we are also interested in the LHG scenario. Now it
is easy to reconstruct the new spectrum from the preceding
reasonings. In this case there is only one copy of the fermions,
i.e. in this scenario it must be imposed $Q_i=U_i=D_i=0$ for
$i\neq 0$ as well as \Eq{eq:chirality}. With this, $\mathcal{L}_F$
is the usual kinetic term, the changes in \Eq{eq:QU} are no longer
required and they get masses through the SM Yukawa term
\rf{eq:yukpi} that now it is written as
\begin{equation}
\mathcal{L}_Y=\overline{Q}_L\widetilde{Y}_u H^c_0U_R +
\overline{Q}_L\widetilde{Y}_d H_0 D_R+\hc
\end{equation}
the term that provides them with masses is the coupling with the SM
Higgs doublet $\widetilde{H}_0$
\begin{equation}
\mathcal{L}_Y=\overline{Q}_L Y_u \widetilde{H}^c_0U_R +
\overline{Q}_L Y_d \widetilde{H}_0 D_R+\hc
\end{equation}

Of course the manipulations performed on the gauge and Higgs sectors
remain the same.}

\subsection{Couplings}
Our aim will be to extract the dominant corrections to some precision
observables and from them to extract bounds on the new physics. We
will concentrate on the corrections proportional to the top-quark mass
$m_t$, thus as a approximation we will identify
$\Phi_{Pi}^\pm=\widetilde{\Phi}_i^\pm$ and $\Phi_{Gi}^\pm =
\widetilde{W}_{5i}^\pm$, which is equivalent to neglect $M_W$. This
will simplify greatly the calculus. In the next lines we will extract
the relevant couplings under the previous approximations.

\comment{The vertices required for our calculus are the ones that couple among
them $\widetilde{W}_{\mu i}^\pm$, $\widetilde{G}^{\pm}_{i}$,
$\widetilde{a}^{\pm}_{i}$, $Z$, $\gamma$ and the fermion fields. It
can be checked explicitly that, in the case our approximations are not
used, these couplings can be extracted from \cite{Buras:2002ej} with
the substitution $m_n\to M_n$, which will enable us to improve some of
our estimations.}

For computing the contributions to the $\rho$ parameter we will work
with the base of fields $\{\widetilde{Q}_{i},\widetilde{U}_{i}\}$
instead of the mass eigenfields $\{Q_{i}^\prime,U_{i}^\prime\}$
because in the former the couplings with $\widetilde{W}_{\mu
0}^{1/3}$, the only ones needed, are rather simple
\begin{equation}
\label{eq:ggaugea} \mathcal{L}_{\rho}=\frac{g}{2}\sum_{i=1}^{N-1}
\widetilde{W}_{\mu 0}^1 \left[
\overline{\widetilde{Q}}_{it}\gamma^\mu
\widetilde{Q}_{ib}+\overline{\widetilde{Q}}_{ib}\gamma^\mu
\widetilde{Q}_{it}\right]+\widetilde{W}_{\mu 0}^3 \left[
\overline{\widetilde{Q}}_{it}\gamma^\mu \widetilde{Q}_{it}\right],
\end{equation}
where we have already used the relation
$g=\widetilde{g}/\sqrt{N}$ \cite{Cheng:2001vd}. So the couplings in
this base are the same as in SM but with the difference that the
propagators of the tilde fields are
\begin{displaymath}
\left[
\begin{array}{cc}
\includegraphics[scale=0.6]{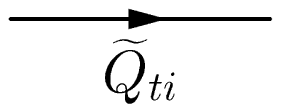}\hspace{3ex}
&
\includegraphics[scale=0.6]{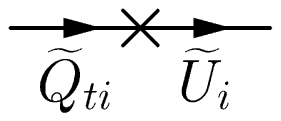}\hspace{3ex}\\
\includegraphics[scale=0.6]{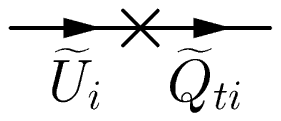}
&
\includegraphics[scale=0.6]{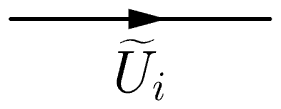}\hspace{3ex}
\end{array}
\right]
=
\left[
\begin{array}{cc}
\;i \frac{\slash{p}+M_i}{p^2-m_Q^2} &i \frac{m_t}{p^2-m_Q^2}\\
i \frac{m_t}{p^2-m_Q^2}&\;i \frac{\slash{p}-M_i}{p^2-m_Q^2}
\end{array}
\right]
\end{displaymath}
From the Yukawa piece, the couplings proportional to $m_t$ are
\begin{equation}
\label{eq:yuklued}
\mathcal{L}_Y=\sum_{i=1}^{N-1} m_t\frac{\sqrt{2}}{v}
V_{tb}\overline{\widetilde{U}}_{tiR} \Phi_i^+b_L+\hc
\end{equation}
The couplings with the $Z$ will be
required, these can be obtained from
\begin{equation}
\mathcal{L}_Z =
\frac{g}{2c_w}Z_\mu[J^{\mu}_{SM} + J^{\mu}_{F} + J^\mu_{\Phi}]
\end{equation}
where $J^{\mu}_{SM}$ is the usual SM current and
\begin{eqnarray}
\label{eq:sourcen}
J^{\mu}_{F}&=&\sum_{i=1}^{N-1}\left(1-\frac{4}{3}
s_w^2\right)\overline{\widetilde{Q}}_{ti} \gamma^\mu
\widetilde{Q}_{ti} -\frac{4}{3}
s_w^2\overline{\widetilde{U}}_{it}\gamma^\mu \widetilde{U}_{it}\qquad\\
J^{\mu}_{\Phi} &=& \sum_{i=1}^{N-1} (-1 +2s_w^2)\widetilde{\Phi}^{+}_i i\partial^\mu\widetilde{\Phi}^{-}_i
+\hc
\end{eqnarray}
The couplings with the photon can be derived similarly.
\comment{
it easy to check that the
electromagnetic current can be written as
\begin{equation}
\mathcal{L}_\gamma= e A_\mu [j^\mu_{SM}+j^\mu_F + j^\mu_Phi]
\end{equation}
where the new currents are
\begin{eqnarray}
\label{eq:emsource}
j^{\mu}_{F}&=&\sum_{i=1}^{N-1}\frac{2}{3}\overline{\widetilde{Q}}_{ti}
\gamma^\mu \widetilde{Q}_{ti} -\frac{1}{3}
\overline{\widetilde{U}}_{it}\gamma^\mu \widetilde{U}_{it}\qquad\\
j^{\mu}_{\Phi} &=& \sum_{i=1}^{N-1} -\widetilde{\chi}^{+}_i
i\partial^\mu\widetilde{\chi}^{-}_i +\hc
\end{eqnarray}
}%endcomment

%%%%%%%%%%%%%%%%%%%%%%%%%%%%%%%%%%%%%%%%%%%%%%%%%%%%%%%%%%%%%%%%%%%%
%%%%%%%%%%%%%%%%%%%%%%%%%%%%%%%%%%%%%%%%%%%%%%%%%%%%%%%%%%%%%%%%%%%%
%%%%%%%%%%%%%%%%%%%%%%%%%%%%%%%%%%%%%%%%%%%%%%%%%%%%%%%%%%%%%%%%%%%%
%%%%%%%%%%%%%%%%       Bounds                 %%%%%%%%%%%%%%%%%%%%%%
%%%%%%%%%%%%%%%%%%%%%%%%%%%%%%%%%%%%%%%%%%%%%%%%%%%%%%%%%%%%%%%%%%%%
%%%%%%%%%%%%%%%%%%%%%%%%%%%%%%%%%%%%%%%%%%%%%%%%%%%%%%%%%%%%%%%%%%%%
%%%%%%%%%%%%%%%%%%%%%%%%%%%%%%%%%%%%%%%%%%%%%%%%%%%%%%%%%%%%%%%%%%%%
\section{Phenomenology} 
\label{Bounds} 
In this section we establish which is the minimum
value for the energy scale of the new physics allowed by
experiments. We set bounds on the first mode, $M_1$, as
defined in \Eq{eq:massdefinition} and called simply $M$ in the
following. To this end, we compute the impact of the new physics on a
set of standard electroweak observables. In particular, we focus
on those which in SM display large radiative corrections due to
their strong dependence on the top-quark mass: the decay rates $b
\rightarrow s\; \gamma$ and $Z\to b\overline{b}$, the $\rho$
parameter and the rates of $B^0\rightleftharpoons \overline{B^0}$.
We expect that they will be also very sensitive to the top mass since the
structure of these kind models is basically the same of the SM
replicated.  

%%%%%%%%%%%%%%%%%%%%%%%%%%%%%%%%%%%%%%%%%%%%%%%%%%%%%%%%%%%%%%%%%%%%%
%%%%%%%%%%%%%%%%%%    Z  ->  b  b    %%%%%%%%%%%%%%%%%%%%%%%%%%%%%%%%
%%%%%%%%%%%%%%%%%%%%%%%%%%%%%%%%%%%%%%%%%%%%%%%%%%%%%%%%%%%%%%%%%%%%%
\subsection{Radiative corrections to the $ Z\to b\overline{b}$
decay} The theory developed in \Sec{sec:Zbb} can be
straightforwardly used here, in particular the new physics is also
parametrized through the modifications to $g_L$. Now these come
from the same set of diagrams displayed in \Fig{fig:ZbbinUED}, where the
tilde fields are now the fields that run inside the loop. The
different contributions are parametrized in exactly the same 
way they were in the continuous scenario, we reproduce them here
for commodity of the reader.
\begin{equation}
i\mathcal{M}_i = i \frac{g}{c_w}
\frac{\sqrt{2}G_Fm_t^2}{(4\pi)^2}f(r_i)
\overline{u}^\prime\gamma^\mu P_L u  \epsilon_\mu
\end{equation}
where $u$ and $u^\prime$ are the spinors of the $b$ quarks and
$\epsilon_\mu$ stands for the polarization vector of the $Z$
boson. The argument of the function $f$ is now $r_i=m_t^2/M_i^2$.
\begin{eqnarray}
f^{(a)}(r_i)& =& \left(1-\frac{4}{3}s_w^2 \right)
\left[\frac{r_i-\log(1+r_i)}{r_i}\right]\\
f^{(b)}(r_i)& = &\left(-\frac{2}{3}s_w^2 \right)\left[\delta_i-1
%\right. \\
%&
+
%&\left.
\frac{2r_i+r_i^2-2(1+r_i^2)\log(1+r_i)}{2r_i^2}\right]
%\nonumber
\\
f^{(c)}(r_i) & = & \left(-\frac{1}{2}+s_w^2\right) \left[ \delta_i
%\right. \\
%&
+
%&\left.
\frac{2r_i+3 r_i^2-2(1+r_i)^2 \log(1+r_i)}{2r_i^2}\right]
%\nonumber
\\ f^{(d)}(r_i)+f^{(e)}(r_i)&=&\left(\frac{1}{2}-\frac{1}{3}s_w^2\right)\left[
\delta_i+ \frac{2r_i+3 r_i^2-2(1+r_i)^2
\log(1+r_i)}{2r_i^2}\right]
\end{eqnarray}
where $\delta_i\equiv 2/\epsilon-\gamma+\log (4\pi) + \log
(\mu^2/M_i^2)$, and $\mu$ is the 't Hooft mass scale. From the above
equations it is straightforward to verify that all the terms
proportional to $\delta_i$ cancel, and so do all terms proportional to
$s_w^2$. Thus, finally, the only term which survives is the term in
$f^{(a)}(r_i)$ not proportional to $s_w^2$, yielding the following
contribution
\begin{equation}
\label{eq:deltagl} \delta g_{Li}
=\frac{\sqrt{2}G_Fm_t^2}{(4\pi)^2}
\left[\frac{r_i-\log(1+r_i)}{r_i}\right]
\end{equation}
The gaugeless limit leads exactly to the same conclusions as it is
done in Ref.~\cite{Oliver:2002up} to derive essentially the same
calculation in UED. Notice also here the absence of logarithms in
\Eq{eq:deltagl} when $r_i\to 0$.
%, in this limit $\delta g_{Li}$ as it should to fulfill decoupling.
The full contribution
%\begin{equation}
$\delta g_L^{NP} = \sum_{i=1}^{N-1}\delta g_{Li}$
%\end{equation}
expressed in terms of $F(a)$ can be written in the form
\begin{equation}
F_{\mathrm{LUED}}(a)=\int_0^1dx\sum_{i=1}^{N-1}\frac{a^2 x}{4 (N-1)^2
\sin^2(i\pi/ 2N)+a^2 x}
\end{equation}
This function captures the correction proportional to $m_t^2$, the
full one loop result could be adapted from Ref.~\cite{Buras:2002ej} by
replacing $m_n\to M_i$ as explained above. We have shown in
\Sec{sec:Zbb} that $F(a)-1<0.39$ at $95 \% \mbox{C.L.}$, from which
the results displayed in \Fig{fig:results} 
%on page \pageref{fig:results} 
follow.

 \comment{The results for UED and HG
\cite{Papavassiliou:2000pq} can be easily derived from
\rf{eq:faued} and \rf{eq:fahg}
\begin{equation}
R^{-1}_{UED}> 230\;\mbox{GeV} \qquad R^{-1}_{HG}> 1\;\mbox{TeV}
\qquad 95\%\;\mbox{CL}
\end{equation}
Observe that in the case of UED the bound is perfectly compatible
with the one obtained from the $\rho$ parameter. Taking into
account the $m_t$ proportional corrections has not improved the
bound although the one it gives is not as loose as previously
believed \cite{Appelquist:2000nn}.

In the case of the latticized scenarios the bound to the first
excited mode as function of $N$ is extracted from \rf{eq:falhg}
and \rf{eq:falued} and it the results are shown in figure
\ref{fig:LHG} in dashed line.}

%%%%%%%%%%%%%%%%%%%%%%%%%%%%%%%%%%%%%%%%%%%%%%%%%%%%%%%%%%%%%%%%%%%%%
%%%%%%%%%%%%%%%%%%      b -> s  photon      %%%%%%%%%%%%%%%%%%%%%%%%%
%%%%%%%%%%%%%%%%%%%%%%%%%%%%%%%%%%%%%%%%%%%%%%%%%%%%%%%%%%%%%%%%%%%%%
\subsection{Radiative corrections to $b\to s \gamma$} Given the
similitude with the continuous case we will not develop all the theory
again, instead we will focus on the computation of the
contribution to the i-th mode to the $C_7$ coefficient,
$C_{7\;i}(M_W)$. Again, it comes from the diagrams of
\Fig{fig:bsginUED}, when the tilde fields are the ones that run inside
the loop and amounts to
\begin{equation}
C_{7\; i}= \frac{m_t^2}{m_t^2+M_i^2} \left[
B\left( \frac{m_t^2+M_i^2}{M_i^2}\right) - \frac{1}{6} A \left(
\frac{m_t^2+M_i^2}{M_i^2} \right) \right]
\end{equation}
where $A(x)$ and $B(x)$ are defined in \Eq{eq:defa} and \Eq{eq:defb}
respectively. Of course, an expansion of $C_{7\;i}$ is free of
logarithms that relate the two different mass scales $M_i$ and $m_t$,
due to the same reasons as in the continuous scenario. The total
result reads
\begin{equation}
C_{7}^{\mathrm{LUED}}(M_W)=C_7^{\mathrm{SM}}(M_W)+\sum_{i=1}^{N-1}
C_{7\;i}(M_W),
\end{equation}
where we have neglected the running between $m_t$ and $M_W$, i.e.
$C_{7\;i}(m_t)\approx~C_{7\;i}(M_W)$. To take into account the QCD
running from $M_W$ to $m_b$ \Eq{eq:running} is used. Again $C_2$ takes
the same value as in SM model $C_2(M_W) = 1$ and the contribution of
$C_8(M_W)$ is neglected. The modifications to $b\to cl\nu$ are also
negligible because LUED corrects it again at the one loop level.

Using the result derived in Ref.~\cite{Agashe:2001xt}
\begin{equation}
\left| \frac{|C_7^{total}(m_b)|^2}{|C_7^{SM}(m_b)|^2} - 1
\right|<0.36 \qquad 95\%\;\mbox{CL}.
\end{equation}
With this the results can be extracted, and they are shown in
\Fig{fig:results}.

%%%%%%%%%%%%%%%%%%%%%%%%%%%%%%%%%%%%%%%%%%%%%%%%%%%%%%%%%%%%%%%%%%%%%
%%%%%%%%%%%%%%%%%%    rho parameter    %%%%%%%%%%%%%%%%%%%%%%%%%%%%%%
%%%%%%%%%%%%%%%%%%%%%%%%%%%%%%%%%%%%%%%%%%%%%%%%%%%%%%%%%%%%%%%%%%%%%
\subsection{Radiative corrections to the $\rho$ parameter}
\begin{figure}
\begin{center}
\includegraphics[scale = 1]{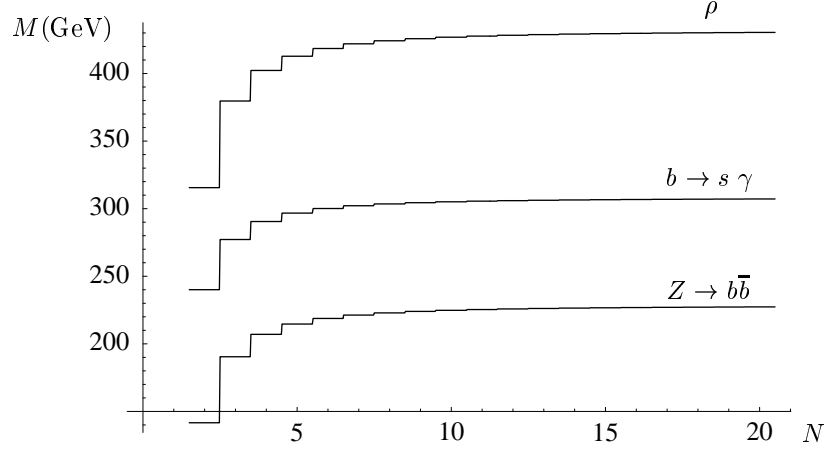}
\caption{The bounds on the mass of the first KK mode, $M$, as a
  function of the number of sites $N$.}
\label{fig:results}
\end{center}
\end{figure}

For this observable the same steps as in the continuous cases must be
done, this leads to
\begin{equation}
\Delta\rho_{i} = \frac{4}{g^2
v^2}\left[\Sigma_{1\;i}(0)-\Sigma_{3\;i}(0)\right] = 2 N_c
\frac{\sqrt{2} G_F m_t^2}{(4\pi)^2}
\left[1-\frac{2}{r_i}+\frac{2}{r_i^2}\log(1+r_i)\right],
\end{equation}
where $r_i=m_t^2/M_i^2$ and the correspondent diagrams are displayed
in \Fig{fig:rhoparaUED}. The total contribution would be found summing
$\Delta\rho^{\mathrm{LUED}}=
\Delta\rho^{\mathrm{SM}}+\sum_{i=1}^{N-1}\Delta\rho_{i}$. Now, for
each value of the number of sites $N$ we can extract the correspondent
bounds, these are displayed in \Fig{fig:results}.

%%%%%%%%%%%%%%%%%%%%%%%%%%%%%%%%%%%%%%%%%%%%%%%%%%%%%%%%%%%%%%%%%%%%%
%%%%%%%%%%%%%%%%%%      B mixing            %%%%%%%%%%%%%%%%%%%%%%%%%
%%%%%%%%%%%%%%%%%%%%%%%%%%%%%%%%%%%%%%%%%%%%%%%%%%%%%%%%%%%%%%%%%%%%%
\subsection{Radiative corrections to the $B^0 - \overline{B}^0$
system} The new physics modifies the value of $S(x_t)$ defined in
\Eq{eq:sxtdef} and again we parametrize this modification through the
function $G(a)$. The diagrams are the ones in \Fig{fig:BBmixinUED}
when the tilde fields run inside the loop and the result is
\begin{equation}
G_{\mathrm{LUED}}(a) = \frac{S_{NP}}{S_{SM}}= 1+2\int_0^1\sum_{n=1}^{N-1}\frac{a^2
x (1-x)\;dx}{4 (N-1)^2 \sin^2(n\pi/2N)+a^2 x}
\end{equation}
\comment{The expression for $G_{\mathrm{UED}}(a)$ is obtained following arguments
very similar to the ones used in Ref.~\cite{Oliver:2002up}. The
full one loop calculation has been previously performed in
Ref.~\cite{Buras:2002ej,Chakraverty:2002qk} and reduces to
\rf{eq:sxt} when only corrections proportional to $m_t$ are
retained. With this, the full 1-loop result for $G_{LUED}(a)$ can
be easily computed by adapting the results in
Ref.~\cite{Buras:2002ej} because in LUED the diagrams are exactly
the same ones as in UED (for each mode) as well as the couplings
that appear in them and the only difference appears in the
propagators, in UED they contain $m_n=n/R$ while in LUED it is
replaced by $M_n$ which is defined in \rf{eq:massdefinition}.
Notice that the contributions in all cases happen to be positive.}

The last experimental determinations \cite{Buras:2002yj} agree with
the SM expectations
\begin{equation}
\label{eq:Sxtexper}
1.3\le S(x_t)\le 3.8  \qquad  95\%\;\mbox{CL}.
\end{equation}
The possible positive contributions have been lowered with respect to
previous determinations \cite{Bergmann:2001pm} allowing better
bounds.
% In \Fig{fig:bbLxdim} we plot $M$ in terms of the number of copies
% $N$. 
%For the case of LUED not only the bound imposed by the
%dominant $m_t$ pieces is shown but in addition it is displayed the
%bound obtained when the full one loop contribution is considered, what
%can be done by using the results provided in
%Ref.~\cite{Buras:2002ej}. It can be seen that 
Despite the enhancement on the upper bound in \Eq{eq:Sxtexper} the
bound on $M$ is still rather weak.
%\begin{figure}
%\begin{center}
%\includegraphics[scale=0.9]{LUED/graphBBxdim.eps}
%\caption{Bounds on $M$ coming from the $B$ sector. The one loop
%  superscript means that the full one loop contributions has been
%  maintained. We also show the result of the LHG scenario that will be
%  studied in chapter \ref{chap:HGLHG}.}
%\label{fig:bbLxdim}
%\end{center}
%\end{figure}
At the end, the bounds one can set on the masses of the new modes are
below the $W$ mass, and are therefore irrelevant compared to
previously discussed bounds.  Given the future experimental
improvements on the determinations of $\sin 2\beta$ by BaBar and BELLE
and in particular of the mass splitting $\Delta M_s$ for the B sector
in LHC and FNAL one may use this observable to predict possible
deviations from the SM predictions. It turns out that to an excellent
accuracy \cite{Buras:2002ej} the mentioned deviations in the case of
$\Delta M_s$ are governed by $G(a)$
\begin{equation}
G_{\mathrm{NP}}(a)=\frac{(\Delta M_s)_{NP}}{(\Delta M_s)_{SM}}>1
\end{equation}
The greater values of $G(a)$ occur for small $N$ but they are
at most $G(a)\le 1.14$ that happens to be a too small deviation to be
discriminated experimentally, in fact it is of the same order as the
deviation studied in Ref.~\cite{Buras:2002ej} and therefore the same
reasonings given there apply here. The virtue of this results is
that the possible existence of extra latticized dimensions will not
pollute the extraction of the CKM matrix parameters from the future
improvements in the determination of the unitarity triangle.

%%%%%%%%%%%%%%%%%%%%%%%%%%%%%%%%%%%%%%%%%%%%%%%%%%%%%%%%%%%%%%%%%%%%%%%%%
%%%%%%%%%%%%%%%%%%%%%%%%%%%%%%%%%%%%%%%%%%%%%%%%%%%%%%%%%%%%%%%%%%%%%%%%%
%%%%%%%%%%%%%%%%%%%%%%%%%%%%%%%%%%%%%%%%%%%%%%%%%%%%%%%%%%%%%%%%%%%%%%%%%

\section{Outlook and conclusions} \label{Conclusions} 
We have studied a five-dimensional extension of the SM in which the
extra spatial dimension is latticized, and all SM fields propagate in
it.  The model has the property that there are no tree-level effects
below the threshold of production of new particles. Therefore, to set
a lower bound on the scale of the new physics one should consider
one-loop processes. We considered a number of well-measured
observables, and which depend strongly on the top-quark mass: the
$\rho$ parameter, $b\to s \gamma$, $Z\to b\overline b$, and the
$B^0\rightleftharpoons\overline{B}^0$ mixing.  The dominant
corrections, i.e. those proportional to the top-quark mass, have been
computed, and compared with the ones obtained when only the SM is
considered.  It is found that the known bounds for the continuous
version (UED) are rapidly reached when the extra dimension is
latticized by only about 10 to 20 (four dimensional) sites. However,
when a smaller number of sites is considered, the bounds on the scale
of new physics is lowered by roughly a factor of $10$\%--$25$\%, as
can be seen in \Fig{fig:results}. This suggests that the phenomenology
of latticized scenarios can be more accessible than in the continuous
cases. Then, the limits on new particles are about $320$--$380$~GeV.
The bounds shown in \Fig{fig:results} correspond to the mass of the
lightest modes, defined in \Eq{eq:massdefinition}.

\comment{
We have also briefly discussed a latticized version of a model in which
fermions are confined to the four-dimensional subspace. In these models, 
there are contributions that modify the muon decay rate already at tree
level, allowing for much stronger bounds (of the order of $1$~TeV).

To summarize, we found that models with one universal latticized 
extra dimension provide new interesting physics which could be
well within the reach of the next generation of accelerators.
}%endcomment

%\documentclass[a4paper,preprint,aps,prd,eqsecnum,showpacs,nofootinbib]{revtex4}
%\usepackage[T1]{fontenc}
%\usepackage[latin1]{inputenc}
%\usepackage{amsmath}
%\usepackage{graphics}
%\usepackage{amssymb}
%\makeatletter
%\newcommand{\db}{\rlap{$\partial$}/}
%\newcommand{\sla}[1]{\rlap{$#1$}/}
%\newcommand{\slap}[1]{\rlap{$#1$}\hspace{0.3ex}/}
%\newcommand{\slam}[1]{\rlap{$#1$}\hspace{-0.3ex}/}
%\newcommand{\comment}[1]{}
%\newcommand{\msb}{\overline{\mathrm{MS}}}
%\newcommand{\tr}[1]{\mathrm{Tr}\left\{ #1 \right\} }
\chapter{Power corrections in models with extra dimensions}  
We revisit the issue of power-law running in models with extra
dimensions\cite{Oliver:2003cy}.  The analysis is carried out in the
context of a higher-dimensional extension of QED, with the extra
dimensions compactified on a torus.  It is shown that a naive $\beta$
function, which simply counts the number of modes, depends crucially
on the way the thresholds of the Kaluza-Klein modes are crossed.  To
solve these ambiguities we turn to the vacuum polarization, which, due
to its special unitarity properties, guarantees the physical
decoupling of the heavy modes. This latter quantity, calculated in the
context of dimensional regularization, is used for connecting the low
energy gauge coupling with the coupling of the $D$-dimensional
effective field theory.  We find that the resulting relation contains
only logarithms of the relevant scales, and no power corrections.  If,
instead, hard cutoffs are used to regularize the theory, one finds
power corrections, which could be interpreted as an additional
matching between the effective higher-dimensional model and some
unknown, more complete theory.  The possibility of estimating this
matching is examined in the context of a toy model. The general
conclusion is that, in the absence of any additional physical
principle, the power corrections depend strongly on the details of the
underlying theory.  Possible consequences of this analysis for gauge
coupling unification in theories with extra dimensions are briefly
discussed.

\section{Introduction}

The study of models with extra dimensions has received a great deal of
attention recently~\cite{Arkani-Hamed:1998nn,Arkani-Hamed:1998rs,Antoniadis:1990ew,
Antoniadis:1994jp}, mainly because  of the plethora of theoretical and
phenomenological ideas associated with  them, and the flexibility they
offer  for  realizing   new,  previously  impossible,  field-theoretic
constructions. One of the  most characteristic features of such models
is  that  of  the  {}``early  unification{}'': the  running  of  gauge
couplings is  supposed to be modified  so strongly by  the presence of
the tower of  KK modes, that instead of  logarithmic it becomes linear,
quadratic,    etc,    depending     on    the    number    of    extra
dimensions~\cite{Bachas:1998kr,Dienes:1998vh,Dienes:1998vg,Ghilencea:1998st,
Carone:1999cb,Delgado:1999ba,Floratos:1999bv,Frampton:1999ue,Kakushadze:1998vr,Kakushadze:1999bb,
Perez-Lorenzana:1999qb,Dumitru:1999ji,Masip:2000yw,Arkani-Hamed:2001vr,
Berezhiani:2001ub,Hebecker:2002vm,Kubo:1999ua}. 
Specifically, it has  been widely argued 
that the gauge couplings
run as \( \mu ^{\delta } \),  where the \( \delta \) is the number of
compact  extra   dimensions.  Thus,   if  the  extra   dimensions  are
sufficiently large, such  a behavior of the couplings  could allow for
their unification at  accessible energies, of the order  of a few TeV,
clearly an exciting possibility.

The assertion  that gauge-couplings display power-law  running is based
on rather intuitive arguments: In \(  \msb \) schemes the QED \( \beta
\) function  is proportional to the number  of {}``active{}'' flavors,
namely  the  number  of  particles lighter  than  the  renormalization
scale.  Using this  argument, and  just counting  the number  of modes
lighter than \( \mu \), one easily finds that the ``$\beta$ function''
of QED  in models  with extra  dimensions grows as  \( \mu  ^{\delta }
\). This  behavior is also  justified by explicit calculations  of the
vacuum polarization of the photon using hard cutoffs; since the cutoff
cannot be removed, due to  the non-renormalizability of the theory, it
is  finally identified  with  the renormalization  scale, a  procedure
which eventually  leads to a similar  
conclusion~\cite{Dienes:1998vh,Dienes:1998vg} (but  with the final
coefficient  adjusted by  hand  in order  to  match the  naive
expectation in \( \msb \)).

Even  though these arguments  are plausible,  the importance  of their
consequences   requires   that  they   should   be  scrutinized   more
carefully~\cite{Contino:2001si}.  In particular, the argument based on
\( \msb \) running is rather tricky.  As it is well known, the \( \msb
\)  scheme,  because  of  its  mass  independence,  does  not  satisfy
decoupling,  already  at   the  level  of  four-dimensional  theories.
Instead, decoupling  has to be  \textit{imposed} by hand every  time a
threshold  is  passed:  one  builds  an  effective  theory  below  the
threshold,  \(  m  \),  and   matches  it  to  the  theory  above  the
threshold.  This  matching  is  carried  out by  requiring  that  some
physical amplitude or Green's  function (i.e. the effective charge) is
the  same  when calculated  using  either  theory,  at energies  where
\textit{both} theories are reliable, namely  at \( Q^{2} \) much below
the threshold.  Then, since the  renormalization scale, \( \mu  \), is
still a free parameter, one chooses \( \mu \) around \( m \), in order
to avoid  large logarithms in the  matching equations. In  the case of
gauge    couplings    and    \(    \msb    \)    schemes    with    \(
\tr{I_{\mathrm{Dirac}}}=4  \)  one  finds  (at one  loop)  that  gauge
couplings are continuous at \(  \mu =m \). This statement is, however,
extremely     scheme     dependent:     just    by     choosing     \(
\tr{I_{\mathrm{Dirac}}}=2^{D/2}  \) it  gets completely 
modified (see for instance \cite{Rodrigo:1993hc}) .  In
addition to  these standard ambiguities, a new  complication arises in
the   context  of  higher-dimensional   models.  In   particular,  the
aforementioned procedure requires that  the different scales be widely
separated in  order  to   avoid  that  higher  dimension  operators,
generated in the process  of matching, become important.  However, the
condition  of having  well-separated thresholds  is  rather marginally
satisfied  in the  case  of an  infinite  tower of  KK  modes with  \(
M_{n}=nM_{c} \) (\( M_{c} \)  is the compactification scale). In fact,
as we will see in detail  later, the results obtained for a \( \beta
\) function  that just counts the number of active modes depend very strongly  
on the prescription chosen  for the
way the various thresholds are crossed.

As has been hinted above, the deeper reason behind these additional
type of ambiguities is the fact that, gauge theories in more than \( 4
\) dimensions, compactified or not, are not renormalizable.  At the
level of the 4-dimensional theory with an infinite number of KK modes
the non-renormalizability manifests itself by the appearance of extra
divergences, encountered when summing over all the modes. If the
theory is not compactified the non-renormalizability is even more
evident, since gauge couplings in theories with \( \delta \) extra
dimensions have dimension \( 1/M^{\delta /2} \).  Therefore, gauge
theories in extra dimensions should be treated as effective field
theories (EFT).  Working with such theories presents several
difficulties, but, as we have learned in recent years, they can also
be very useful. In the case of quantum field theories in extra
dimensions, there is no alternative: basic questions, such as the
calculation of observables or the unification of couplings, can only
be addressed in the framework of the EFT's. However, before attempting
to answer specific questions related to the running of couplings in
the extra-dimensional theories, one should first clarify the type of
EFT one is going to use, since there are, at least, two types of
EFT~\cite{Georgi:1993qn}: In one type, known as ``Wilsonian EFT''
(WEFT)~\cite{Wilson:1974jj}, one keeps only momenta below some scale
\( \Lambda \), while all the effects of higher momenta or heavy
particles are encoded in the couplings of the effective theory.  This
method is very intuitive and leads, by definition, to finite results
at each step; however, the presence of the cutoff in all expressions
makes the method cumbersome to use, and in the particular case of
gauge theories difficult to reconcile with gauge-invariance.  The WEFT
approach has already been applied to the problem of running of
couplings in theories with compact extra dimensions, but only for the
case of scalar theories \cite{Kubo:1999ua}. Within the context of
another type of EFT, often termed ``continuum effective field
theories'' (CEFT) (see for instance
\cite{Weinberg:1979kz,Weinberg:1980wa,Hall:1981kf,Georgi:1993qn,
Leutwyler:1994iq,Manohar:1996cq,Pich:1998xt}), one allows the momenta
of particles to vary up to infinity, but heavy particles are removed
from the spectrum at low energies. As in the WEFT case the effects of
heavier particles are absorbed into the coefficients of higher
dimension operators. Since the momenta are allowed to be infinite,
divergences appear, and therefore the CEFT need to undergo both:
regularization and renormalization.  In choosing the specific scheme
for carrying out the above procedures particular care is
needed. Whereas in principle one could use any scheme, experience has
shown that the most natural scheme for studying the CEFT is
dimensional regularization with minimal subtraction
\cite{Weinberg:1979kz,Weinberg:1980wa,Hall:1981kf,Georgi:1993qn,
Leutwyler:1994iq,Manohar:1996cq,Pich:1998xt}. CEFT are widely used in
Physics: for example, when in the context of QCD one talks about 3, 4
or 5 active flavors, one is implicitly using this latter type of
effective theories~\cite{Witten:1977kx,Weinberg:1980wa}.  Moreover,
most of the analyses of Grand Unification
\cite{Georgi:1974yf,Hall:1981kf} resort to CEFT-type of constructions:
one has a full theory at the GUT scale, then an effective field theory
below the GUT scale (SM or MSSM) is built, and then yet another
effective field theory below the Fermi scale (just QED+QCD).  In these
cases the complete theory is known, and the CEFT language is used only
in order to simplify the calculations at low energies and to control
the large logarithms which appear when there are widely separated
scales. Nevertheless, CEFT's are useful even when the complete theory
is not known, or when the connection with the complete theory cannot
be worked out; this is the case of Chiral Perturbation theory (\( \chi
PT \)) \cite{Callan:1969sn,Coleman:1969sm,
Weinberg:1968de,Gasser:1984yg} (for more recent reviews see also
\cite{Leutwyler:1994iq,Pich:1995bw,Ecker:1995gg}).

It is important  to maintain a sharp distinction  between the two types
of  EFT   mentioned  above,  i.e.  Wilsonian   or  continuum,  because
conceptually  they are quite  different. However,  perhaps due  to the
fact that the language is in part common to both types of theories, it
seems  that they  are often  used interchangeably  in  the literature,
especially  when  employing cutoffs  within  the  CEFT framework.   In
particular, since the  couplings \( \alpha _{i} \)  have dimensions \(
[\alpha _{i}]=M^{-n}  \), when  computing loops one  generally obtains
effects which  grow as \(  (\Lambda ^{n}\alpha _{i})^{m} \),  where \(
\Lambda \) is the formal CEFT  cutoff, and as such is void of physics.
As a  consequence, physical observables should be  made as independent
of these  cutoffs as possible  by introducing as many  counterterms as
needed   to    renormalize   the   answer.    Not   performing   these
renormalizations correctly, or identifying naively formal cutoffs with
the physical cutoffs  of the effective theory, can  lead to completely
non-sensible         results         (see         for         instance
\cite{Burgess:1992va,Burgess:1993gx}).  This  type of pitfalls  may be
avoided  by simply using  dimensional regularization, since  the latter
has  the  special property  of  not  mixing  operators with  different
dimensionalities.

The usual way to treat  theories with compactified extra dimensions is
to define them as a 4-dimensional  theory with a truncated tower of KK
modes at some  large but otherwise arbitrary \(  N_{s} \), a procedure
which effectively amounts to using a hard cutoff in the momenta of the
extra dimensions. Thus, physical  quantities calculated in this scheme
depend explicitly  on the  cutoff \( N_{s}  \), which  is subsequently
identified with  some physical cutoff. However,  as already commented,
\(  N_{s} \)  plays the  role  of a  formal cutoff,  and is  therefore
plagued with  all the  aforementioned ambiguities.  Identification of
this  formal cutoff  with a  universal  physical cutoff  can give  the
illusion of predictability, 
making us forget that we are
dealing  with  a non-renormalizable  theory  with  infinite number  of
parameters, which can be predictive only at low energies, where higher
dimension operators may be neglected.

In this paper we want to  analyze the question of the running of gauge
couplings  in   theories  with   compact  dimensions  from   the  CEFT
{}``canonical{}'' point of view. We  hasten to emphasize that even the 
CEFT
presents   conceptual   problems   in   theories   with   compactified
dimensions. Specifically, as mentioned above, in the CEFT approach the
(virtual) momenta are allowed to vary up to infinity; however, momenta
related to the compactified extra  dimensions turn out to be KK masses
in the  4-dimensional compactified theory,  where it is  supposed that
one  only  keeps particles  lighter  than  the  relevant scale.  Thus,
truncating the  KK series  amounts to cutting  off the momenta  of the
compactified  dimensions.   Therefore,  in  order  to  define  a  true
{}``non-cutoff{}'' CEFT scheme we are forced to keep all KK modes. Our
main motivation is to seriously explore this approach, and investigate
both its virtues and its limitations  for the problem at hand. We hope
that this  study will help  us identify more clearly  which quantities
can  and  which  cannot  be computed  in  effective  extra-dimensional
theories.

In section \ref{sec:thresholds} we discuss the usual arguments in
favor of power-law running of gauge couplings and show that they
depend crucially on the way KK thresholds are crossed.  In particular
we show that, a one-loop $\beta$ function which simply counts the
number of modes, diverges for more than \( 5 \) dimensions, if the
physical way of passing thresholds dictated by the vacuum polarization
function (VPF) is imposed.

In section \ref{sec:model} we introduce  a theory with one fermion and
one  photon in  4+$\delta$ dimensions,  with the  extra  $\delta$ ones
compactified. This theory, which is  essentially QED in 4+\( \delta \)
dimensions, serves  as toy model  for studying the issue  of power
corrections and the running of the coupling in a definite framework.

In section  \ref{sec:effectiveKK} we study the  question of decoupling
KK modes in the aforementioned theory by analyzing the behavior of the
VPF of  the (zero-mode) photon. Since, as  commented above, decoupling
the KK modes  one by one is problematic, we study  the question of how
to decouple all  of them at once.  To accomplish  this we consider the
VPF of the photon with all KK modes included, and study how it reduces
at \(  Q^{2}\ll M_{c} \) to the  standard QED VPF with  only one light
mode.  Since the entire KK tower is kept untruncated, the theory is of
course non-renormalizable;  therefore, to compute  the VPF we  have to
regularize and renormalize it in the  spirit of the CEFT, in a similar
way that observables are  defined in \( \chi PT \).  As  in \( \chi PT
\),  it is  most  convenient to  use  dimensional regularization  with
minimal  subtraction, in  order to  maintain a  better control  on the
mixing  among  different  operators.  However,  at the  level  of  the
4-dimensional theory the non-renormalizability manifest itself through
the  appearance of  divergent sums  over  the infinite  KK modes,  and
dimensional regularization does no  seem to help in regularizing them.
The dimensional  regularization of the VPF  is eventually accomplished
by exploiting  the fact  that its UV
behavior coincides to that found when the 
\( \delta \)  extra dimensions have not been compactified\footnote{This
is in a way expected,
since for  very large \(Q^{2}\gg M_{c}=1/R_{c} \)
the compactification effects should be negligible. Note, however, that
this is not always the case; a known exception is provided by the orbifold
compactification~\cite{Contino:2001si}.}.
To explore this point
we first resort to  the standard unitarity relation (optical theorem),
which relates the imaginary part of the VPF to the total cross section
in the  presence of  the KK  modes; the latter is finite  because the
phase-space truncates the  series.  For $Q^2\gg M_c^2$ the uncompactified
result for the imaginary part of the VPF is rapidly reached, i.e. after 
passing a few thresholds. 
We then compute the real part of the one-loop VPF in
the non-compact theory in \(  4+\delta \) dimensions, where, of course
we can use directly dimensional regularization to regularize it (since
no  KK reduction has  taken place).   For later  use we  also present
results in which the same quantity is evaluated by using hard cutoffs.
Finally,  we show  that the  UV divergences  of the  one-loop  VPF are
indeed the  same in  both the (torus)-compactified  and uncompactified
theories.   Therefore,   in  order  to  regularize  the   VPF  in  the
compactified  theory  with  an  infinite  number of  KK  modes  it  is
sufficient  to split the  VPF into  two pieces,  an ``uncompactified''
piece,  corresponding  to the  case  where  the  extra dimensions  are
treated at the same footing as  the four usual ones, and a piece which
contains all compactification effects.  We show that this latter piece
is  UV  and  IR finite  and  proceed  to  evaluate  it, while  all  UV
divergences remain in the  former, which we evaluate using dimensional
regularization.

The     results     of    previous     sections     are    used     in
section~\ref{sec:matching}  to define  an effective  charge  \( \alpha
_{\mathrm{eff}}(Q) \)  which can be continuously  extrapolated from \(
Q^{2}\ll  M_{c} \) to  \( Q^{2}\gg  M_{c} \).   We use  this effective
charge to study the matching  of couplings in the low energy effective
theory (QED) to the couplings of the theory containing an infinite  of KK 
modes.  In
the context  of dimensional regularization we find  that this matching
contains only the standard logarithmic running from \( m_{Z} \) to the
compactification scale \( M_{c} \), with no power corrections.  On
the other hand, if hard cutoffs  are used to regularize the VPF in the
non-compact  space,  one  does  find power corrections,
which may  be interpreted  as  an  additional
matching between  the effective \( D=4+\delta \) dimensional field  theory and
some  more complete  theory. We discuss the possibility of estimating this
matching in the EFT without knowing the details of the full theory. This
point is studied in a simple extension 
of our original toy-model, by endowing 
the theory considered (QED in  4+\( \delta \) compact dimensions) with
an  additional  fermion with  mass  \(  M_{f}\gg  M_{c} \),  which  is
eventually integrated out.

\section{Crossing thresholds\label{sec:thresholds}}

The simplest argument (apart from the purely dimensional ones)
in favor of power-law running in theories with extra dimensions is based on
the fact that in \( \msb  \)-like schemes the $\beta$ function is
proportional to the number of active modes.
Theories with \( \delta  \) extra 
compact dimensions contain, in general, 
a tower of KK modes. In particular, if we embed QED in extra dimensions
we find that electrons (also photons) 
have a tower of KK modes 
with masses \( M^{2}_{n}=\left( n^{2}_{1}+n^{2}_{2}+
\cdots +n^{2}_{\delta }\right)  \)
\( M^{2}_{c} \) with \( n_{i} \) integer values and \( M_{c}=1/R_{c} \) the
compactification scale. The exact multiplicity of the spectrum depends on the
details of the compactification procedure (torus, orbifold, etc). As soon as
we cross the 
compactification scale, the KK modes begin to contribute,
and therefore one expects that the $\beta$ function 
of this theory will start to
receive additional contributions from them. In a general renormalization scheme
satisfying decoupling one can naively write 
\begin{equation}
\label{eq:beta-naive}
\beta =\sum _{n}\beta _{0}f\left( \frac{\mu }{M_{n}}\right)~, 
\end{equation}
 where \( \mu  \) is the renormalization scale, 
\( \beta _{0} \) is the contribution
of a single mode, and \( f(\mu /M) \) is a general step-function that decouples
the modes as \( \mu  \) crosses the different thresholds, namely \( f(\mu /M)\rightarrow 0\, \, \, \, \, \mu \ll M \)
and \( f(\mu /M)\rightarrow 1\, \, \, \, \, \mu \gg M \). For instance in \( \msb  \) schemes \( f(\mu /M)\equiv \theta (\mu /M-1) \) 
where \( \theta (x) \) is
the step-function. Then one finds \( \left( \Omega _{\delta }=2\pi ^{\delta /2}/\Gamma (\delta /2)\right)  \)
\begin{equation}
\label{eq:beta-naive-dr}
\beta =\sum _{n<\mu /M_{c}}\beta _{0}\approx \beta _{0}\int d\Omega _{\delta
}n^{\delta -1}dn=\beta _{0}\frac{\Omega _{\delta }}{\delta }\left( \frac{\mu ^{2}}{M^{2}_{c}}\right) ^{\delta /2}~.
\end{equation}
 This argument, simple and compelling as it may seem, cannot be trusted completely
because in \( \msb  \) schemes the decoupling is put in by hand.
Therefore, other types of schemes, in which decoupling seems natural, have been
studied in the literature. For instance, in Ref.~\cite{Dienes:1998vg} the
VPF of the photon at \( Q^{2}=0 \) was calculated in the presence of the infinite
tower of KK modes by using a hard cutoff in proper time, and the result was
used to compute the $\beta$ function; in that case the modes decouple smoothly.
In addition, after adjusting the cutoff by hand one can reproduce the aforementioned
result obtained in \( \msb  \). One can easily see that this procedure is equivalent
to the use of the function \( f(\Lambda /M)\equiv e^{-\frac{M^{2}_{n}}{\Lambda ^{2}}} \)
to decouple the KK modes \begin{equation}
\label{eq:beta-naive-kk}
\beta =\sum _{n}\beta _{0}e^{-\frac{M^{2}_{n}}{\Lambda ^{2}}}\approx \beta _{0}\left( \pi \frac{\Lambda ^{2}}{M^{2}_{c}}\right) ^{\delta /2}\, .
\end{equation}
 If one chooses by hand 
 \( \mu ^{\delta }=\Gamma (1+\delta /2)\Lambda ^{\delta } \),
the sum in Eq.~(\ref{eq:beta-naive-kk}) agrees exactly with the sum obtained
if one uses a sharp step-function. Even though this particular way of decoupling
KK modes appears naturally in some string 
scenarios~\cite{Quiros:2001yi,Hamidi:1987vh,Dixon:1987qv,Antoniadis:1994jp},
it hardly appears compelling from the field theory point of view; this procedure
is not any better conceptually than the sharp step-function decoupling of modes:
one obtains a smooth $\beta$ function because one uses a smooth function to decouple
the KK modes.

These two ways of decoupling KK modes, due to the very sharp step-like behavior
they impose, lead to a finite result in (\ref{eq:beta-naive}) for any number
of extra dimensions. One is tempted to ask, however, what would happen if one
were to use a more physical way of passing thresholds. In fact, heavy particles
decouple naturally and smoothly in the VPF, because they cannot be
produced physically. Specifically, in QED in 4-dimensions at the one-loop level,
the imaginary part, \( \Im m\Pi (q^{2}) \), of the VPF \( \Pi (q^{2}) \) is
directly related, via the optical theorem, to the tree level cross sections
\( \sigma  \) for the physical processes \( e^{+}e^{-}\rightarrow f^{+}f^{-} \),
by 
\begin{equation}
\label{sigmaff1}
\Im m\Pi (s)=\frac{s}{e^{2}}\, \sigma (e^{+}e^{-}\rightarrow f^{+}f^{-})~.
\end{equation}
Given a particular contribution to the spectral function \( \Im m\Pi (s) \),
the corresponding contribution to the renormalized vacuum polarization function
\( \Pi _{R}(q^{2}) \) can be reconstructed via a once--subtracted dispersion
relation. For example, for the one--loop contribution of the fermion \( f \),
choosing the on--shell renormalization scheme, one finds (if \( q \) is the
physical momentum transfer with \( q^{2}<0 \), as usual we define \( Q^{2}\equiv -q^{2} \)):
\[
\Pi _{R}(Q)=Q^{2}\, \int _{4m_{f}^{2}}^{\infty }ds\frac{1}{s(s+Q^{2})}\frac{1}{\pi }\, \Im m\Pi (s)\]

\begin{equation}
\label{PiRffQED}
=\frac{\alpha }{\pi }\times \left\{ \begin{array}{lr}
\displaystyle {\frac{1}{15}\frac{Q^{2}}{m_{f}^{2}}+\mathcal{O}\left( \frac{Q^{4}}{m_{f}^{4}}\right) } & \, \, \, \, Q^{2}/m_{f}^{2}\rightarrow 0\\
{\frac{1}{3}\, \ln \left( \frac{Q^{2}}{m^{2}_{f}}\right) -\frac{5}{9}+\mathcal{O}\left( \frac{m_{f}^{2}}{Q^{2}}\right) } & \, \, \, \, Q^{2}/m_{f}^{2}\rightarrow \infty \, ,
\end{array}\right. 
\end{equation}
where \( \alpha \equiv e^{2}/(4\pi ) \). The above properties can be extended
to the QCD effective charge \cite{Binosi:2002vk}, with the
appropriate modifications to take into account the non-Abelian nature of the
theory, and provide a physical way for computing the matching equations between
couplings in QCD at quark mass thresholds. One computes the VPF of QCD with
\( n_{f} \) flavors and that of QCD with \( n_{f}-1 \) flavors, and requires
that the effective charge is the same for \( Q^{2}\ll m_{f} \) in the two theories.
This procedure gives the correct relation between the couplings in the two theories
~\cite{Bernreuther:1982sg,Rodrigo:1993hc,Rodrigo:1998zd}. However, one can
easily see that this cannot work for more than one extra dimension. To see that,
let us consider the decoupling function \( f(\mu /M) \) provided by the
one-loop VPF, which, as explained, captures correctly the physical thresholds.
The corresponding \( f(\mu /M) \) may be obtained by differentiating \( \Pi
_{R}(Q) \)
once with respect to \( Q^{2} \); it is known \cite{Brodsky:1998mf} that the
answer can be well-approximated by a simpler function of the form \( f(\mu /M)=\mu ^{2}/(\mu ^{2}+5M^{2}) \).
We see immediately that if we insert this last function in
Eq.~(\ref{eq:beta-naive})
and perform the sum over all KK modes the result is convergent only for one
extra dimension (with a coefficient which is different from the one obtained
with the renormalization schemes mentioned earlier), while it becomes highly divergent for
several extra dimensions. We conclude therefore that the physical way of decoupling
thresholds provided by the VPF seems to lead to a divergent $\beta$ function in
more than one extra dimension. As we will see, this is due to the fact that,
in order to define properly 
the one-loop VPF for \( \delta >1 \), more than one subtraction is needed.

\section{A toy model\label{sec:model}}

To be definite we will consider a theory 
with one fermion and one photon in
\( 4+\delta  \) dimensions, 
in which the \( \delta  \) extra dimensions are
compactified on a torus of 
equal radii \( R_{c}\equiv 1/M_{c} \). 
The Lagrangian is given by
\begin{equation}
\label{eq:DeffLagrangian}
{\mathcal{L}}_{\delta }=-\frac{1}{4}F^{\alpha\beta}F_{\alpha\beta}+i\bar{\psi }\gamma
^{\alpha}D_{\alpha}\psi +\mathcal{L}_{\mathrm{ct}}~,
\end{equation}
where \( \alpha=0,\cdots ,3,\cdots ,3+\delta \). We will also use
Greek letters to denote four-dimensional indices \( \mu =0,\cdots \, 3
\). \( D_{\alpha}=\partial _{\alpha}-ie_{D}A_{\alpha} \) is the
covariant derivative with \( e_{D} \) the coupling in \( 4+\delta \)
dimensions which has dimension \( [e_{D}]=1/M^{\delta /2} \).  After
compactification, the dimensionless gauge coupling in four-dimensions,
\( e_{4} \), and the dimensionfull \( 4+\delta \) coupling are related
by the compactification scale \footnote{ Note that the factors \(
2\pi \) depend on the exact way the extra dimensions are compactified
(on a circle, orbifold, etc). }
\begin{equation} 
e_{4}=e_{D}\left( \frac{M_{c}}{2\pi }\right) ^{\delta /2}  \,. 
\label{relcoup}
\end{equation}
Evidently 
\( e_{D} \) is determined from 
the four-dimensional gauge coupling and the compactification
scale, but in the uncompactified space we can regard it  
as a free parameter
(as \( f_{\pi } \) in \( \chi PT \)). 
Finally \( \mathcal{L}_{\mathrm{ct}} \)
represents possible gauge invariant operators 
with dimension \( 2+D \) or higher,
which are in general needed for renormalizing the theory; they can be computed
only if a more complete theory, from which our effective theory originates,
is given. For instance, by computing the VPF we will see that 
a $\mathcal{L}_{\mathrm{ct}}$ of the form 
\begin{equation}
\label{eq:lct}
\mathcal{L}_{\mathrm{ct}}=\frac{c_{1}}{M^{2}_{s}}D_{\alpha}F^{\alpha\beta}D^{\lambda}F_{\lambda\beta}+\cdots
\end{equation}
is needed to make it finite. 

The spectrum after compactification contains a photon (the zero mode of the
four-dimensional components of the gauge boson), the \( \delta  \) extra 
components
of the gauge boson remain in the spectrum as \( \delta  \) massless real scalars,
a tower of massive vector bosons with masses \( M^{2}_{n}=\left( n^{2}_{1}+n^{2}_{2}+\cdots +n^{2}_{\delta }\right)  \)
\( M^{2}_{c} \), \( n_{i}\in \mathbb {Z},\, n_{i}\neq 0 \) , \( 2^{[\delta /2]} \)
massless Dirac fermions (here the symbol \( [x] \) represents the closest integer
to \( x \) smaller or equal than \( x \)), and 
a tower of massive Dirac fermions
with masses given also by the above mass formula. Note that this theory does
not lead to normal QED at low energies, first because the \( \delta  \) extra
components of the gauge boson remain in the spectrum, and second because in
\( 4+\delta  \) dimensions the fermions have \( 4\cdot 2^{[\delta /2]} \)
components, which remain as zero modes, leading at low energy to a theory with
\( 2^{[\delta /2]} \) Dirac fermions. In the \( D=4+\delta  \) theory these
will arise from the trace of the identity of the $\gamma$ matrices, which just
counts the number of components of the spinors. To obtain QED as a low energy
one should project out the correct degrees of 
freedom by using some more appropriate
compactification (for instance, orbifold compactifications can remove the extra
components of the photon from the low energy spectrum, and leave just one Dirac
fermion). However this is not important for our discussion of the VPF, we just
have to remember to drop the additional factors \( 2^{[\delta /2]} \) to make
contact with usual QED with only one fermion. Theories of this type, with all
particles living in extra dimensions are called theories with {}``universal
extra dimensions{}''~\cite{Appelquist:2000nn} and have the characteristic
that all the effects of the KK modes below the compactification scale cancel
at tree level due to the conservation of the KK number. 
In particular, and contrary to what happens in 
theories where gauge and scalar fields 
live in the bulk and fermions in the 
brane~\cite{Pomarol:1998sd,Delgado:1999sv}, 
no divergences associated to summations over KK towers 
appear at tree level.  
Finally, the couplings
of the electron KK modes to the standard zero-mode photon are universal and
dictated by gauge invariance. The couplings among the KK modes can be found
elsewhere \cite{Papavassiliou:2001be,Muck:2001yv}; they will not be important
for our discussion of the VPF that we present here.

\section{The vacuum polarization in the presence of KK modes\label{sec:effectiveKK}}

In this section we will study in detail the behavior of the one-loop
VPF in the theory defined above for general values of the number
$\delta$ of extra dimensions. The main problems we want to address are:
i) the general divergence structure of the VPF, ii) demonstrate that
it is possible to regulate the UV divergences using dimensional
regularization, iii) the appearance of non-logarithmic (power)
corrections, and, iv) their comparison to the analogous terms obtained
when resorting to a hard-cutoff regularization.

\subsection{The imaginary part of the vacuum polarization}

One can try to compute directly the VPF of the zero-mode photon in a theory
with infinite KK fermionic modes. However, one immediately sees that, 
in addition
to the logarithmic divergences that one finds in QED, new divergences
are encountered when summing over the infinite number of KK modes. One can understand
the physical origin of these divergences more clearly by resorting to the unitarity
relation (here \( s \) denotes the center-of-mass energy 
available for the production process):
\begin{eqnarray}
\Im m\Pi ^{(\delta )}(s) & = & \frac{s}{e_{4}^{2}}\, \sum _{n}\sigma (e^{+}e^{-}\rightarrow f_{n}^{+}f_{n}^{-})\nonumber \label{eq:first} \\
 & = & \frac{\alpha _{4}}{3}\sum _{n<n_{\mathrm{th}}}\left( 1+\frac{2M_{n}^{2}}{s}\right) \sqrt{1-4M_{n}^{2}/s}\, \, ,
\end{eqnarray}
where \( n<n_{\mathrm{th}} \) represents the sum over all the electron
KK modes that fulfill the relation $4 \left( n^{2}_{1} + n^{2}_{2} +
\cdots n^{2}_{\delta }\right) M^{2}_{c} < s$, and \( \alpha
_{4}=e^{2}_{4}/(4\pi ) \).  This sum can be evaluated approximately
for \( s\gg M^{2}_{c} \) by replacing it by an integral; then we
obtain

\begin{equation}
\label{QWER}
\Im m\Pi ^{(\delta )}(s)\approx \frac{\alpha _{4}}{2^{3+\delta }}\frac{(\delta +2)\, \pi ^{(\delta +1)/2}}{\Gamma \left( (\delta +5)/2\right) }\left( \frac{s}{M^{2}_{c}}\right) ^{\delta /2} \,.
\end{equation}

\begin{figure}
\begin{displaymath}
\begin{array}{ccc}
\includegraphics[scale=0.41]{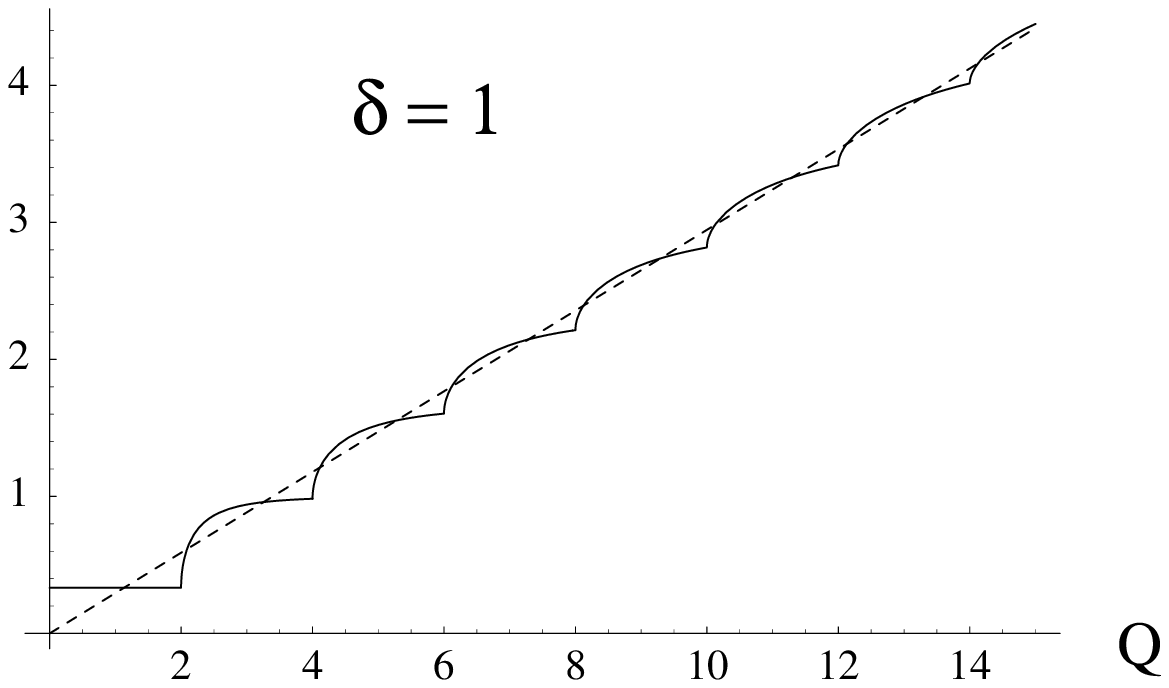} &
\includegraphics[scale=0.41]{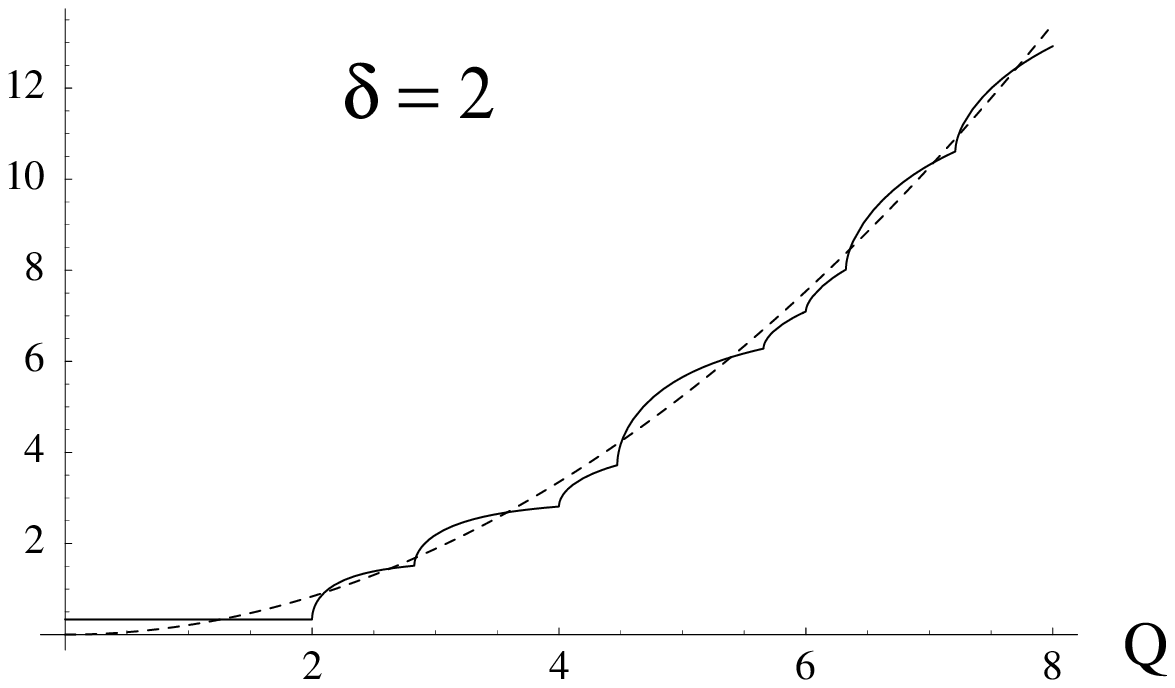} &
\includegraphics[scale=0.41]{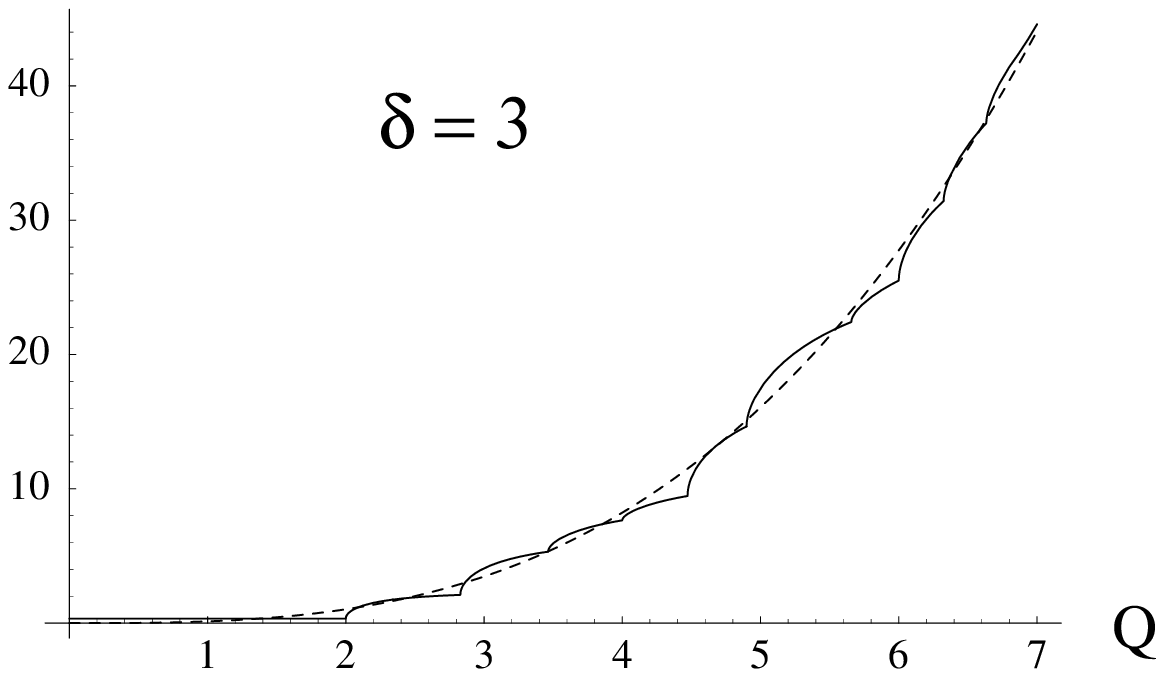} 
\end{array}
\end{displaymath}
%{\par\centering \includegraphics{running/imag1.eps} \par}
%
%{\par\centering \includegraphics{running/imag2.eps} \par}
%
%{\par\centering \includegraphics{running/imag3.eps} \par}
%
%
\caption{\label{fig:imaginary}\protect\( \Im m\Pi ^{(\delta )}(Q)\protect \) as compared
with the asymptotic value (\protect\( \delta =1,2,3\protect \)). $Q$ is
given in units of $M_c$.}
\end{figure}

It turns out that this last result 
captures the behavior of the 
same quantity when
the extra dimensions are not compact; this is so  
because, at high energies, the effects of the compactification
can be neglected. 
In fact, this result may be deduced on simple dimensional grounds: as
commented, the gauge coupling in \( 4+\delta  \) dimensions has dimension \( 1/M^{\delta /2} \);
therefore one expects that \( \Im m\Pi ^{(\delta )}(s) \) will grow with \( s \)
as \( \left( s/M^{2}\right) ^{\delta /2} \), which is what we obtained from
the explicit calculation. To see how rapidly one reaches this regime we can
plot the exact result of \( \Im m\Pi ^{(\delta )}(s) \) together with the asymptotic
value.
As we can 
see in Fig.\ref{fig:imaginary}, the asymptotic limit is reached very
fast, especially for higher dimensions. 
For practical purposes one can reliably use the asymptotic
value soon after passing the first threshold, \( Q>2M_{c} \), 
incurring errors which are below 10\%.

Now we can try to obtain the real part by using a dispersion relation
as the one used in 4-dimensional QED,
i.e. Eq.(\ref{PiRffQED}). However, one immediately sees that it will
need a number of subtractions which depends on the value of \(
\delta\).  Thus, for just one extra dimension, as in 4-dimensional
QED, one subtraction is enough, for \( \delta =2 \) and \( \delta =3
\) two subtractions are needed, see \Eq{QWER}, and so on. This just
manifests the non-renormalizability of the theory, and in the
effective field theory language, the need for higher dimension
operators acting as counterterms.  Even though this ``absorptive''
approach is perfectly acceptable, it would be preferable to have a way
of computing the real part directly at the Lagrangian level (by
computing loops, for instance).  As commented in the introduction, to
accomplish this we will use dimensional regularization.

\subsection{The vacuum polarization in uncompactified \protect\( 4+\delta \protect \)
dimensions}

When using dimensional regularization to compute the VPF in uncompactified
space, to be denoted $\Pi_{\mathrm{uc}}$, simple dimensional arguments suggest
that one should typically obtain contributions of the form
\[
\Pi _{\mathrm{uc}}(Q)\propto e^{2}_{4}\, \left( 2\pi \frac{Q}{M_{c}}\right)
^{\delta }\, ~,
\]
since the two vertices in the loop provide a factor \( e^{2}_{D} \), whose
dimensions must be compensated by the only 
available scale in the problem, 
namely
\( Q^{2} \). In the above formula we have 
used the relation of Eq.(\ref{relcoup})
in order to trade off \( e_{D} \)
for  \( e_{4} \). The omitted 
coefficient in front will be generally divergent, 
and will be regularized by letting
\( \delta \rightarrow \delta -\epsilon  \).

Let us compute the VPF \( \Pi ^{\alpha\beta}_{\mathrm{uc}}(q) \) 
in uncompactified space, assuming that, if necessary, 
the dimensions will be continued to complex values.
We have that 
\begin{equation}
\Pi ^{\alpha\beta}_{\mathrm{uc}}(q)=ie^{2}_{D}\int \frac{d^{4+\delta }k}{(2\pi
)^{4+\delta }}\mathrm{Tr}\left\{ \gamma ^{\alpha}\frac{1}{\sla k}\gamma
^{\beta}\frac{1}{\sla k+\sla q}\right\}~,
\end{equation}
which, by gauge-invariance assumes the standard form 
\[
\Pi ^{\alpha\beta}_{\mathrm{uc}}(q)=\left(
q^{2}g^{\alpha\beta}-q^{\alpha}q^{\beta}\right) \Pi
_{\mathrm{uc}}(q)~.
\]
If we now were to use that, in \( D \)-dimensions,  
Tr\( [\gamma ^{\alpha}\gamma ^{\beta}]=2^{[D/2]}g^{\alpha\beta} \), 
we would find that the low energy limit has an extra \( 2^{[\delta /2]} \)
factor, which, as commented, is an artifact of the torus compactification: there
are \( 2^{[\delta /2]} \) too many fermions in the theory. Therefore we simply
drop this factor by hand. Moreover, 
we use Eq.(\ref{relcoup})
and employ the proper-time parametrization in intermediate steps,
thus arriving at:
\begin{eqnarray}
\Pi _{\mathrm{uc}}(Q) & = & \frac{e^{2}_{4}}{2\pi ^{2}}\left( \frac{\pi }{M_{c}^{2}}\right) ^{\delta /2}\int _{0}^{1}dxx(1-x)\int ^{\infty }_{0}\frac{d\tau }{\tau ^{1+\frac{\delta }{2}}}\exp \left\{ -\tau \, x(1-x)Q^{2}\right\} \nonumber \\
 & = & \frac{e^{2}_{4}}{2\pi ^{2}}\frac{\pi ^{\delta /2}\, \Gamma
^{2}(2+\frac{\delta }{2})}{\Gamma (4+\delta )}\, \Gamma \left( -\frac{\delta }{2}\right) \, \left( \frac{Q^{2}}{M_{c}^{2}}\right) ^{\delta /2}~.
\label{EFC1}
\end{eqnarray}

A simple check of this result may be obtained by computing its imaginary part.
To that end we let \( Q^{2}\rightarrow -q^{2}-i\epsilon  \) with \( q^{2}>0 \).
Then \[
\Im m\left\{ -q^{2}-i\epsilon \right\} ^{\delta /2}=-\left( q^{2}\right) ^{\delta /2}\sin \frac{\delta \pi }{2}\, .\]
 Now we can use that \( \Gamma (-\delta /2)\Gamma (1+\delta /2)=-\pi /\sin (\delta \pi /2) \)
to write

\[
\Im m\left\{ \Pi _{\mathrm{uc}}(q)\right\} =\alpha _{4}\frac{2\pi
  ^{\delta /2}\, \Gamma ^{2}(2+\frac{\delta }{2})}{\Gamma (4+\delta
  )\Gamma (1+\delta /2)}\left( \frac{q^{2}}{M_{c}^{2}}\right) ^{\delta
  /2}=\frac{\alpha _{4}}{2^{3+\delta }}\frac{(\delta +2)\, \pi
  ^{(\delta +1)/2}}{\Gamma \left( (\delta +5)/2\right) }\left(
\frac{q^{2}}{M^{2}_{c}}\right) ^{\delta /2}\, , \]
which agrees with our previous result of Eq.(\ref{QWER}).

For odd values of \( \delta \), the one-loop \( \Pi _{\mathrm{uc}}(Q)
\) computed above is finite, since the \( \Gamma (-\frac{\delta }{2})
\) can be calculated by analytic continuation. This result is in a way
expected, since in odd number of dimensions, by Lorentz invariance,
there are no appropriate gauge invariant operators able to absorb any
possible infinities generated in the one-loop VPF; this would require
operators which give contributions that go like \( Q^{\delta } \).
Notice, however, that at higher orders \( \Pi _{\mathrm{uc}}(Q) \)
will eventually become divergent. For instance, in five dimensions at
two loops, the VPF should go as \( Q^{2} \), since there are four
elementary vertices. The divergences generated by these contributions
could be absorbed in an operator such as the one considered in the
previous section, namely \(
D_{\alpha}F^{\alpha\beta}D^{\lambda}F_{\lambda\beta} \).  On the other
hand, when \( \delta \) is even, \( \Gamma (-\frac{\delta }{2}) \) has
a pole, and subtractions are needed already at one loop. To compute
the divergent and finite parts in a well-defined way we will use
dimensional regularization, i.e. we will assume that \( \delta
\rightarrow \delta -\epsilon \) .  Notice however that, unlike in
4-dimensions, we do not need to introduce an additional scale at this
point, i.e. the equivalent of the 't Hooft mass scale $\mu$: \( M_{c}
\) plays the role of \( \mu \), and can be used to keep \( e_{4} \)
dimensionless.  After expanding in \( \epsilon \) we find a simple
pole accompanied by the usual logarithm
\begin{equation}
\label{eq:uc-divergent}
\Pi _{\mathrm{uc}}(Q)\propto \left( \frac{Q^{2}}{M_{c}^{2}}\right) ^{\delta /2}\left\{ -\frac{2}{\epsilon }+\ln (Q^{2}/M_{c}^{2})+\cdots \right\} \, .
\end{equation}
 Here the ellipses represent a finite constant. Now, to renormalize
this result we must introduce higher dimension operators (for
instance, if \( \delta =2 \) the operator \(
D_{\alpha}F^{\alpha\beta}D^{\lambda}F_{\lambda\beta} \) will do the
job) which could absorb the divergent piece. The downside of this,
however, is that we also have to introduce an arbitrary counterterm,
\( \kappa \), corresponding to the contribution of the higher
dimension operator; thus we obtain a finite quantity proportional to
\( \log (Q^{2}/M^{2}_{c})+\kappa \). Note that, since \( \kappa \) is
arbitrary, we can always introduce back a renormalization scale and
write \( \log (Q^{2}/M^{2}_{c})+\kappa =\log (Q^{2}/\mu ^{2})+\kappa
(\mu ) \) with \( \kappa (\mu )=\kappa +\log (\mu ^{2}/M^{2}_{c})
\). It is also important to remark that, in the case of odd number of
dimensions, although at one loop we do not need any counterterm to
make the VPF finite, higher dimensional operators could still be
present and affect its value.

In the case of uncompactified space, 
it is interesting to compare the above result with that 
obtained  
by regularizing the integral using a hard cutoff. To study this it
is enough to carry out 
the integral of Eq.~(\ref{EFC1}), with a cutoff in 
\( \tau_{0}=1/\Lambda ^{2} \): 
\begin{equation}
\label{eq:PiCutoff}
\Pi_{\mathrm{uc}}(Q)=\frac{e_{4}^{2}}{2\pi ^{2}}\left( \frac{\pi }{M_{c}^{2}}\right) ^{\delta /2}\int _{0}^{1}dx\, x(1-x)\int _{\tau _{0}}^{\infty }\frac{d\tau }{\tau ^{1+\delta /2}}\exp \left\{ -\tau \, x(1-x)Q^{2}\right\} .
\end{equation}
 Then, for \( \delta =1,2,3 \) we obtain
\begin{equation}
\Pi ^{(1)}_{\mathrm{uc}}(Q)=\frac{e_{4}^{2}}{2\pi ^{2}}\left( -\frac{3\pi
^{2}Q}{64M_{c}}+\frac{\sqrt{\pi }Q^{2}}{15M_{c}\Lambda }+\frac{\sqrt{\pi
}\Lambda }{3M_{c}}\right) \, ~,
\label{d1}
\end{equation}
\begin{equation}
\Pi ^{(2)}_{\mathrm{uc}}(Q)=\frac{e_{4}^{2}}{2\pi ^{2}}\left( \frac{\pi
\Lambda ^{2}}{6M_{c}^{2}}+\frac{\pi Q^{2}}{30M_{c}^{2}}\left( \log
(Q^{2}/\Lambda ^{2})+\gamma -\frac{77}{30}\right) \right) \, ~,
\label{d2}
\end{equation}
\begin{equation}
\Pi ^{(3)}_{\mathrm{uc}}(Q)=\frac{e_{4}^{2}}{2\pi ^{2}}\left( \frac{5\pi
^{3}Q^{3}}{768M_{c}^{3}}-\frac{\pi ^{3/2}Q^{2}\Lambda
}{15M_{c}^{3}}+\frac{\pi ^{3/2}\Lambda ^{3}}{9M_{c}^{3}}\right) \, ~.
\label{d3}
\end{equation}

As we can see, the pieces which are independent of the cutoff are
exactly the same ones we obtained using dimensional
regularization. But, in addition, we obtain a series of contributions
which depend explicitly on the cutoff.  For instance we find
corrections to the gauge coupling which behave as \( \Lambda ^{\delta
} \), and just redefine the gauge coupling we started with
\cite{Taylor:1988vt}.  In the case of five dimensions we also generate
a term linear in \( Q^{2} \); however it is suppressed by 1/\( \Lambda
\), and therefore it approaches zero for large \( \Lambda \). In the
case of six dimensions we obtain the same logarithmic behavior we
found with dimensional regularization, and the result can be cast in
identical form, if the cutoff is absorbed in the appropriate
counterterm.  For seven dimensions we also find divergent
contributions which go as \( Q^{2} \).  This means that, when using
cutoffs, higher dimension operators in the derivative expansion
(e.g. operators giving contributions as \( Q^{2} \) or higher) are
necessary to renormalize the theory and must be included. In the case
of dimensional regularization this type of operators is not strictly
needed at one loop; however, nothing forbids them in the Lagrangian,
and they could appear as ``finite counterterms''. If one were to
identify the \( \Lambda \) in the above expressions with a physical
cutoff, one might get the impression that, contrary to the dimensional
regularization approach where arbitrary counterterms are needed, one
could now obtain all types of contributions with only one additional
parameter, namely \( \Lambda \).  This is however not true: the
regulator function is arbitrary, we simply have chosen one among an
infinity of possibilities.  By changing the regulator function we can
change the coefficients of the different contributions at will, except
for those few contributions which are independent of \( \Lambda
\). These latter are precisely the ones we have obtained by using
dimensional regularization. Thus, even when using cutoffs one has to
add counterterms from higher dimension operators, absorb the cutoff,
and express the result in terms of a series of unknown
coefficients. The lesson is that with dimensional regularization we
obtain all calculable pieces, while the non-calculable pieces are
related to higher dimensional terms in the Lagrangian.

What we will demonstrate next is that the one-loop VPF in the compactified theory
on a torus can be renormalized exactly as the VPF in the uncompactified theory;
this will allow us to compute it for any number of dimensions, and examine its
behavior for large and for small values of the \( Q^{2} \).

\subsection{The vacuum polarization in \protect\( \delta \protect \) compact dimensions}

From the four-dimensional point of view the vacuum polarization tensor in the
compactified theory is
\[
\Pi^{\mu \nu }(q^2)=\sum _{n}ie_{4}^{2}\int \frac{d^{4}k}{(2\pi )^{4}}\mathrm{Tr}\left\{ \gamma ^{\mu }\frac{1}{\sla k-m_{n}}\gamma ^{\nu }\frac{1}{\sla k+\sla q-m_{n}}\right\} 
\]
with 
\( m^{2}_{n}=\left( n^{2}_{1}+n^{2}_{2}+\cdots +n^{2}_{\delta }\right) M^{2}_{c} \);
for simplicity we have assumed a common compactification radius \( R=1/M_{c} \)
for all the extra dimensions. The sum over \( n \) denotes collectively the
sum over all the modes \( n_{i}=-\infty ,\cdots ,+\infty  \). 
The contribution of each mode
to this quantity seems quadratically divergent, like in ordinary QED; however,
we know that gauge invariance converts it to only logarithmically divergent.
But, in addition, the sum over all the modes makes the above expressions highly
divergent. Instead of attempting to compute it directly,
we will add and subtract
the contribution of the vacuum polarization function of the uncompactified theory
in \( 4+\delta  \) dimensions:
\begin{equation}
\label{eq:VPFSplitting}
\Pi ^{\mu \nu }(q)=
\left[\Pi ^{\mu \nu }(q)-\Pi _{\mathrm{uc}}^{\mu\nu }(q)\right]+
\Pi_{\mathrm{uc}}^{\mu\nu }(q)=
\Pi _{\mathrm{fin}}^{\mu \nu }(q)+\Pi _{\mathrm{uc}}^{\mu\nu }(q)~.
\end{equation}
Here we have taken already into account the relation between the coupling in
\( 4+\delta  \) dimensions and the four-dimensional coupling and have restricted
the external Lorentz indices to the 4-dimensional ones. Depending on the value
of \( \delta  \) the vacuum polarization can be highly divergent 
(naively as \( \Lambda ^{\delta +2} \), 
and after taking into account gauge invariance
as \( \Lambda ^{\delta } \)). However, we can use dimensional regularization
(or any other regularization scheme) to make it finite. The important point
is that the quantity \( \Pi _{\mathrm{fin}}^{\mu \nu }(Q) \) is UV and IR finite
and can unambiguously computed.

Instead of doing the two calculations from scratch, we will do the following:

i) We will first compute the compactified expression
by using Schwinger's proper
time,~$\tau$, to regularize the UV divergences.

ii) We will show that the UV behavior of the compactified theory, \( \tau \rightarrow 0, \)
is just the behavior of the uncompactified theory.

iii) Therefore, to compute \( \Pi _{\mathrm{fin}}^{\mu \nu }(Q) \) it is sufficient
to compute \( \Pi ^{\mu \nu }(Q) \) and then subtract its most divergent contribution
when \( \tau \rightarrow 0 \). 
We will see that it is sufficient to make it finite.

After a few manipulations \( \Pi ^{\mu \nu }(Q) \) can be written as 
\[
\Pi ^{\mu \nu }(q)=\left( q^{2}g^{\mu \nu }-q^{\mu }q^{\nu }\right) \Pi (q) ~,
\]
 where \cite{Dienes:1998vg}
\[
\Pi (Q)=\frac{e_{4}^{2}}{2\pi ^{2}}\sum _{n}\int ^{1}_{0}dxx(1-x)\int
^{\infty }_{0}\frac{d\tau }{\tau }\exp \left\{ -\tau \left( x(1-x)Q^{2}+m^{2}_{n}\right) \right\} ~.
\]
 \( \Pi (Q) \) can be written in terms of the function
\[
\bar{\theta }_{3}(\tau )\equiv \sum ^{+\infty }_{n=-\infty }e^{-n^{2}\tau }=\sqrt{\frac{\pi }{\tau }}\bar{\theta }_{3}\left( \frac{\pi ^{2}}{\tau }\right) 
\]
 as
\[
\Pi (Q)=\frac{e_{4}^{2}}{2\pi ^{2}}\int ^{1}_{0}dxx(1-x)\int ^{\infty
}_{0}\frac{d\tau }{\tau }\exp \left\{ -\tau
x(1-x)\frac{Q^{2}}{M^{2}_{c}}\right\} \bar{\theta }^{\delta }_{3}(\tau )~,
\]
where we have rescaled \( \tau  \) in order to remove \( M_{c} \) from the
\( \bar{\theta }_{3}(\tau ) \) function. This last expression for \( \Pi (Q) \)
is highly divergent in the UV (\( \tau \rightarrow 0) \), because in that limit
the \( \bar{\theta }_{3}(\tau ) \) function goes as \( \sqrt{\pi /\tau } \).
Then, if we define, as in Eq.~(\ref{eq:VPFSplitting}),
$\Pi_{\mathrm fin}=\Pi-\Pi_{\mathrm uc}$, we have
\[
\Pi _{\mathrm{fin}}(Q)=\frac{e_{4}^{2}}{2\pi ^{2}}\int ^{1}_{0}dxx(1-x)\int
^{\infty }_{0}\frac{d\tau }{\tau }\exp \left\{ -\tau
x(1-x)\frac{Q^{2}}{M^{2}_{c}}\right\} \left( \bar{\theta }^{\delta }_{3}(\tau )-\left( \frac{\pi }{\tau }\right) ^{\delta /2}\right)~,
\]
which is completely finite for any number of dimensions. In fact,
the last term provides a factor 
\[
F_{\delta }(\tau )\equiv \bar{\theta }^{\delta }_{3}(\tau )-\left( \frac{\pi
}{\tau }\right) ^{\delta /2}\stackrel{\tau \rightarrow 0}{\longrightarrow
}2\delta \left( \frac{\pi }{\tau }\right) ^{\delta /2}\exp \left\{ -\frac{\pi ^{2}}{\tau }\right\}~,
\]
that makes the integral convergent in the UV, while for large \( \tau  \) 
this function
goes to 1 quite fast. In this region the integral is cut off by the exponential
of momenta; so we can think of the exponential \( \exp \left\{ -\tau x(1-x)\frac{Q^{2}}{M^{2}_{c}}\right\}  \)
as providing a cutoff for \( \tau >4M^{2}_{c}/Q^{2} \), 
and \( F_{\delta }(\tau ) \)
as providing a cutoff for \( \tau <\pi ^{2} \). 
With this in mind, we can estimate \( \Pi _{\mathrm{fin}}(Q) \) as
\begin{equation}
\Pi _{\mathrm{fin}}(Q)\approx \frac{e_{4}^{2}}{2\pi ^{2}}\sum _{n}\int
^{1}_{0}dxx(1-x)\int ^{4M^{2}_{c}/Q^{2}}_{\pi ^{2}}\frac{d\tau }{\tau
}=-\frac{e^{2}}{2\pi ^{2}}\frac{1}{6}\log \frac{Q^{2}\pi ^{2}}{4M^{2}_{c}} ,\qquad Q^{2}<4M^{2}_{c}/\pi ^{2}
;\end{equation}
it is just the ordinary running of the zero mode. As \( Q^{2} \) grows,
the upper limit of integration is smaller than the lower limit,
and then we expect
that \( \Pi _{\mathrm{fin}}(Q) \) should vanish. 
\begin{figure}
\vspace{0.3cm}
{\par\centering \includegraphics{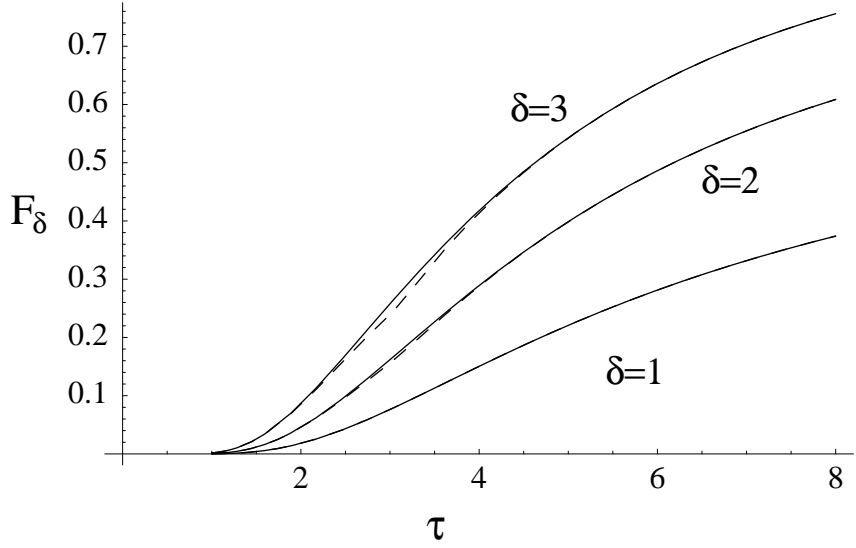} \par}
\vspace{0.3cm}
\caption{Exact values of \protect\( F_{\delta }(\tau )\protect \) (solid) as compared
with the approximation discussed in the text (dashed) for \protect\( \delta =1,2,3\protect \).\label{fig:approximation}}
\end{figure}
In that region \( \Pi (Q) \)
will be dominated completely by \( \Pi _{\mathrm{uc}}(Q) \).

Let us evaluate \( \Pi _{\mathrm{fin}}(Q) \) for any number 
of extra dimensions.
To this end we will approximate the function \( F_{\delta }(\tau ) \) as follows
\begin{equation}
\label{eq:approximation}
F_{\delta }(\tau )=\left\{ \begin{array}{ll}
2\delta \left( \frac{\pi }{\tau }\right) ^{\delta /2}\exp \left\{ -\frac{\pi ^{2}}{\tau }\right\} \quad \tau <\pi  & \\
1+2\delta \exp \left\{ -\tau \right\} -\left( \frac{\pi }{\tau }\right) ^{\delta /2}\quad \tau >\pi 
\end{array}\right. ~. 
\end{equation}
 The matching point in \( \tau =\pi  \) makes the function continuous. 
In Fig.~\ref{fig:approximation}
we display the exact function \( F_{\delta }(\tau ) \) (solid) and the approximation
above (dashed) for \( \delta =1,2,3 \).
The approximation is very good except at a small region around the matching
point \( \tau =\pi  \). This can be further improved by adding more terms from
the expansions of the \( \bar{\theta }(\tau ) \) functions.

The approximate expression of Eq.(\ref{eq:approximation}) can be used to 
obtain semi-analytical expansions for \( \Pi ^{\delta }_{\mathrm{fin}}(Q) \)
for small \( Q^{2} \) (we define \( w\equiv Q^{2}/M^{2}_{c} \)) 
\begin{eqnarray}
\Pi ^{(1)}_{\mathrm{fin}}(Q) & = & \frac{e_{4}^{2}}{2\pi ^{2}}
\left( -0.335-0.167\log (w)+0.463\sqrt{w}-0.110w+\cdots \right)
\label{eq:approx}~,\\
\Pi ^{(2)}_{\mathrm{fin}}(Q) & = & \frac{e_{4}^{2}}{2\pi ^{2}}
\left( -0.159-0.167\log (w)-0.105w\left( \log (w)-1.75\right) +\cdots
\right) ~, \\
\Pi ^{(3)}_{\mathrm{fin}}(Q) & = & \frac{e_{4}^{2}}{2\pi ^{2}}
\left( -0.0937-0.167\log (w)+0.298w-0.202\sqrt{w^{3}}+\cdots \right)~.
\end{eqnarray}

To see how good these approximate results are, we can compare with the exact
results that can be obtained when \( \delta =1 \). In this case we
have
\begin{eqnarray}
\Pi ^{(1)}_{\mathrm{fin}}(Q) &= &\frac{e_{4}^{2}}{2\pi ^{2}}\int
^{1}_{0}dxx(1-x)\sum ^{\infty }_{n=1}\int ^{\infty }_{0}\frac{d\tau
}{\tau }\exp \left\{ -\tau x(1-x)w\right\} 2\left( \frac{\pi }{\tau
}\right) ^{1/2}\exp \left\{ -\frac{n^{2}\pi ^{2}}{\tau }\right\} \nonumber \\
 & = & \frac{e_{4}^{2}}{2\pi ^{2}}\int ^{1}_{0}dxx(1-x)\sum ^{\infty
 }_{n=1}\frac{2}{n}\exp \left\{ -2\pi n\sqrt{x(1-x)w}\right\} \nonumber \\
%\begin{equation}
\label{PI1full}
 & = &\frac{e_{4}^{2}}{2\pi ^{2}}\int ^{1}_{0}dxx(1-x)\left( -2\log \left( 1-\exp
\left( -2\pi \sqrt{x(1-x)w}\right) \right) \right)~.
\end{eqnarray}
This last expression 
can be expanded for \( w\ll 1 \), and the integral over \( x \) can then
be performed analytically, yielding 
\begin{eqnarray}
\Pi ^{(1)}_{\mathrm{fin}}(Q) &\approx& 
\frac{e_{4}^{2}}{2\pi ^{2}}\left( \frac{1}{18}
\left( 5-6\log (2\pi )\right) -\frac{1}{6}\log w+
\frac{3\pi ^{2}}{64}\sqrt{w}-\frac{\pi ^{2}}{90}w+\cdots \right)~, 
\end{eqnarray}
which is in excellent agreement with our approximation (see
Eq.~(\ref{eq:approx})).

Eq.~(\ref{PI1full}) can also be used to obtain the behavior of \( \Pi
^{(1)}_{\mathrm{fin}}(Q) \) for \( w\gg 1. \) In this limit we obtain
\[
\Pi ^{(1)}_{\mathrm{fin}}(Q)\approx \frac{e_{4}^{2}}{2\pi ^{2}}\frac{3\zeta
(5)}{\pi ^{4}}\frac{M_{c}^{4}}{Q^{4}}\, ,\qquad Q^{2}\gg M^{2}_{c}~.
\]
 For higher dimensions things are more complicated, but the behavior is the
same, and we find 
\[
\Pi ^{(\delta )}_{\mathrm{fin}}(Q)\approx \frac{e_{4}^{2}}{2\pi
^{2}}\frac{4\delta \Gamma (2+\delta /2)}{\pi ^{4+\delta /2}}K_{\delta
}\frac{M^{4}_{c}}{Q^{4}}\, ,\qquad Q^{2}\gg M^{2}_{c}~,
\]
 where \( K_{\delta } \) is of the order of unity and is determined numerically
(\( K_{1}=\zeta (5)=1.037, \) \( K_{2}=1.165, \) \( K_{3}=1.244 \)). However,
since the uncompactified contribution grows as \( \left( Q^{2}/M^{2}_{c}\right) ^{\delta /2} \)
it is obvious that the contributions to \( \Pi ^{(\delta )}(Q) \) from \( \Pi ^{(\delta )}_{\mathrm{fin}}(Q) \)
will be completely irrelevant for \( Q^{2}\gg M^{2}_{c}. \)

Adding the finite and the uncompactified contributions we find that for \( Q^{2}\ll M^{2}_{c} \)
the uncompactified contribution exactly cancels the corresponding piece obtained
from the expansion of \( \Pi ^{(\delta )}_{\mathrm{fin}}(Q) \) (the \( \sqrt{w} \)
piece for \( \delta =1 \), the \( w\log (w) \) piece for \( \delta =2 \),
or the \( \sqrt{w^{3}} \) for \( \delta =3 \)). Then, for \( Q^{2}\ll M^{2}_{c} \)
and choosing \( \mu =M_{c} \) we finally find:

\begin{equation}
\label{PIfinal}
\Pi ^{(\delta )}(Q)=\frac{e_{4}^{2}}{2\pi ^{2}}\left( a^{(\delta )}_{0}-\frac{1}{6}\log \left( \frac{Q^{2}}{M^{2}_{c}}\right) +a^{(\delta )}_{1}\frac{Q^{2}}{M^{2}_{c}}+\cdots \right) \, ,\qquad Q^{2}\ll M^{2}_{c}
\end{equation}
 with the coefficients for 1,2 and 3 extra dimensions given by

\medskip{}
{\centering \begin{tabular}{|c|c|c|c|}
\hline 
\( \delta  \)&
 \( 1 \)&
 \( 2 \)&
 \( 3 \)\\
\hline 
\( a^{(\delta )}_{0} \)&
 \( -0.335 \)&
 \( -0.159 \)&
 \( -0.0937 \)\\
\hline 
\( a^{(\delta )}_{1} \)&
 \( -0.110 \)&
 \( 0.183 \)&
 \( 0.298 \) \\
\hline 
\end{tabular}\par}
\medskip{}

As we will see below, in general the coefficients \( a^{(\delta )}_{1} \) can
be affected by non-calculable contributions from higher dimension operators
in the effective Lagrangian, which we have not included.
 
The following comments related to Eq.~(\ref{PIfinal}), which is only valid
in the dimensional regularization scheme we are using,  are now in order:

({\bf i}) 
From Eq.~(\ref{PIfinal}) we see that for small \( Q^{2} \), as expected,
we recover the standard logarithm with the correct coefficient, independently
of the number of extra dimensions. 
In addition, interestingly enough, we can compute
also the constant term. 
Thus, although the full theory in \( 4+\delta  \) dimensions
is non-renormalizable and highly divergent, the low energy limit 
of the VPF calculated in our dimensional regularization scheme
is actually finite: when seen from low energies the compactified extra 
dimensions seem to act as an ultraviolet regulator for the theory. 

({\bf ii})
When the energy begins to grow, we start
seeing effects suppressed by \( Q^{2}/M^{2}_{c} \),
which are finite, at one
loop, for any number of dimensions except for $\delta=2$. 
This is so because
the gauge couplings have dimensions \( 1/M^{\delta /2} \), and therefore, the
one-loop VPF goes like \( 1/M^{\delta } \). 

({\bf iii})
For \( \delta =1 \) one finds that,
because of gauge and Lorentz invariance, 
there are no possible counterterms
of this dimension. The VPF must be finite, and that is precisely the result
one obtains with dimensional regularization. This of course changes if higher
loops are considered: for instance, two-loop diagrams go like \( 1/M^{2} \),
and, in general, we expect that they will have divergences, which, in turn, 
should be absorbed in the appropriate counterterms. 
In principle the presence of these
counterterms could pollute our result; 
however, the natural size of these counterterms,
arising at two loops, should be suppressed compared to 
the finite contributions
we have computed. 

({\bf iv})
For \( \delta =2 \) one finds that the
VPF goes as \( 1/M^{2} \), already at one loop, and that the result is 
divergent. The divergences
have to be absorbed in the appropriate counterterm coming from higher dimension
operators in the higher dimensional theory. 
The immediate effect of this, is that 
the coefficient of the \( Q^{2} \) term in \( \Pi ^{(2)}(Q) \) 
becomes arbitrary, its value depending 
 on the underlying physics beyond
the compactification scale.

({\bf v}) For \( \delta >2 \) all loop contributions to the
\( Q^{2} \) term are finite, simply because of the dimensionality of the couplings.
This, however, does not preclude the 
existence of finite counterterms, which could
be generated by physics beyond the compactification scale, 
that is, contributions
from operators suppressed by two powers of the new physics scale like 
the operator in Eq.~(\ref{eq:lct}).

For \( Q^{2}\gg M^{2}_{c} \) the full VPF is completely 
dominated by the uncompactified contribution:

\begin{eqnarray*}
\Pi ^{(1)}(Q) & = & -\frac{e_{4}^{2}}{2\pi ^{2}}\frac{3\pi
^{2}}{64}\sqrt{\frac{Q^{2}}{M^{2}_{c}}}~,\\
\Pi ^{(2)}(Q) & = & \frac{e_{4}^{2}}{2\pi ^{2}}\frac{\pi
}{30}\frac{Q^{2}}{M^{2}_{c}}\log \left( \frac{Q^{2}}{M^{2}_{c}}\right) \,
\qquad Q^{2}\gg M^{2}_{c}~,\\
\Pi ^{(3)}(Q) & = & \frac{e_{4}^{2}}{2\pi ^{2}}\frac{5\pi ^{3}}{768}\left(
\frac{Q^{2}}{M^{2}_{c}}\right) ^{\frac{3}{2}}~.
\end{eqnarray*}

As before, the VPF could also receive
non-calculable contributions from higher dimension
operators which we have not included; in fact, for \( \delta =2 \), these
are needed to renormalize the VPF. How large can these non-calculable
contributions be?
Since our D-dimensional theory is an effective theory valid only for
$Q^2 \ll M_s^2$, even above \( M_{c} \) the results will be dominated by the
lowest power of $Q^2$. 
In the case of \( \delta =1 \), the first operator that one can write down
goes as \( Q^{2} \); therefore we expect that the one-loop contribution,
of order \( \sqrt{Q^{2}} \),
that we have computed, will dominate completely the result, as long as we do not 
stretch it beyond the applicability of the effective Lagrangian approach.
For \( \delta =2 \),
counterterms are certainly needed at order \( Q^{2} \); still one can hope that
the result will be dominated by the logarithm (as happens with chiral 
logarithms in \( \chi PT \)). 
For \( \delta =3 \) (and higher), the one loop result grows as 
\( \left(Q^{2}\right) ^{\delta/2} \);
however there could be operators giving contributions of  order \( Q^{2} \) with
unknown coefficients (in fact although in dimensional regularization those are
not needed, they must be included if cutoffs are used to regularize the theory).
Therefore,
unless for some reason they are absent from the theory, the result will be 
dominated
by those operators.

\section{Matching of gauge couplings\label{sec:matching}}

Using the VPF constructed in the previous section we can define 
a higher dimensional analogue of the conventional QED effective charge 
\cite{Itzykson:1980rh,Peskin:1995ev},
which will enter in any process involving off-shell photons, e.g.  
\begin{equation}
\frac{1}{\alpha _{\mathrm{eff}}(Q)}\equiv \frac{1}{\alpha _{4}}\left. \left( 1+\Pi ^{(\delta )}(Q)\right) \right| _{\msb _{\delta }}\, ,
\label{eq:DefinitionEffectiveCharge}
\end{equation}
where \( \alpha _{4}=e^{2}_{4}/(4\pi ) \). 
We remind the reader that \( e_{4} \)
denotes the (dimensionless) coupling of the four-dimensional theory including
all KK modes; it is directly related to the gauge coupling in the theory with
\( \delta  \) extra dimensions by Eq.(\ref{relcoup}). 
The subscript \( \msb _{\delta } \) means that the VPF has been regularized
using dimensional regularization in \( D=4+\delta -\epsilon  \) dimensions,
and that divergences, when present, are subtracted according to 
the \( \msb  \) procedure. 

To determine the relation between \( \alpha_{4} \) and the low energy coupling in
QED, we have to identify the effective charge computed in the compactified theory
with the low energy effective charge, at some low energy scale (for instance
\( Q^{2}=m^{2}_{Z}\ll M^{2}_{c} \)), where both theories are valid.
In that limit we can trust our approximate results of Eq.~(\ref{PIfinal}),
and write
\begin{equation}
\label{eq:matching}
\frac{1}{\alpha _{\mathrm{eff}}(m_{Z})}=\frac{1}{\alpha _{4}}+\frac{2}{\pi
}a^{(\delta )}_{0}-\frac{2}{3\pi }\log \left( \frac{m_{Z}}{M_{c}}\right)~.
\end{equation}
This equation connects the low energy QED coupling with the coupling in the
compactified D-dimensional theory, regularized by dimensional regularization. 
Note that this equation is completely independent of the way subtractions 
are performed
to remove the poles in \( 1/\epsilon  \). These poles only appear 
(and only for even number of dimensions) in the contributions proportional 
to \( Q^{\delta } \),
which vanish for \( Q\rightarrow 0 \). Eq.~(\ref{eq:matching}) contains,
apart from a finite constant, the standard logarithmic running from \( m_{Z} \)
to the compactification scale \( M_{c} \). 
It is interesting to notice that, 
in this approach, the logarithm
comes from the finite piece, and should therefore be considered as an infrared
(IR) logarithm. When seen from scales smaller than \( M_{c} \), these logarithms
appear to have an UV origin, while, when seen from scales above \( M_{c} \),
appear as having an IR nature.

It is important to emphasize that, in this scheme, 
the gauge coupling does not run any more above the compactification
scale. This seems counter-intuitive,
but it is precisely what happens in \( \chi PT \) when using dimensional regularization:
\( f_{\pi } \) does not run, it just renormalizes higher dimensional
operators~\cite{Weinberg:1979kz}. 

Now we can use Eq.~(\ref{eq:matching}) to write the effective charge at all
energies in terms of the coupling measured at low energies:
\begin{equation}
\label{eq:fullEffectiveCharge}
\frac{1}{\alpha _{\mathrm{eff}}(Q)}\equiv \frac{1}{\alpha
_{\mathrm{eff}}(m_{Z})}+\left. \frac{1}{\alpha _{4}}\left( \Pi ^{(\delta )}(Q)-\Pi ^{(\delta )}(m_{Z})\right) \right| _{\msb _{\delta }}~.
\end{equation}
 Note that the last term is independent of \( \alpha _{4} \) due to the implicit
dependence of \( \Pi ^{(\delta )} \) on it. Eq.~(\ref{eq:fullEffectiveCharge})
has the form of a momentum-subtracted definition of the coupling; in fact, in
four dimensions it is just the definition of the momentum-subtracted running
coupling. For \( \delta =1 \) and at one loop, 
\( \Pi ^{(\delta )}(Q)-\Pi ^{(\delta )}(m_{Z}) \)
is finite, and \( \alpha _{\mathrm{eff}}(Q) \) 
can still be interpreted as a momentum-subtracted 
definition of the coupling. 
For \( \delta >1 \), however, Eq.~(\ref{eq:fullEffectiveCharge})
involves additional subtractions, a fact which thwarts such an interpretation.

\begin{figure}
\begin{center}
\includegraphics[scale=0.8]{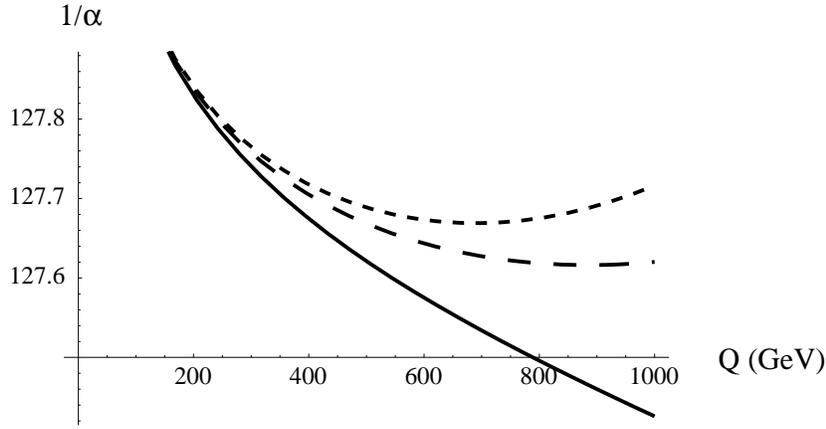}
%{\par\centering \includegraphics{running/new-running.eps} \par}
\caption{The ``effective charge'' against the energy scale for
\protect\( \delta =1\protect \) (solid), \protect\( \delta =2\protect
\) (short dash), \protect\( \delta =3\protect \) (long
dash). Contributions from counterterms have not been
considered.\label{fig:effective-running}}
\end{center}
\end{figure}

For \( Q^{2}\ll M^{2}_{c} \) we can expand \( \Pi ^{(\delta )}(Q) \) and obtain

\[
\frac{1}{\alpha _{\mathrm{eff}}(Q)}\equiv \frac{1}{\alpha
_{\mathrm{eff}}(m_{Z})}-\frac{2}{3\pi }\log \left( \frac{Q}{m_{Z}}\right)
+\mathcal{O}\left( \frac{Q^{2}}{M^{2}_{c}}\right)~, 
\]
which is nothing but the standard expression of the effective charge in QED,
slightly modified by small corrections of order \( Q^{2}/M^{2}_{c} \). However,
as soon as \( Q^{2}/M^{2}_{c} \) approaches unity, the effects of the 
compactification scale start to appear in \( \alpha _{\mathrm{eff}}(Q) \), 
forcing it to deviate dramatically from the logarithmic behavior, 
as shown in Fig.~\ref{fig:effective-running}.

The crucial point, however, is that this effective charge
\emph{cannot} be interpreted anymore as the running coupling (as can
be done in four dimensions) since it may receive contributions from
higher dimension operators; in fact some of them are needed to define
this quantity properly. These contributions have nothing to do with
the gauge coupling which is defined as the coefficient of the operator
\( F^{2}\). In particular, one should not use this quantity to study
gauge coupling unification.  Instead, one could use
Eq.~(\ref{eq:matching}), which relates the coupling measured at low
energies with the one appearing in the \( D \)-dimensional Lagrangian
valid at energies \( M_{c}<Q<M_{s} \). This relation involves a
logarithmic correction, which is the only contribution that can be
reliably computed without knowing the physics beyond \( M_{s} \).

It is instructive to see what happens if instead of dimensional
regularization we use hard cutoffs to regularize the uncompactified part of 
the VPF as in
Eqs.~(\ref{eq:PiCutoff})--(\ref{d3}). Then, when using cutoffs,
one can define an ``effective
charge'' as in Eq.~(\ref{eq:DefinitionEffectiveCharge})
\begin{equation}
\frac{1}{\alpha _{\mathrm{eff}}(Q)}\equiv \frac{1}{\alpha _{4}(\Lambda)}\left. \left( 1+\Pi ^{(\delta )}(Q)\right) 
\right| _{\Lambda}\, ,
\label{eq:DefinitionEffectiveChargeLambda}
\end{equation}
where $\alpha_4(\Lambda)$ now is the coupling constant in the theory
regularized with cutoffs and the subscript $\Lambda$ indicates that the VPF
has been regularized with cutoffs. The use of a $\Lambda$ dependent coupling
obviously implies the WEFT
formulation, in which the cutoff is not removed from the theory. 
On the other hand, in the CEFT
formulation one should renormalize the coupling constant by adding the
appropriate counterterms and then take the limit $\Lambda\rightarrow
\infty$. This usually brings in a new scale at which the coupling is
defined, and which effectively replaces $\Lambda$ in the previous equation.
Notice also that for $\delta=2$ in Eq.~(\ref{d2}) there are logarithmic
contributions proportional to $Q^2$, which cannot be removed when
$\Lambda\rightarrow \infty$. 
The same is true for $\delta>2$, but with
dependencies which are proportional to $\Lambda^{(\delta-2)}$. This just
manifests the need of higher dimensional operators, as was already clear in
the dispersive approach, to define properly the effective charge. 
As one can see, the full VPF
contains a term
that goes as \( \Lambda ^{\delta } \) and is independent of \( Q \). This
piece survives when \( Q\rightarrow 0 \), and thus  we obtain (we assume
$m_Z^2\ll Q^2$) 
\begin{equation}
\frac{1}{\alpha _{\mathrm{eff}}(m_{Z})}=
\frac{1}{\alpha _{4}(\Lambda )}+\frac{2}{3\pi \delta }
\left( \sqrt{\pi }\frac{\Lambda }{M_{c}}\right) ^{\delta }
+\frac{2}{\pi }a^{(\delta )}_{0}-\frac{2}{3\pi }
\log \left( \frac{m_{Z}}{M_{c}}\right) ~.
\end{equation}
Since $\alpha_{\mathrm{eff}}(m_Z)$ should be the same in the two schemes, 
we find the following relation between $\alpha_4$ and $\alpha_4(\Lambda)$
\begin{equation}
\frac{1}{\alpha _{4}}=\frac{1}{\alpha _{4}(\Lambda )}+\frac{2}{3\pi \delta
}\left( \sqrt{\pi }\frac{\Lambda }{M_{c}}\right) ^{\delta }~.
\label{match}
\end{equation}
If one identifies \(\Lambda  \) with the onset of a more complete theory
beyond the compactification scale, but at which the EFT treatment is still
valid, i.e. if one assumes that
\(\Lambda \sim M_G \ll M_{s} \), $M_G$ being this new scale,  Eq.(\ref{match})
could be reinterpreted as a matching equation
between the coupling \( \alpha _{4} \)  of our effective theory 
and the coupling
of the theory at scales \( M_G \), \( \alpha _{4}(M_G) \). 
Eq.~(\ref{match}) 
generically tells us that one expects 
corrections which go as \( \left( M_{G}/M_{c}\right) ^{\delta } \).
However, without knowledge of 
the full theory beyond \( M_{G} \), the meaning of
\( M_{G} \) (or even \( \alpha _{4}(M_{G}) \)) is unclear. In particular, if 
the new theory is some Grand Unified Theory in extra dimensions, 
\( M_{G} \) will be, in general, not just one single  
mass, but several
masses of the same order of magnitude, related by different coefficients. In
the case of logarithmic running those coefficients can be neglected, because
they give small logarithms 
next to the large logarithms containing the common scale. However,
in the case of contributions which depend on powers of the new physics scale
the situation is completely different, 
and the presence of several masses could change
completely the picture of unification. 
Cutoffs can give an indication of the
presence of power corrections, but the coefficients of these corrections cannot
be computed without knowing the details of the full theory. 

To see this point more clearly,
we add to our \( 4+\delta  \) dimensional theory an additional fermion 
with mass $M_f$ satisfying \( M_s\gg M_{f}\gg M_{c} \), 
such that  compactification corrections may be neglected,
and compute its effects on the coupling constant for 
\( M^2_{c}\ll Q^{2}\ll M^2_{f} \), using dimensional regularization. We have
\begin{equation}
\Pi ^{(\delta )}_{f}(Q)=\frac{e_{4}^{2}}{2\pi ^{2}}\left( \frac{\pi }{M_{c}}\right) ^{\delta /2}\Gamma (-\delta /2)\int ^{1}_{0}dxx(1-x)\left( M^{2}_{f}+x(1-x)Q^{2}\right) ^{\delta /2}\, .
\end{equation}
By expanding for \( Q^{2}\ll M_{f} \) and integrating over \( x \) 
we obtain
\begin{equation}
\Pi ^{(\delta )}_{f}(Q)=\frac{e_{4}^{2}}{2\pi ^{2}}\left( \sqrt{\pi }\frac{M_{f}}{M_{c}}\right) ^{\delta }\Gamma \left( -\frac{\delta }{2}\right) \left( \frac{1}{6}+\frac{\delta }{60}\frac{Q^{2}}{M_{f}^{2}}\right) \, .
\label{extraferm}
\end{equation}
 For odd values of \( \delta  \) we can use the analytic continuation of the
$\Gamma$ 
function to obtain a finite result. For even values of \( \delta  \)
we will allow a slight departure of the integer value in order to dimensionally
regularize the integral. 
Clearly, integrating out  
the heavy fermion  gives power corrections to the gauge
coupling. In addition, it 
also generates contributions 
to the higher dimension operators, e.g. 
contributions proportional \( Q^{2} \) and higher powers. 
As can be seen by comparing with Eqs.~(\ref{eq:PiCutoff})--(\ref{d3})
these power corrections are qualitatively similar 
to those calculated using a hard cutoff. 
Evidently, in the context of a more complete theory  
(in this case, given the existence of a heavy fermion), 
power corrections may be encountered 
even if the dimensional regularization is employed.
However, as one can easily see by setting, for example,  
$\delta = 1$ in Eq.(\ref{extraferm}),
the coefficients
of the power corrections obtained 
knowing the full theory 
are in general different from those 
obtained 
using a hard cutoff, 
e.g. Eq.(\ref{d1}). In fact, no choice of $\Lambda$ in 
Eq.(\ref{d1}) can reproduce all the coefficients appearing 
in Eq.(\ref{extraferm}). 

The situation is somewhat similar to what happens when $\chi PT$ with
$SU(2)\otimes SU(2)$ is matched to $\chi PT$ with $SU(3)\otimes SU(3)$.
In the $SU(2)\otimes SU(2)$ theory, just by dimensional arguments, one
can expect corrections like $m_K^2/f_\pi^2$. But, can one compute them
reliably without even knowing that there are kaons?

\section{Conclusions}

We have attempted a critical discussion of  
the arguments in favor of power-law running of coupling constants
in models with extra dimensions. We have shown that the naive arguments lead
to an arbitrary \( \beta  \) function depending on the particular way chosen
to cross KK thresholds. In particular, if one chooses 
the physical way of passing thresholds provided by the vacuum polarization 
function of the photon, a \( \beta  \)
function that counts the number of modes is divergent for more 
than \( 5 \) dimensions.

We have studied the question of decoupling of KK modes in QED with 4+ \( \delta  \)
(compact) dimensions by analyzing the behavior of the VPF of the photon. We
have computed first the imaginary part of the VPF by using unitarity
arguments,
and found that it rapidly reaches the value obtained in a non-compact theory
(only a few modes are necessary). We also showed that it grows as \(
(s/M^{2}_{c})^{\delta /2} \),
exhibiting clearly the non-renormalizability of theories in extra dimensions.
To obtain the full VPF, one can use an appropriately subtracted dispersion relation.
Instead, we use the full quantum effective field theory, with the expectation, 
suggested
by the calculation of the imaginary part of the VPF, that the bad UV behavior
of the theory is captured 
by the behavior of the uncompactified theory. To
check this idea, we have computed the VPF in the uncompactified theory,
regularized
by dimensional regularization (\( \delta \rightarrow \delta -\varepsilon  \)).
We have found that, after analytical continuation, the one loop VPF is finite,
and proportional to \( Q^{\delta } \) for odd number of dimensions, and has
a simple pole, proportional to \( Q^{\delta } \), for even number of dimensions.
This result can be understood easily, because there are no possible Lorentz and
gauge invariant operators in the 
Lagrangian able to absorb a term like \( Q^{\delta } \)
for odd \( \delta  \). For \( \delta  \) even it shows that higher dimension
operators are needed to regularize the theory. As a check we also recovered
the imaginary part of the VPF in the limit of 
infinite compactification radius.

For comparison with other approaches, we have also obtained the VPF in the case
that a hard cutoff is used to regularize it. We found that the pieces that do
not depend on the cutoff are exactly the same as those 
obtained by dimensional regularization,
while the cutoff dependent pieces are arbitrary, and can be changed at will by
changing the cutoff procedure.

Next we have computed the VPF in the compactified theory, and showed that it
can be separated 
into a UV and IR finite contribution and the VPF calculated in the 
uncompactified theory;  
as was shown previously, the latter 
can be controlled using dimensional regularization.
The finite part is more complicated, but can be computed numerically for any
number of dimensions. Also, some analytical approximations have been obtained
for the low and the high energy limits (\( Q\ll M_{c} \) and \( Q\gg M_{c} \)
respectively). Adding these two pieces, 
together with the counterterms coming from higher
dimension operators, we obtain 
a finite expression 
for an effective charge which can be extrapolated
continuously from \( Q\ll M_{c} \) to \( Q\gg M_{c} \); 
however, its value does depend 
on higher
dimension operator couplings.

Decoupling  of all KK  modes in  this effective  charge is  smooth and
physically  meaningful, and  the  low energy  logarithmic running  is
recovered.  We use  this effective  charge to  connect the  low energy
couplings (i.e. \( \alpha _{\mathrm{eff}}(m_{Z}) \)) with the coupling
of  the   theory  including  all   KK  modes,  regularized   by  dimensional
regularization. We find that  this matching only involves the standard
logarithmic running from \( m_{Z}  \) to the compactification scale \(
M_{c}  \).  In  particular, no  power corrections  appear  in this
matching.  However, if  cutoffs are used to regularize  the VPF in the
non-compact  space,  one  does  find  power  corrections,  exactly  as
expected from naive dimensional analysis.  In the EFT language one could
interpret  these corrections  as  an additional  matching between  the
effective  \( D  \) dimensional  field theory  and some  more complete
theory. The  question is how  reliably can this matching be estimated  
without knowing  the complete theory. By adding to our  theory 
an additional fermion with  \( M_{f}\gg  M_{c} \), 
and  integrating it  out, we
argue that  power corrections cannot  be computed without  knowing the
details of the complete theory, in which the \( D \) dimensional theory
is embedded. Some  examples in which this matching  can, in principle,
be        computed         are        some        5D GUT's and string  models
\cite{Ghilencea:1998st,Kakushadze:1999bb,Choi:2002wx,Hebecker:2002vm,PaccettiCorreia:2003en,PaccettiCorreia:2002xs},
and the recently
proposed            de-constructed           extra           dimensions
\cite{Arkani-Hamed:2001ca,Cheng:2001vd,Pokorski:2001pv,Chankowski:2001hz,Arkani-Hamed:2001vr}.     For   the
question  of  unification  of   couplings  this  result  seems  rather
negative,  at   least  when  compared  with   standard  
Grand Unified Theories,
where gauge coupling unification can be tested without knowing 
their details.
Alternatively, one  can approach this  result from a  more optimistic
point of view,  and regard the requirement of  
low-energy unification  of couplings as a stringent constraint 
on the possible extra-dimensional extensions of the SM.

\comment{

}%endcomment

\chapter{A model with a non-universal extra dimension}
\label{chap:HGLHG}
\comment{
{\sffamily \scshape
\begin{center}
\begin{tabular}{|l|r|}
\hline
Text         & Ok      \\\hline
Orthography  & Ok      \\\hline
Figures      & Ok      \\\hline
Links        & Ok      \\\hline
Cites        & Ok      \\\hline
Meaning      & --      \\\hline
Makindex     & --      \\\hline
Date         & \today  \\\hline
\end{tabular}
\end{center}
} } 

Before the universal extra dimensions were studied many other
scenarios had been proposed. These models were justified in some
string scenarios\cite{Arkani-Hamed:1998nn,Arkani-Hamed:1998rs1,Antoniadis:1990ew,Antoniadis:1994jp,Arkani-Hamed:1998vp,Dienes:1998sb,Rizzo:2001cy,Pilaftsis:1999jk}. Initially,
they were proposed to solve the hierarchy problem by introducing a new
scale near the electroweak scale, but their consequences extended to
many other fields: the value of the cosmological constant
\cite{Arkani-Hamed:2000eg,Sundrum:1997js}, supersymmetry breaking
\cite{Antoniadis:1993fh}, fermion masses
\cite{Bando:2000it,Ioannisian:1999cw}, etc...  These models are
radically different with respect to the universal ones because they do
not conserve the fifth component of the momentum, hence the KK number
is no longer conserved.  As a consequence, the bound on this new
physics is set above the TeV. In what follows we will study a model in
which only the Higgs and gauge bosons propagate through one continuous
extra dimension; the latticized version is also considered. We show
briefly the main features as well as how to study the phenomenology by
using the formalism developed in the preceding chapters.

\section{One continuous extra dimension}
\subsection{The model}
This case can be straightforwardly studied by retracing the steps done
in Chapter~\ref{sec:SMinUED}. The Lagrangian has the same pieces,
\Eq{eq:primera}, and the Eqs.~(\ref{eq:segunda} - \ref{eq:yukdefUED})
remain the same. The topology of the extra dimension is an orbifold,
$S^1/Z_2$, but now only the Higgs and gauge fields propagate through
the bulk, this is the main feature of this model, that is why we will
refer to it as \emph{HG}.  The expansions in \Eq{eq:uno} and
\Eq{eq:two} are still valid while $Q$ and $U$ do not present now any
mode.

Upon compactification, we find that the spectrum for the fermions is
exactly the same as in SM. On the other hand, there are again two
towers of scalars, $\Phi_G^{(n)}$ and $\Phi_P^{(n)}$, one unphysical
that is eaten to give mass to the tower of gauge bosons and the other
physical with mass $m_n$. In particular the conclusion about the
dominant contributions coming from $\Phi_P^{(n)}$ running inside the
loop is again valid.

The absence of extra modes for the fermions has as a first consequence
that the coupling between the fundamental modes of the gauge bosons
and the fermionic part of the theory, $\mathcal{L}_\rho$, \Eq{Lagrho}
is exactly the same as in SM. Important differences appear, though, in
the Yukawa piece of the Lagrangian
\begin{equation}
\label{eq:YukawaHG} \mathcal{L}_Y = \int_0^{\pi R}dx^5
(-\overline{Q} \widetilde{Y}_u H^c U - \overline{Q}
\widetilde{Y}_d H D + \hc)\delta(x^5),
\end{equation}
where the delta stands to force the interactions only in the brane
defined by the condition $x^5=0$, this brane is taken to be a rigid
one, therefore it breaks translation invariance in the fifth
dimension. Due to this, the fifth component of the momentum is not
conserved and the KK-number conservation (as well as all its
implications) no longer applies. Plugging the expansion of the Higgs
doublet, \Eq{eq:uno}, in \Eq{eq:YukawaHG}, one obtains the expression
of the Yukawa couplings in this scenario
\begin{equation}
\label{eq:higgscouplingshg} \mathcal{L}_Y = \frac{2}{v} m_t V_{tj}
\overline{t} P_L d_{j} \Phi^{+(n)}+ \hc ,
\end{equation}
where we have already particularized for the $m_t$ proportional
couplings, $t$ and $d_j$ are the usual spinor fields of the top and
down-like quarks. Note the presence of the extra $\sqrt{2}$ factor
with respect to the UED and SM case \Eq{eq:higgscouplings}, it is
present because in the expansion for $H$ the zero-th mode and the rest
have a $\sqrt{2}$ factor of difference. In UED it was absorbed by the
tower of modes of $U$ in \Eq{eq:higgscouplings} that here are
substituted by the delta function.

All the part that has to do with SSB and the zero gauge modes, ranging
from \Eq{eq:Higgspot} to \Eq{eq:SSBlast}, goes exactly the same
way. The couplings with the $Z$ boson can be found by using the same
procedure as in the UED case and the outcome shows that the current
that couples to $Z$ is the same as in UED, i.e. \Eq{eq:couplingZ},
with the sole difference of the lacking of the $J^{\mu(n)}$ piece
\begin{equation}
\mathcal{L}_Z = \sum_{n=1}^{\infty}\frac{g}{2c_w}Z_\mu^{(0)}[J^{\mu(0)} +
J^{\mu(n)}_{\Phi}].
\end{equation}
$J^{\mu(n)}_{\Phi}$ is the same as in UED, given by
\Eq{eq:couplingZHiggs}, it does not get any extra factor due the
presence of two Higgs fields in the current.

\subsection{Bounds}
To know which is the minimum possible value of the scale of this model
we will  use experimental determinations of some  observables, some of
them are  the same as in the  universal scenarios but as  we will show
the tightest restrictions  come from the modification of  the value of
$G_F$, that  is modified already  at tree level. Modifications  at the
tree-level are now possible because  of the non-conservation of the KK
number. As usual we will define  $a=m_t \pi R$ and denote the lightest
KK mode by $M = R^{-1}$.
%, in this case, HG, it happens to be $M=R^{-1}$.

%%%%%%%%%%%%%%%%%%    $rho$ parameter
\subsubsection{The $\rho$ parameter}
The corrections to the $\rho$ parameter proportional
to the $m_t$ mass are the same as in SM, essentially because the
Higgs field propagates in the fifth dimension while the fermions do
not.
%\comment{\cite{Marciano:1999ih, Masip:1999mk, Rizzo:1999br,
%Carone:1999nz}}: 
Since the top quark has no KK tower associated, the only possible
source of terms proportional to $m_t$ are the Yukawa couplings, but
these do not contribute at one-loop to the self-energies of the $Z$
and $W$ masses.

%%%%%%%%%%%%%%%%% Z --> bb 
\subsubsection{Radiative corrections to the $Z\to b\overline{b}$ decay}
The contributions to this process will be parametrized using the
$F(a)$ function defined in \Eq{eq:deffa}. The diagrams that contribute
can be obtained from the diagrams in \Fig{fig:ZbbinUED} replacing the
fermionic modes inside the loops by the lines of the SM fermions
\begin{equation}
\label{eq:thechangeHG}
Q_t^{(n)}\to t_L \qquad  U^{(n)}\to t_R.
\end{equation}  
The result turns out to be \cite{Papavassiliou:2000pq}
\begin{equation}
F(a)=-1+2a\int_0^\infty\;dx \frac{x^2}{(1+x^2)^2} \coth(ax)\approx
\left(\frac{2}{3}\log(\pi/a)-\frac{1}{3}-\frac{4}{\pi^2}\zeta^\prime(2)\right)
a^2
%F(a) = 2a\int_0^\infty\;dx \frac{x^2}{(1+x^2)^2} \coth(ax)\approx
%1-\left(\frac{1}{3}+\frac{4}{\pi^2}\zeta^\prime(2)-\frac{2}{3}\ln\pi\right)
%a^2 - \frac{2}{3}a^2\ln(a) .
\end{equation}
In the expansion can be found a logarithm that relates the two scales:
$m_t$, the only mass we maintain in the SM and $M$, the mass of the
first KK mode. It is because the KK-number is not conserved, what
permits in this kind of model the appearance of effective operators at
the tree level that modify the decay at the one-loop level in the
effective theory\footnote{For a more detailed explanation see
Ref.~\cite{Papavassiliou:2000pq}.}. Using the bound $F(a)\leq 0.39$ at
$95 \%~\mbox{CL}$ derived in \Sec{sec:Zbb}, one can find the next
bound to $M$
\begin{equation}
M=R^{-1}_{HG}> 1\;\mbox{TeV}\qquad 95\%\;\mbox{CL}.
\end{equation}

%%%%%%%%%%%%%%%%%  b --> s photon
\subsubsection{Radiative corrections to $b\to s \gamma$}
\begin{figure}
  \begin{center}
    \includegraphics[scale = 0.8]{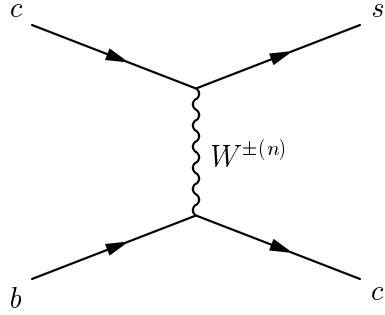}
  \end{center}
\caption{The virtual exchange of the KK modes $W^{\pm (n)}$ modifies
  the value of $C_2$ with respect to the SM value.}
\label{fig:c2}
\end{figure}

The contributions in this scenario came from diagrams that can be
obtained from the ones in \Fig{fig:bsginUED} by performing the change
of \Eq{eq:thechangeHG}. In addition, each vertex between the fermions
and the Higgs modes (the ones dotted in the figure) get an extra
$\sqrt{2}$ factor, and as a consequence all diagrams get an overall
factor of $2$. The contribution per mode to the value of the $C_7$
coefficient defined in \Eq{eq:effhamiltonian} when also the new
spectrum is taken into account is
\begin{equation}
\label{eq:c7inHG} C_{7}^n=2\left[ B\left(
\frac{m_t^2}{m_n^2}\right) - \frac{1}{6} A \left(
\frac{m_t^2}{m_n^2} \right)\right]
\end{equation}
and the value of $C_7$ in this model is therefore
\begin{equation}
\label{eq:c7totalhg}
C_7^{\mathrm{HG}}(M_W)  =  C_7^{\mathrm{SM}}(M_W)+\sum_{n=1}^\infty
C_{7}^n(M_W) \approx -0.195+0.265
a^2-\frac{2}{9}a^2\ln(a).
\end{equation}
It is easy to check that $C_7^n$ vanishes for each mode when its mass
is taken to infinity, as it should be because of the decoupling
theorem. Notice again the presence of the logarithm relating the two
scales.  But what we need, is $C_7^{\mathrm{HG}}(m_b)$, that can be
obtained by using \Eq{eq:running} provided we know
$C_2^{\mathrm{HG}}(M_W)$. This receives contributions at tree level
from the virtual exchange of $W$ modes that could in principle modify
appreciably its value, see \Fig{fig:c2}.
Although the full tower must be considered, the modification to the SM
result happens to be finite and small
\begin{equation}
C_2^{HG}(M_W)=1-\frac{M_W^2R^2\pi^2}{3},
\end{equation}
which represents a modification at the level of 2 \%, much lower than
the error we are committing ignoring the contribution of the
$W^{(n)}_\mu$ and $W^{(n)}_5$ fields inside the loops.  The process
$b\to cl\nu$ will also affected at the tree level by an amount of the
same order, therefore we will again ignore it. In addition,
%$C_8^{HG}(m_t)\approx-0.014$, 
$C_8^{HG}(M_W)$ will be of the same order and its contribution will be
neglected. When experimental bound on the value of $C_7(m_b)$ is used,
\Eq{eq:boundonc7}, one can find the next bound for $M$,
Ref.~\cite{Oliver:2003cc}
\begin{equation}
R^{-1}_{HG}> 1.2~\mbox{TeV} \qquad 95 \%\;\mbox{CL}.
\end{equation}

%%%%%%%%%%%%%%%%%%%%%%%%   B0 - B0 system
\subsubsection{Radiative corrections to the $B^0 - \overline{B}^0$ system}
This case is fully studied in Ref.~\cite{Oliver:2003cc} were all the
details are given. The diagrams can be obtained from the ones in
\Fig{fig:BBmixinUED} with the same modifications as in the previous
sections and also the correspondent factors have to be considered. The
contributions to $S(x_t)$ can be parametrized in $G(a)$, defined in
\Eq{eq:defga}
\begin{equation}
G_{\mathrm{HG}}(a) = 2 a^2 \int_0^\infty dx \frac{x^3}{(1+x^2)^2}\coth^2(ax)
\approx 1 -1.143 a^2-\frac{4}{3} a^2\log(a)+ 2 a^2 \ln(n_s),
\end{equation}
where $a= m_t \pi R$ and $n_s=\Lambda_s/M$. $\Lambda_s$ comes from the
cut-off  that  has  been  performed  in  the sum  of  the  modes,  see
Ref.~\cite{Papavassiliou:2000pq}. $\Lambda_s$  is the scale  where the
scale  where the  more  complete  theory starts  to  be important.

Again the effective field theory  point of view helps us to understand
the results: the reason for the presence of $\ln(a)$ is the same as in
the previous  observables.  There is  now a new $\ln(n_s)$  term, that
cames from cutting the summation on  the KK modes, which at the end is
just cutting  the integration of  the fifth component of  the momentum
from the point of view of the five dimensional theory. The presence of
this logarithm implies that a  full calculation in the theory in which
HG is supposedly embedded contains these logarithmic pieces. But now,
we cannot directly check this because the details of the corresponding
full theory are unknown.

Finally,    using   the    last    experimental   determinations    of
Ref.~\cite{Buras:2002yj,Bergmann:2001pm}    the    bounds    can    be
extracted. The one coming from this observable is
\begin{equation}
R^{-1}_{HG} > [560,900]~\mbox{GeV} \qquad 95\%\;\mbox{CL},
\end{equation}
the  variations are  present because  we have  to choose  a  value for
$n_s$. We have taken $n_s=10$ and $n_s=100$ respectively.

\comment{
Conversely if one takes the
bounds coming from the other observables, specially the $\rho$
parameter, one would find big contributions to the $S(x_t)$
function compatible with experiments. That would be hard to
explain within the SM, in other words, a signature of the extra
dimensions would be an appreciable disagreement between the
experimental value and the SM prediction to $S(x_t)$ and the
absence of such disagreements for the other observables.

In the case of the latticized scenarios this system gives the
poorest bounds of the ones studied here. For the sake of
completeness we have written down the expressions for the
correspondent $S(x_t)$ functions in \rf{eq:sxtlued} and
\rf{eq:sxtlhg} but we do not show these bounds in the plot.}

%%%%%%%%%%%%%%%%%%%%%%   Z --> l l
\subsubsection{$\Gamma( Z \rightarrow \ell \overline\ell )$ 
restrictions. $G_F$ modifications.}  In UED and LUED this observable
receives contributions at the one loop level and therefore it will be
dominated by the SM tree level contributions, that is why it was not
studied in those cases. On the contrary in HG (and later on in LHG) it
receives contributions already a the tree level and could be
substantially modified, see Ref.~\cite{Masip:1999mk,Nath:1999fs}. In
the presence of KK modes, the relation between the coupling constant
$g$, $M_W$ and the Fermi constant as extracted from the muon lifetime,
$G_F$, is modified at tree level due to the virtual exchange of the KK
tower of $W^{(n)}$, \Fig{fig:gfmodifs}:
\begin{figure}
  \begin{center}
    \includegraphics[scale = 0.8]{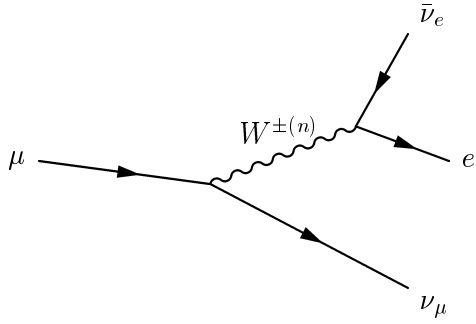}
    \caption{The decay of the muon through virtual KK modes modifies
    the value of $G_F$.}
    \label{fig:gfmodifs}
  \end{center}
\end{figure}
\begin{equation}
\frac{G_F}{\sqrt{2}}=\frac{g^2}{8 M_W^2}+\sum_{n=1}^\infty
\frac{(\sqrt{2}g)^2}{8 m_n^2} =\frac{g^2}{8 M_W^2}\left[ 1+\frac{(M_W \pi
R)^2}{3} \right],
\end{equation}
while for  SM the  relation can  be obtained by  setting $R=0$  in the
previous formula. The fact that  this relation is different for HG and
SM    has    as   a    consequence    that    the   predictions    for
$\Gamma(Z\rightarrow\ell\overline{\ell})$, when  expressed in terms of
$G_F$,       are      different\footnote{The       prediction      for
$\Gamma(Z\rightarrow\ell\overline{\ell})$ when written in terms of $g$
is exactly the same for HG and SM
\begin{equation}
\Gamma(Z\rightarrow\ell\overline{\ell}) = \frac{M_Z}{24
  \pi}\frac{g^2}{c_W^2}(g_R^2 + g_L^2).
\end{equation}
It  is the  different  relationship  between $g$  and  the low  energy
parameter $G_F$ what causes the discrepancy.}
\begin{equation}
\Gamma^{\mathrm{HG}}=\Gamma^{\mathrm{SM}}\left[1-\frac{(M_W \pi R)^2}{3} \right].
\end{equation}
Using the experimental bound \cite{Hagiwara:2002pw} $
%\begin{equation}
\left|\frac{\Gamma^{\mathrm{exp}}}{\Gamma^{SM}}-1 \right| < 0.0028$ at $95\% \;\mbox{CL}
%\end{equation}
$ one gets
\begin{equation}
R^{-1}_{HG}> 2.8~\mbox{TeV} \qquad\qquad 95\%\; \mbox{CL}.
\end{equation}
Including  radiative  corrections  reduces  a  bit  this  number,  see
Ref.~\cite{Masip:1999mk}.

Summarizing, the results of all the observables are collected in table
\ref{HGtable}, where it is clearly shown that the scale of the new
physics in this scenario is above the TeV.

\begin{table}
 \begin{center}\begin{tabular}{|c|c|c|c|c|}
            \hline
            \  &$Z\to b \overline{b}$ & $b \to s + \gamma$ &
            $\overline{B}^0-B^0$ & $G_F$\\
            \hline
            \mbox{M(TeV)}  & 1.0  & 1.2 & [0.5,0.9] & 2.8 \\
            \hline   
 \end{tabular}\end{center}
\caption{Bounds coming from the different observables to the mass of
  the first KK mode $M$.}
\label{HGtable}
\end{table}

%%%%%%%%%%%%%%%%%%%%%%%%%%%%%%%%%%%%%%%%%%%%%%%%%%%%%%%%%%%%%%%%%%%%
%%%%%%%%%%%%%%%%%%%%%%%%%%%%%%%%%%%%%%%%%%%%%%%%%%%%%%%%%%%%%%%%%%%

\section{One latticized extra dimension}
\subsection{The model}
The steps are nearly the same as in the previous section. It can be found
that the spectrum is the same provided one changes $m_n\to M_i$, defined in
\Eq{eq:massdefinition}, and considers the finiteness of the spectrum, i.e.
$i=0,1,\ldots,N-1$.

Again for the $\rho$ parameter the $m_t$ proportional corrections are the same
as in SM. The Yukawa sector is the
usual one, but with the fermions coupled only to the zero-th copy of
the Higgs doublet
\begin{equation}
\mathcal{L}_Y = -\overline{Q_L}\widetilde{Y}_u H_0^{c} u_R + \hc,
\end{equation}
%where we have concentrated only in the terms with $\widetilde{Y}_u$
%and neglected the ones with $\widetilde{Y}_u$, substituting here
%\rf{eq:basechange} one gets
%
%\begin{eqnarray}
%\label{eq:YukHG}
%\mathcal{L}_Y& =& -\overline{u_L}Y_u
%\widetilde{H}^{c\uparrow}_0 u_R-\overline{d_L}Y_u
%\widetilde{H}^{c\downarrow}_0 u_R + \hc\\
%\label{eq:estamismo}
% &-& \sqrt{2}
%\sum_{n=1}^{N-1}\overline{d_L}Y_u\cos\left(\frac{n \pi}{N 2}\right)
%\widetilde{H}^{c\downarrow}_n u_R + \hc\\ &-&\sqrt{2}
%\sum_{n=1}^{N-1}\overline{u_L}Y_u\cos\left(\frac{n \pi}{N 2}\right)
%\widetilde{H}^{c\uparrow}_n u_R + \hc
%\end{eqnarray}
%
%where we have used $Y_u=\widetilde{Y}_u/\sqrt{N}$ because we are yet
%identifying $\widetilde{H}_0$ with the SM Higgs doublet. Observe that
%under the limit of large $N$ one recovers the HG continuum result
%\cite{Papavassiliou:2000pq}. When the VEV of $\widetilde{H}_0$ is
%considered a mass term for the up quarks is generated in the first
%term of \rf{eq:YukHG}, exactly the one of SM, performing the usual
%change of base and concentrating on the $m_t$ proportional terms that
%couple the charged component of the Higgs doublet \rf{eq:estamismo},
%because these are the relevant ones for our observables, one gets
and by resorting to the mass fields we find the important part for us
\begin{equation}
\mathcal{L}_Y = \sum_{n=1}^{N-1}\frac{2 m_t}{v}V_{tj}\cos\left(\frac{n \pi}{2N}\right)\overline{t}
\widetilde{\Phi}_n^+ P_L d_{j} + \hc
\end{equation}
%where we have used
%$\widetilde{H}^{c\downarrow}_n=\widetilde{\Phi}_n^+$. And
The $\widetilde{\Phi}_n^+$ are the physical degrees of freedom. Notice that the
coupling is the SM one with the addition of an extra $\sqrt{2}$ factor
and the presence of a cosine that vanishes in the limit of large
$N$. The cosine appears because there is only one Higgs field. Finally the
couplings with the $Z$ boson can be obtained 
straightforwardly
% . These should be derived from the 4D covariant
%derivative \rf{eq:covderi}, which we can safely use again here because
%this definition is independent of whether the fermions propagate or
%not in the bulk. The steps are: express the gauge fields in terms of
%the mass definite gauge fields through \rf{eq:basechange}, then
%particularize for the zeroth modes $\widetilde{W}_{\mu,0}^3$ and
%$\widetilde{B}_{\mu,0}$, resort to the base of $Z_\mu$ and $A_\mu$
%\rf{eq:weakangle} and concentrate on the $Z$ piece, now express the
%fields $H_i$ in terms of the mass definite ones $\widetilde{H}_i$
%using again \rf{eq:basechange} and finally concentrate on the charged
%component
\begin{equation}
\label{eq:lcouplingZHiggs}
\mathcal{L}_Z =
\frac{g}{2c_w}Z_\mu \sum_{i=1}^{N-1}
%J^{\mu}_i= (-1
(-1+2s_w^2)[\widetilde{\Phi}^+_i i\partial^\mu\widetilde{\Phi}^{-}_i]
+ \hc
\end{equation}
The absence of any cosine is because there are two Higgs fields. Thus
the couplings of the $Z$ are the same as for HG and the differences
are encoded in the spectrum.
%$
%\begin{equation}
 % \begin{array}{l}
%    m(\widetilde{\Phi}_n^{\pm})
    %=M_n \qquad n=1,\ldots,N-1
 % \end{array}
%\end{equation}
%$
With this we are prepared for extracting the different
contributions.

\subsection{Bounds}
To extract the bounds, Ref.~\cite{Oliver:2003cc}, we can use the
results derived for HG. The corresponding results can be found from
the ones in HG by adding an extra factor of $\cos^2(n\pi/2N)$. The
cosines appear only in the Yukawa vertices, and in every diagram that
we consider there are always two of these vertices. In the case of the
the radiative corrections to the $Z\to b\overline{b}$ decay the result
is
\begin{eqnarray}
F_{\mathrm{LHG}}(a) & = & 1 + \int_0^1 dx
\sum_{n=1}^{N-1}\frac{2(1-x)r_n\cos^2 ({n \pi}/{2 N})}{(1-x)r_n+x}\\
&=&1+\int_0^1 dx \sum_{n=1}^{N-1}\frac{2(1-x)a^2\cos^2 ({n \pi}/{2
N})}{(1-x)a^2+4(N-1)^2 x\sin^2 ({n \pi}/{2 N})}
\end{eqnarray}
from the bound $F(a)<0.39$ the bounds on $M$ can be extracted as a
function of $N$. 

\begin{figure}
  \begin{center}
    \includegraphics[scale = 1]{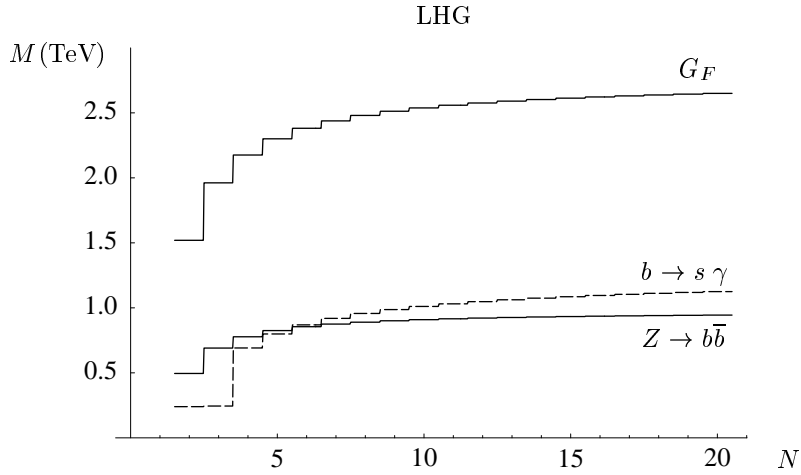}
    \caption{The bounds on $M$ as a function of the number of sites, $N$,
      in the LHG scenario.}
    \label{fig:LHGresults}
  \end{center}
\end{figure}

The radiative corrections to $b\to s \gamma$ can be easily found from
the HG ones
\begin{equation}
C_7^{\mathrm{LHG}}(M_W)=C_7^{\mathrm{SM}} + 2 \sum_{n=1}^{N-1}
\cos^2(n\pi/2N)C_{7}^n,
\end{equation}
where $C_{7}^n$ is defined in \Eq{eq:c7inHG}. The function that
parametrizes the corrections to the $B^0 - \overline{B}^0$ system is
now
\begin{equation}
G_{\mathrm{LHG}}(a) = 1+2\int_0^1\sum_{n=1}^{N-1}\frac{2 a^2 x
(1-x)\cos^2(n\pi/2N)\;dx}{4 (N-1)^2 \sin^2(n\pi/2N)(1-x)+a^2 x}.
\end{equation}
Finally, the branching ratio $\Gamma(Z\rightarrow\ell\overline{\ell})$
is modified to
\begin{equation}
 \Gamma^{\mathrm{LHG}}=\Gamma^{\mathrm{SM}}\left[1-\frac{1}{2} \left(\frac{M_W \pi
 R}{N-1}\right)^2 \sum_{n=1}^{N-1}\frac{\cos^2(n\pi/2N)}{\sin^2(n\pi/2N)}
 \right].
\end{equation}
With these expressions it is possible to extract bounds to the first
mode as a function of the number of sites $N$. These bounds are displayed
in \Fig{fig:LHGresults}, where it is shown that the continuous limit
is rapidly reached. 
If the number of sites is small, the bound can be reduced by a factor
of about 20\%-40\%, thus allowing new modes ranging between
$1.5-2.0\;\mbox{TeV}$.

%%%%%%%%%%%%%%%%%%%%%%%%%%%%%%%%%%%%%%%%%%%%%%%%%%%%%%%%%%%%%%%%%%%
%Tue Dec  2 09:08:44 CET 2003

\chapter{Outlook and conclusions}

The possibility of the existence of more dimensions different than
those we can experience directly with our senses has been for a long
time a very active field of interest for physicists. They were
initially introduced to achieve a more elegant formulation of the
equations of nature, soon the extra dimensions became a key piece in
important theories like supergravity or string theory, which tried to
give a quantum treatment to gravity. It was realized in the latter
that, under some assumptions, the extra dimensions could be of the
size as low as the TeV without contradicting any of the experimental
results. Extra dimensions have been also studied using the quantum
field theory formalism, and many extensions of the SM have been
proposed. In general, the scale associated with the extra dimensions
in these extensions is again of the order of TeV, what offers a
relatively accessible, and rich, phenomenology for future experiments.
Placing a new scale at the TeV has important consequences: it could
solve the hierarchy problem, scenarios of grand unification are
modified because the running of the coupling constants is greatly
modified above this scale, the possible localization of the wave
functions for the different fermions in the space of the extra
dimensions could be the origin of the fermion masses, even the value
of the cosmological constant could be modified, just to cite
some. This variety of phenomena explains the interest in extra
dimensional theories in the last years.

Among the above mentioned extensions of the SM, there is one of
special interest in which a single extra dimension is considered and
all the fields propagate through it, a fact that gives it the name
``\emph{universal extra dimension}''. In this scenario there are no
tree-level modifications to the SM results, therefore all the
corrections are, at least, one-loop suppressed. This outcome is a
consequence of the \emph{local} extra-dimensional Lorentz symmetry of
the theory, which implies the existence of a ``conserved'' number
called Kaluza-Klein number. Due to this suppression in the SM
modifications, the scale of the new physics can be as low as hundreds
of GeV, clearly a challenging situation for the next generation of
accelerators.

In the first part of this work we have studied in detail the
phenomenology of this scenario. In particular, we have concentrated on
observables that display a strong dependence on the mass of the
top-quark, $m_t$, because this enhances the corrections coming from
the new physics: $Z\to b \overline{b}$, $\rho$ parameter and
$B^0-\overline{B}^0$ mixing. We have also studied the process $b\to s
\gamma$; although it does not present a strong dependence with $m_t$,
the relative importance of the new physics is also enhanced because
the SM contribution to this transition is suppressed. This is so
because this process is forbidden at tree level due to gauge
invariance and is only allowed through radiative corrections. The
details of the study are given in Chapter \ref{sec:SMinUED} and the
bounds on the size of the compactification scale or, equivalently, on
the mass of the first Kaluza-Klein mode are given in table
\ref{thetable}.

We have also studied a non-universal scenario, in which only the Higgs
and gauge bosons of the SM are allowed to propagate through the extra
dimension, we have given this scenario the name of HG. We have
calculated the new physics contributions to several one loop processes
which depend strongly on the top-quark mass and used them to set
bounds on the mass of the lightest KK mode, $M$. The results are
summarized in table \ref{HGtable} and compared with other relevant
bounds found in the literature. 

In the HG case the new physics scale should be above 1 TeV but
surprisingly in the UED case it can be well below the TeV. One might
think that the scale for UED should be bigger than for HG since UED
represents a bigger modification to SM than HG (because in HG only
Higgs and gauge bosons propagate in the extra
dimension). Nevertheless, the contributions of the fermionic KK modes
in the UED case are such that produce several cancellations in the
amplitudes. The reason is the above mentioned KK number conservation
due to the local Lorentz symmetry. The lack of this symmetry in the HG
case permits some observables to be modified already at the tree
level; as a consequence the most restrictive bound comes from a very
well measured quantity, $G_F$.

These extra dimensional models are non-renormalizable because the
coupling constants have canonical dimensions of mass to some negative
power. This does not mean at all that these models should be wrong, on
the contrary, this indicates that they must be regarded as effective
field theories that are valid only below some scale at which new
physics should enter to correct the bad behaviour of the theory. There
is a class of string theories which could provide such a new physics,
however, this is only one among the possible ultraviolet
completions. Another kind of completion, proposed initially in
Ref.~\cite{Arkani-Hamed:2001ca,Hill:2000mu,Cheng:2001vd}, used a
discretization or \emph{lattization} of the fifth dimension. In these
models, the number of modes is finite and their masses are
modified with respect to the continuous cases. As a possible extension
of the SM, it is also interesting to study the phenomenology that
stems from these scenarios. In this work we have studied the
latticized versions of UED and HG, called LUED and LHG. Among the
results we obtained, we found that the predictions are the same as in
the respective continuous cases when the extra dimension is latticized
in a relatively small number of (four dimensional) sites. For a very
small number, the results show that in all cases the natural scale
could be further lowered with respect to the continuous scenarios what
makes more accessible the new physics for future experiments.
%\begin{figure}
%  \begin{center}
%    \begin{displaymath}
%      \begin{array}{cc}
%	\includegraphics[scale = 0.7]{LUED/graphresultsLUED.eps} &
%	\includegraphics[scale = 0.7]{HG_LHG/graphresultsLHG.eps}
%      \end{array}
%    \end{displaymath}
%    \caption{The plots show the different bounds to the scale of the
%    new physics when one lattiziced extra dimension is present. The
%    one in the left stands for LUED and the one on the right for LHG.}
%    \label{fig:resultslattice}
%  \end{center}
%\end{figure}
The scale can be reduced by factor of about 10\%-25\%, thus allowing
new modes ranging between $320-380\;\mbox{GeV}$ in the case of UED and
of about 20\%-40\%, with the new modes ranging between
$1.5-2.0\;\mbox{TeV}$ in the case of LHG. The details can be found in
\Fig{fig:results} and \Fig{fig:LHGresults}.

On the other hand, it has been suggested that the existence of extra
dimensions could make the gauge couplings to run as a power of the
energy scale instead of the usual logarithmic running.  This is a very
interesting property because Grand Unification Theories in extra
dimensions could achieve unification at much lower energies (few TEV)
than the usual four dimensional GUT's. This would allow us to test
unification at the planned accelerators, clearly a very exciting
possibility. This scenario is also challenging from the theoretical
point of view because gauge couplings in extra dimensions are
dimensionful and gauge theories in extra dimensions are not
renormalizable. Therefore, the behaviour of couplings can depend on
the approach to renormalization of non-renormalizable theories:
Wilsonian effective field theory or continuum effective theory. We
have attempted a pure continuum effective field theory approach based
in dimensional regularization.  In this approach the running of
couplings, at most, logarithmic and power corrections are recovered as
matching conditions at the (extra-dimensional) GUT scale. Therefore, to
compute them one needs to know the complete details (full spectrum and
pattern of symmetry breaking) of the GUT model. This is quite
different from the usual four-dimensional GUT's that can be tested to
good accuracy without knowing the details of the GUT model.

To conclude, we have studied the phenomenology of two
extra-dimensional models with all or only part of the SM particles
propagating in the extra dimensions. We have also studied their
latticized versions. We have focused on one-loop effects that display
a strong dependence on the top-quark mass which contribute to
observables which are measured with good precision. For the models
considered, those effects provide the best limits on the
compactification radius.  We have also studied the question of
power-law running in extra dimensions by using the continuum effective
field theory framework with dimensional regularization and discussed
its impact in (extra-dimensional) GUT's.  We found that, at difference
from the four-dimensional case, unification can depend strongly on the
details of the GUT model if it is dominated by power corrections.

\comment{
of a number of different
extra-dimensional scenarios. We have focused on observables that
display a strong dependence on the mass of top because we expect them
to be the first place where extra dimensions can be found. In
addition, we have also studied a more theoretical issue, as the
running of the coupling constants in this type of theories. To this
end, we have applied the effective field theory machinery and found
that its use is very useful when trying to disentangle many of the
subtleties related with the presence of extra dimensions in nature.
}%endcomment

%endcomment

% LocalWords:  renormalizable latticized lattiziced LHG renormalizability VPF
% LocalWords:  KK

%%%%%%%%%%%%%%%%%%%%%%%%%%%%%%%%%%%%%%%%%%%%%%%%%%%%%%%%%%%%%%%%%%%%%%%
%%%%%%%%%%%%%%%%%%%      BIBLIOGRAPHY AND INDEX     %%%%%%%%%%%%%%%%%%%
%%%%%%%%%%%%%%%%%%%%%%%%%%%%%%%%%%%%%%%%%%%%%%%%%%%%%%%%%%%%%%%%%%%%%%%
%\thebibliography
\bibliographystyle{unsrt}%alpha
\bibliography{articles}
\thispagestyle{empty}
\printindex
\end{document}